\documentclass[Crown,sagev,times]{sagej}

\usepackage{moreverb,url}
\usepackage{float} 

\usepackage[colorlinks,bookmarksopen,bookmarksnumbered,citecolor=red,urlcolor=red]{hyperref}
\usepackage{dsfont}
\usepackage{amssymb}
\usepackage{amsmath}
\usepackage{mathtools}
\usepackage{epstopdf}
\usepackage{epsfig}
\usepackage{graphicx}
\usepackage{enumerate}
\usepackage[english]{babel}
\usepackage{booktabs}
\usepackage{pdflscape}
\usepackage{natbib} 
\usepackage{microtype} 
\usepackage{subcaption}
\usepackage{rotating} 
\usepackage{dblfloatfix}
\usepackage{xcolor}
\definecolor{ao(english)}{rgb}{0.0, 0.5, 0.0}

\newcommand{\R}{\mbox{$\mathbb{R}$}}

\newcommand{\vX}{\boldsymbol{X}}
\newcommand{\vZ}{\boldsymbol{Z}}
\newcommand{\vA}{\boldsymbol{A}}
\newcommand{\vD}{\boldsymbol{D}}
\newcommand{\vSigma}{\boldsymbol{\Sigma}}
\newcommand{\whsigma}{\widehat{\sigma}}
\newcommand{\bqan}{\begin{eqnarray}}
\newcommand{\eqan}{\end{eqnarray}}

\newtheorem*{assumption*}{\assumptionnumber}
\providecommand{\assumptionnumber}{}
\makeatletter
\newenvironment{assumption}[2]
{%
	\renewcommand{\assumptionnumber}{Assumption #1$\mathcal{#2}$}%
	\begin{assumption*}%
		\protected@edef\@currentlabel{#1$\mathcal{#2}$}%
	}
	{%
	\end{assumption*}
}
\makeatother

\theoremstyle{plain}
\newtheorem{prop}{Proposition}[section]
\newtheorem{thm}{Theorem}[section]

\newcommand\BibTeX{{\rmfamily B\kern-.05em \textsc{i\kern-.025em b}\kern-.08em
T\kern-.1667em\lower.7ex\hbox{E}\kern-.125emX}}

\begin{document}
\definecolor{ao(english)}{rgb}{0.0, 0.5, 0.0}
\definecolor{bronze}{rgb}{0.8, 0.5, 0.2}
\definecolor{byzantine}{rgb}{0.74, 0.2, 0.64}

\runninghead{Amro, Pauly, and Ramosaj}

\title{Asymptotic based bootstrap approach for matched pairs with missingness in a single-arm}

\author{Lubna Amro\affilnum{1}, Markus Pauly\affilnum{1} and Burim Ramosaj\affilnum{1} }

\affiliation{\affilnum{1}Mathematical Statistics and Applications in Industry, Faculty of Statistics, Technical University of Dortmund, Germany}
\corrauth{Lubna Amro, Mathematical Statistics and Applications in Industry, Faculty of Statistics, Technical University of Dortmund, Germany.}

\email{lubna.amro@tu-dortmund.de}

\begin{abstract}
 The issue of missing values is an arising difficulty when dealing with paired data. Several test procedures are developed in  the literature to tackle this problem. Some of them are even robust under deviations and control type-I error quite accurately. However, most these methods are not applicable when missing values are present only in a single arm. For this case, we provide asymptotic correct resampling tests that are robust under heteroscedasticity and skewed distributions. The tests are based on a clever restructuring of all observed information in a quadratic form-type test statistic. An extensive simulation study is conducted exemplifying the tests for finite sample sizes under different missingness mechanisms. 
In addition, an illustrative data example based on a breast cancer gene study is analyzed.

\end{abstract}


\maketitle

\section{Introduction}
\label{Intro}

Conducting statistical tests on units measured repeatedly requires the consideration of the dependence structure of the resulting random vector. The simplest design is the matched pairs model, where units are measured at two endpoints of the same subject. This design has experienced a large field of application, including industrial and life sciences. In Biomedicine for example, several studies have been focused on identifying genes for up- or down regulated effects in head and neck squamous, prostate, lung  or breast cell carcinoma.\cite{kuriakose2004selection, lapointe2004gene, feng2008dna}In common statistical analysis, testing the equality of means in matched pairs design is conducted using the paired t-test. Even for non-normal data, the procedure is asymptotically exact, i.e. for sufficiently large samples, the test procedure is correctly reflecting type-I-error. However, first limitation of the paired t-test arises when data is only partially observed. Deleting observations with missing values is a sub-optimal solution, since variance or mean estimation based only on complete case analysis can be biased leading to incorrect statistical inference. This is especially the case when complete samples are small.

To tackle this issue,  a simple approach is to impute missing values singly (or multiply) and to carry out statistical tests as if there were no missing values so far\cite{schafer1999multiple, rubin2004multiple, sterne2009multiple}.
However, although leading to good imputation error \cite{stekhoven2011missforest, waljee2013comparison, Ramosaj2019}, such approaches may lead to inflated type-I error rate or remarkably low power in small to moderate sample sizes \cite{ramosaj2018caution, van2018flexible}.  Therefore, we do not follow this approach here.


Differing to imputation, several test procedures that (only) use all observed information in the matched pairs design have been proposed in the literature \cite{mehta1969testing, lin1973procedures, morrison1973test, lin1974difference, little1976inference, ekbohm1976comparing, bhoj1978testing, looney2003method, kim2005statistical, samawi2014notes, uddin2017testing}. These tests, however, rely on specific model assumptions such as symmetry or even bivariate normality, which are hard to verify in practice. Moreover, these procedures are usually non robust to deviations and may results in inaccurate decisions caused by possibly conservative or inflated type-I error rates \cite{samawi2014notes,amro2017permuting, amro2019multiplication,qi2018testing}.

 To overcome these problems, the typical recommendation is to use the method based on combining separate results of adequate test statistics for the underlying paired and unpaired portions of the data using either weighted test statistics\cite{samawi2014notes, amro2017permuting}, a multiplication combination test \cite{amro2019multiplication},   or combined p-values  \cite{rempala2006asymptotic, samawi2011nonparametric,  yu2012permutation, kuan2013simple}. 
 However, all these methods are only applicable for matched pairs with missingness in both arms. Thus, these methods cannot be used to analyze data on pathological stage I breast cancer patients from the Cancer Genome Atlas (TCGA) project. This data set consists of observations from $90$ patients of which $74$ had entries in one component of it, only $16$ were complete, see Section \ref{Secdata} below for details. The question is now how to analyze such data?.

 \begin{figure}[h!]
 	\begin{center}
 		\includegraphics[scale=0.8]{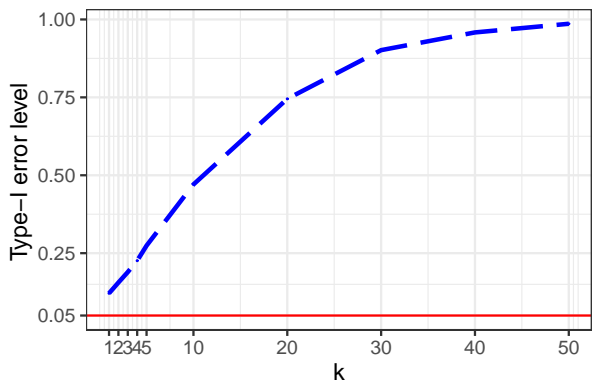}	
 	\end{center}
 	\caption{Type-I error simulation results ($\alpha=0.05$) of the nonparametric combination test  $T_{NC}$ $({\color{blue} \textendash\textendash \quad \textendash\textendash})$ for exponential distribution under correlation factor ($\rho=0.7$) and a heteroscedastic set-up with variances $1$ and $2$ respectively for increasing sample sizes  $k \cdot (n_c,n_u)=(k\cdot 10, k\cdot30)$ under the MCAR framework.}
 	\label{fig:kmultnExp}	
 \end{figure}
In contrast to the above methods,  barely any work can be found that is potentially applicable in this special missing pattern, requires no parametric assumptions and also leads to valid inferences in case of heteroscedasticity or skewed distributions. One exception is given by the recent proposals of Qi et. al. \cite{qi2018testing} who recommended a so-called nonparametric combination test (NCT). It is based on merging results from {\it Sign test} and {\it Wilcoxon Mann-Whitney test}.  In situations where these two nonparametric procedures show their efficiency, their proposed combination is indeed tempting. However, neither the  {\it Sign test} is known to be very powerful  for metric data nor is the {\it Mann-Whitney test} known for being robust against heteroscedasticity.   
  In fact, our simulation studies demonstrate that their NCT inherits these unsatisfying properties to some extend. It show that, under heteroscedasticity and/or skewed distributions, the test statistic (NCT) tends not to maintain the pre-assigned type-I error level. The degree of variance heterogeneity, skewness, and sample sizes can all affect the type-I error rate control level. An overview of the type-I error control of NCT when heteroscedasticity coincides with a skewed error distribution is displayed in Figure ~\ref{fig:kmultnExp}. It reveals that, under heteroscedasticity and exponential distribution, the NCT type-I error rate function becomes surprisingly analogous to the power function where the type-I error rate increases dramatically with an increase in sample sizes. \\

 The aim of this paper is therefore bilateral: First, we aim to provide a statistical test that is capable of treating single arm missing values in matched pairs which drops the common assumptions such as homoscedasticity and normality and secondly, is able to satisfactorily control type-I-error while maintaining good power properties. To this end, we propose three different test statistics, analyze their asymptotic behaviors under the null hypothesis and  
 equip them with an asymptotically correct parametric bootstrap procedure for calculating critical values. 
 In doing so, we structured the paper by firstly introducing the statistical model and the hypothesis of interest. In Section \ref{SecStatistics}, we provide different test statistics of quadratic form-type that either converge to a $\chi^2$ or a weighted $\chi^2$ distributions. Proofs presenting theoretical guarantees of the proposed methods are delivered in the supplement. In Section \ref{SecPBtsrtp}, we introduce a parametric bootstrap technique to calculate critical values and prove it's theoretical correctness. Section \ref{Secpooling} is devoted to already existing methods for statistical inference in matched pairs with single arm missingness while in Section \ref{SimStudy} and \ref{Secdata}, novel and existing methods are compared based on an extensive simulation study and a real data example from a breast cancer gene study. The supplement contains additional analyses.

\section{Statistical Model and Hypotheses \label{sec2}}

We consider  matched pairs given by a sample $\mathcal{D}_n := \{ \mathbf{X}_1, \dots , \mathbf{X}_n \}$, where $\mathbf{X}_j = (X_{1j}, X_{2j})^T \in \R^2$ are i.i.d. random vectors with mean vector $\mathbb{E}[\mathbf{X}_1]  = \boldsymbol{\mu} = [\mu_1, \mu_2]^\top \in \R^2 $ and an arbitrary covariance matrix $   0<\boldsymbol{\Gamma} \in \R^{2 \times 2}$. To incorporate missingness in one arm (says, the second) only denote  with $R_{2j} \in \{0,1\}$, $j = 1, \dots , n$  the vector whose $j$-th component indicates whether $X_{2j}$ is observed $(R_{2j} = 1)$ or missing $(R_{2j} = 0)$ for $j = 1, \dots, n$. If $*$ denotes the component-wise multiplication of vectors, then in practice, one observes $\mathbf{X}^{(o)} :=  \{  \mathbf{X}_j * \mathbf{R}_j \}_{j = 1}^n$ where $\mathbf{R}_j = [1, R_{2j}]^\top  \in \R^2$  , $j = 1, \dots , n$, and a "$0$" entry is interpreted as missing. Hence our framework has the following form, where $---$ stands as  a placeholder for $R_{2j} = 0$: 

\begin{equation}
\label{ModelMis}
\begin{gathered}
\underbrace{ \begin{bmatrix*}[c]
	X_{11}^{(c)} \\
	X_{21}^{(c)}
	\end{bmatrix*},  \dots , 	 \begin{bmatrix*}[c]
	X_{1n_c}^{(c)} \\
	X_{2n_{c}}^{(c)} 
	\end{bmatrix*} }_{ \mathbf{X}^{(c)} },  \underbrace{ 
	\begin{bmatrix*}[c]
	X_{11}^{(i)}\\
	\text{ }_{---}
	\end{bmatrix*}, \dots , \begin{bmatrix*}[c]
	X_{1n_{u}}^{(i)} \\
	\text{ }_{---}
	\end{bmatrix*} }_{ \mathbf{X}^{(i)} }
\end{gathered}
\end{equation}

Rubin defines the missing mechanism through a parametric distributional model on $\mathbf{R} = \{ \mathbf{R}_j \}_{j = 1}^n$ and classifies their presence through Missing Completely at Random (MCAR), Missing at Random (MAR) and Missing not at Random (MNAR) schemes.\cite{rubin2004multiple} In our work, we first assume a MCAR mechanism, in that $\vX^{c}$ is independent of $\vX^{(i)}$.
However, we will also study MAR mechanisms in simulations and relate to the supplement for the explicit definition of the missing mechanisms. For notational purposes, let $I_{n_c}$ denote the index set of $\lvert I_{n_c} \rvert =n_c$ complete pairs, i.e. $\mathbf{R}_{j} = [1,1]^\top$ for all $j \in I_{n_c}$. Similarly, $I_{n_u}$ is the index set of observations with second component missing  $(\mathbf{R}_j = [1,0]^\top, j \in I_{n_u} )$ and  $\lvert I_{n_u} \rvert =n_u$. Thus, there are in total $N=2n_c+n_u$ observations from $n = n_c + n_u$ subjects.\\
 In this framework, we like to use all the available data to test the null hypothesis $H_0:$ $\{\mu_1 = \mu_2\}$ of equal means against the alternative $H_1:$  \{$\mu_1 \neq \mu_2$\}.\\ 

To construct our paper test statistics, we first fix estimators of the population means $\mu_1$, and $\mu_2$. 
 For estimating  $\mu_1$, we consider two estimators; the sample mean of the first components of the completed data set $\bar{X}_{1.}^{(c)}=\frac{1}{n_c}\sum_{i=1}^{n_c}X_{1i}^{(c)}$, and the sample mean of the first components of the unpaired data $\bar{X}_{1.}^{(i)}=\frac{1}{n_u}\sum_{j=1}^{n_u}X_{1j}^{(i)}$. For estimating the population mean $\mu_2$, we use  the sample mean of the second components of the complete data $\bar{X}_{2.}^{(c)}=\frac{1}{n_c}\sum_{i=1}^{n_c}X_{2i}^{(c)}$.\\

Next, we define the normalized vector $\vZ_n$ that aggregates the difference between the mean values $\boldsymbol{\mu}=[\mu_1,\mu_2]^\top$ and their empirical estimators $[\bar{X}_{1.}^{(c)},\bar{X}_{2.}^{(c)},\bar{X}_{1.}^{(i)} ]^\top$  
\begin{equation}
\vZ_n=\sqrt{n}[\bar{X}_{1.}^{(c)}-\mu_1, \bar{X}_{2.}^{(c)}-\mu_2, \bar{X}_{1.}^{(i)}-\mu_1]^\top
\label{Z}
\end{equation}

and take their correlations into account in the covariance matrix $\vSigma_n=Cov(\vZ_n)$.\\

For subsequently asymptotic analyses, we need to set up the following assumption regarding  sample sizes, which we assume throughout 
\begin{assumption}{1}{}\label{Assump1}
	For $\min\{n_c, n_u \} \to \infty$ we require that
	\begin{itemize}
		\item $\frac{n_c}{n_c+n_u} \to \kappa_1 \in  (0,1)$ 
		
		\item $\frac{n_u}{n_c+n_u} \to \kappa_2 = (1- \kappa_1) \in (0,1)$ 
	\end{itemize}

\end{assumption}

\begin{prop}
	\label{prop1}
	Let $\sigma_1^{2}=Var(X_{11}^{(c)}) = Var(X_{11}^{(i)})$, $\sigma_2^{2}=Var(X_{21}^{(c)})$ and $\rho=corr(X_{11}^{(c)},X_{21}^{(c)})$. 
	 The statistic $\vZ_n$ has, asymptotically, as $n\to \infty$, a multivariate normal distribution with expectation $0$ and covariance matrix given by\\
\begin{align} 
		\vSigma=\lim_{n \to \infty}\vSigma_n= \begin{bmatrix} \kappa_1^{-1}\sigma_1^{2}&\kappa_1^{-1}\rho\sigma_1\sigma_2&0\\ \kappa_1^{-1}\rho\sigma_1\sigma_2 &\kappa_1^{-1}\sigma^{2}_2 & 0\\
	0&0&\kappa_2^{-1}\sigma^{2}_1
	\end{bmatrix}.
	\label{sigma}
\end{align}
\end{prop}

$\vSigma_n$ can be consistently estimated by  

\begin{flalign}
\hat{\vSigma}_n= \begin{bmatrix} \hat{\kappa}_1^{-1}\hat{\sigma}^{2}_1&\hat{\kappa}_1^{-1}\hat{\rho}\hat{\sigma}_1\hat{\sigma}_2 & 0\\ \hat{\kappa}_1^{-1}\hat{\rho}\hat{\sigma}_1\hat{\sigma}_2 & \hat{\kappa}_1^{-1}\hat{\sigma}^{2}_2 & 0\\
0&0&\hat{\kappa}_2^{-1}\hat{\sigma}^{2}_1
\end{bmatrix},
\label{sigmah}
\end{flalign}
where $\hat{\kappa}_1 = n_c/n$, $\hat{\kappa}_2 = n_u/n$ and $\hat{\sigma}_1^2 $ is the unbiased empirical variance estimator based on the pooled first components $ X_{11}^{(c)}, \dots, X_{1n_c}^{(c)}, X_{11}^{(i)}, \dots, X_{1n_u}^{(i)}$, $\hat{\sigma}_2^2 = \frac{1}{n_c - 1} \sum\limits_{i = 1}^{n_c} (X_{2i}^{(c)} - \bar{X}_{2\cdot}^{(c)})^2$ and the correlation factor $\rho$ is estimated through the empirical correlation $\hat{\rho}$ calculated from the paired data $\mathbf{X}^{(c)}$. \\

To test the null hypothesis $H_0:\mu_1-\mu_2=0$, we define the two estimators $\bar{X}_{1.}^{(c)} - \bar{X}_{2.}^{(c)}$ and $\bar{X}_{1.}^{(i)} - \bar{X}_{2.}^{(c)}$ for $\mu_1-\mu_2$. Their joined asymptotic behavior under the null hypothesis $H_0$ is studied below.

\begin{prop}
		\label{prop2}
	Set $f_{\vA}(\mathbf{x})=\vA \mathbf{x}$,  for the matrix 
	${\vA}=\begin{bmatrix}
	1&-1&0\\
	0&-1&1
	\end{bmatrix}$ $\in \R^{2X3}$. Then, under the null hypothesis $H_0$, the composite statistic
	\begin{equation}
f_{\vA}\circ \vZ_n= \mathbf{A} \mathbf{Z}_n = \sqrt{n}[\bar{X}_{1.}^{(c)} - \bar{X}_{2.}^{(c)}, \bar{X}_{1.}^{(i)} - \bar{X}_{2.}^{(c)}]^\top,
\label{fz}
	\end{equation}
is asymptotically $N_2(\mathbf{0}, \mathbf{A} \boldsymbol{\Sigma} \mathbf{A}^\top)$ distributed as $n \rightarrow \infty$.
\end{prop}


\section{Statistics and Asymptotics}
\label{SecStatistics}
In this section, we propose three different quadratic forms for testing $H_0$: a Wald-type statistic (WTS), an ANOVA-type statistic (ATS), and a modified ANOVA-type statistic (MATS). To introduce the WTS, denote by $\mathbf{B}^+$ the Moore-Penrose inverse of a matrix $\mathbf{B}$. Then, the WTS is given by 
\begin{equation}
T_W=(\vA\vZ_n )^\top(\vA\hat{\vSigma}_n \vA^\top)^+(\vA\vZ_n).
\label{Wald}
\end{equation}

Thanks to the introduced studentization by $(\vA\hat{\vSigma}_n \vA^\top)^+$, the WTS is asymptotically distribution-free as studied below.   

\begin{thm}
	\label{thmwald}
The statistic $T_W$ has under the null hypothesis $H_0:\{\mu_1=\mu_2\}$, asymptotically, as $n\to \infty$, a central $\chi^2_2$ distribution. \\
\end{thm}

Similar WTS versions are also studied in the context of heteroscedastic ANOVA or MANOVA  \cite{krishnamoorthy2010parametric, xu2013parametric, konietschke2015parametric, friedrich2018mats}. From these settings, it is known that the convergence to its limiting $\chi^2$-distribution is rather slow \cite{vallejo2010analysis, konietschke2015parametric, smaga2017bootstrap}, which leads to several refinements regarding bootstrapping for the calculations of critical values (see Section \ref{SecPBtsrtp} below) or other structures of test statistics. In particular, Brunner (2001) \cite{brunner2001asymptotic} proposed an alternative quadratic form  by deleting the Moore-Penrose inverse involved in the computation of the WTS, resulting in the following ATS:
\begin{equation}
T_A= \frac{1}{tr(\mathbf{A} \hat{\boldsymbol{\Sigma}}_n \mathbf{A}^\top )}(\mathbf{A}\vZ_n )^\top( \mathbf{A}\vZ_n), 
\label{ANOVA}
\end{equation}
which is also applicable in case of $|\hat{\boldsymbol{\Sigma}}_n|=0$. 

\begin{thm}
	\label{thmanova}
	Under the null hypothesis $H_0:\{\mu_1=\mu_2\}$, the test statistic $T_A$ has asymptotically,    as $n\to \infty$, the same distribution as the random variable
	\begin{equation}
	Y=\sum_{i=1}^{2}\lambda_iY_i/ tr(\mathbf{A} \boldsymbol{\Sigma} \mathbf{A}^\top ),
	\end{equation}
	where $Y_i \overset{\text{i.i.d}}{\sim} \chi^2_1 $ and the weights $\lambda_i$ are the eigenvalues of $\vA\vSigma \mathbf{A}^\top$ where $\vSigma$ is given in  (\ref{sigma}).\\
\end{thm}



 Another possible test statistic would be the modified version of the ANOVA type-statistic (MATS) that was developed by Friedrich and Pauly (2017) \cite{friedrich2018mats} for MANOVA models. Here, it is given by 

\begin{equation}
T_M=(\vA\vZ_n )^\top\hat{\vD}_n(\vA\vZ_n), 
\label{MATS}
\end{equation}
where $\hat{\vD}_n=diag(( \mathbf{A} \hat{\boldsymbol{\Sigma}}_n \mathbf{A}^\top )^+_{ii})$. \\

\begin{thm}
	\label{themmats}
	Under the null hypothesis $H_0:\{\mu_1=\mu_2\}$, 
	the test statistic $T_M$ has asymptotically,as $n\to \infty$, the same distribution as the random variable
	\begin{equation}
	\tilde{Y}=\sum_{i=1}^{2}\tilde{\lambda}_i\tilde{Y}_i ,
	\end{equation}
	where $\tilde{Y}_i \overset{\text{i.i.d}}{\sim} \chi_1^2 $ and the weights $\tilde{\lambda}_i $ are the eigenvalues of $\mathbf{D} \mathbf{A} \vSigma \mathbf{A}^\top$ and $\mathbf{D} =diag(( \mathbf{A} \vSigma \mathbf{A}^\top )^+_{ii})$.\\
\end{thm}

As, the weights $\lambda_i$, and $\tilde{\lambda}_i$ in Theorems \ref{thmanova} and \ref{themmats} are unknown and the $\chi_2^2-$approximation to $T_W$ is rather slow, we will develop  adequate and asymptotically correct testing procedures based on bootstrap versions of $T_W, T_A,$ and  $T_M$ in the subsequent section.

\section{Parametric Bootstrapping}
\label{SecPBtsrtp}
To estimate critical values, we apply  an asymptotic model based bootstrap approach
 which  has , e.g. been applied in the context of (M)ANOVA  factorial designs \cite{friedrich2018mats, konietschke2015parametric}. To this end, first, we generate parametric bootstrap variables as 
\begin{equation}
\mathbf{X}^*_{j}=\begin{bmatrix}
X_{1j}^{*}\\
 X_{2j}^{*}
\end{bmatrix}
\overset{\text{i.i.d}}{\sim} N(0,\hat{\boldsymbol{\Gamma}}), j=1,...,n.
\end{equation}
Here,
$\hat{\boldsymbol{\Gamma} }=\begin{bmatrix}
\hat{\sigma}_1^{2}& \hat{\rho}\hat{\sigma}_1\hat{\sigma}_2\\
\hat{\rho}\hat{\sigma}_1\hat{\sigma}_2&\hat{\sigma}_2^{2}
\end{bmatrix}$
is an empirical covariance matrix estimator, where $\hat{\sigma}_1^{2}$, $\hat{\sigma}_2^{2}$ and $\hat{\rho}$ are as in Proposition \ref{prop1}. The idea is to reflect the original covariance structure to obtain more accurate finite sample approximation. Next, we generate missing values under the MCAR scheme by randomly inserting them to  the  second component of the bivariate vector $\mathbf{X}^*_{j}$ until a fixed amount of missing values of size $n_u$ is achieved. This results into the following bootstrapped data set
\begin{equation}
\label{PBMiss}
\begin{gathered}
\underbrace{ \begin{bmatrix}
	X_{11}^{*(c)} \\
	X_{21}^{*(c)}
	\end{bmatrix},  \dots , 	 \begin{bmatrix}
	X_{1n_c}^{*(c)} \\
	X_{2n_{c}}^{*(c)} 
	\end{bmatrix} }_{ \mathbf{X}^{*(c)} },  \underbrace{ 
	\begin{bmatrix}
	X_{11}^{*(i)}\\
	\text{ }_{---}
	\end{bmatrix}, \dots , \begin{bmatrix}
	X_{1n_{u}}^{*(i)} \\
	\text{ }_{---}
	\end{bmatrix} }_{ \mathbf{X}^{*(i)} }
\end{gathered}
\end{equation}
and the combined vector $(f\circ \vZ_n)^*=\vA\vZ_n^*=\sqrt{n}(\bar{X}_{1.}^{*(c)}-\bar{X}_{2.}^{*(c)}, \bar{X}_{1.}^{*(i)}-\bar{X}_{2.}^{*(c)}) $. From this, the bootstrapped versions of the quadratic forms, i.e. the Wald-type statistic $T_W^*$, the ANOVA-type statistic $T_A^*$  and the modified ANOVA-type statistic $T_M^*$ are computed 
\begin{align}
T_W^* & =(\vA\vZ_n^* )^\top(\mathbf{A}\hat{\vSigma}^{*}_n\mathbf{A}^T)^+ (\vA\vZ_n^* ),  \label{PBWald}\\
T_A^* &=\frac{1}{tr(\mathbf{A} \hat{\vSigma}^*_n \mathbf{A}^\top )}(\vA\vZ_n^* )^\top(\vA\vZ_n^* ),   \label{ANOVABoot} \\
T_M^* &=(\vA\vZ_n^* )^\top\hat{\vD}^*_n(\vA\vZ_n^* ), \label{MATSBoot}
\end{align} 

where $\hat{\vSigma}^{*}_n=\hat{\vSigma}_n(\mathbf{X}^{*(c)}, \mathbf{X}^{*(i)})$ and $\hat{\vD}^*_n=diag(( \mathbf{A} \hat{\boldsymbol{\Sigma}}_n^* \mathbf{A}^\top )^+_{ii})$.\\

The next theorem proves that all three bootstrapped test statistics can be used to approximate the null distribution of the respective test statistic.

\begin{thm}\label{ThmPBWald}
For any choice $[-] \in \{A,M,W\}$, the conditional distribution of $T_{[-]}^*$ converges weakly to the null distribution of $T_{[-]}$ in probability for any choice of $\boldsymbol{\mu}\in\R^2$ and $\boldsymbol{\mu}_0\in H_0$. In particular we have
	\begin{align*}
 	\sup_{x\in\R}|P_{\boldsymbol{\mu}}(T_{[-]}^*\leq x|\vX)-P_{\boldsymbol{\mu}_0}(T_{[-]}\leq x)| \xrightarrow{\text{p}} 0.
 \end{align*}
\end{thm}

From Theorem \ref{ThmPBWald}, we thus obtain the asymptotically correct bootstrap tests $\varphi_{W}^* =\mathds{1}\{T_W > c_W^*\}, \varphi_{A}^* =\mathds{1}\{T_A > c_A^*\}$, and $\varphi_{M}^* =\mathds{1}\{T_M > c_M^*\}$  where $c_W^*, c_A^*,$ and $c_M^*$ denote the conditional $(1-\alpha)$- bootstrap quantiles of $T_W^*, T_A^*$, and $T_M^*$ respectively.\\

To analyze their finite sample performance, we below conduct extensive simulations (Section \ref{SimStudy}). Before that, we will first discuss other possible candidates from the literature that should or should not be included in our simulation study.

\section{Comparison with existing methods}
\label{Secpooling}
We briefly review the existing literature on methods that can deal with the case of matched pairs with missing values in one arm only. As outlined in the introduction, there only exists a few which we can summarize them as follows

\begin{enumerate}[(a)]
	\item Simple methods such as: using the paired t-test while excluding the unpaired data OR using the independent t-test while ignoring the covariance structure of the data.
	
	\item Tests based on modified maximum likelihood estimators. \cite{morrison1973test, ekbohm1976oncomparing, little1976inference}
	\item Tests based on simple mean difference estimator. \cite{lin1973procedures, mehta1969testing, mehta1973test, ekbohm1976oncomparing}
	\item P-values pooling methods.\cite{qi2018testing}
	
	\item Weighted linear and nonlinear combination tests.\cite{qi2018testing}
\end{enumerate}

However, none of the methods is free from distributional assumptions and at the same time robust against deviations such as heteroscedasticity and skewed distributions. In particular,  the recent paper by Qi et al.\cite{qi2018testing} already included a simulation study to compare  several of the tests mentioned in (a) - (e). As a conclusion, they recommended a so-called non-parametric combination test (NCT). 

 Therefore, we will mainly focus on 
 the non-parametric combination method proposed in Qi et al.\cite{qi2018testing} 
 As additional competitor for these bootstrap procedures proposed in Section \ref{SecPBtsrtp}, we choose the test of Little. \cite{little1976inference} 
  The latter assumes that the data follows a bivariate normal distribution and the test statistic is given by 

\begin{align}
	T_{LT} &= \frac{ \bar{X}_{1\cdot} - \bar{X}_{2 \cdot }^{(c)}  - \frac{ \hat{\rho} \hat{\sigma}_1^{(c)} \hat{\sigma}_2 }{(\hat{\sigma}_1^{(c)})^2} ( \bar{X}_{1\cdot} - \bar{X}_{1\cdot}^{(c)} ) }{\hat{\sigma}_{\mathbf{LT}}},
\end{align}

where $\bar{X}_1 := 1/n( n_c \bar{X}_{1\cdot}^{(c)}  + n_u \bar{X}_{1 \cdot}^{(i)}) $ and $\hat{\sigma}_1^{(c)}$ is the empirical standard deviation of $\{ X_{11}^{(c)}, \dots, X_{1n_c}^{(c)} \}$. Moreover, setting $\hat{\sigma}_{22\cdot 1}^2 = \hat{\sigma}_2^2 - ( \hat{\rho} \hat{\sigma}_1^{(c)} \hat{\sigma}_2 / ( \hat{\sigma}_1^{(c)} )^2 )$ and $\hat{\sigma}_{\mathbf{X}} = \hat{\sigma}_{22\cdot 1}^2 + \frac{(\hat{\rho} \hat{\sigma}_1^{(c)} \hat{\sigma}_2 )^2}{(\hat{\sigma}_1^{(c)})^4} \hat{\sigma}_1^4$, the denominator is given by \cite{little1976inference} 

\begin{align*}
	\hat{\sigma}_{\mathbf{LT}}^2 &= \frac{\hat{\sigma}_{\mathbf{X}}^2}{n} + \left(\frac{1}{n_c} - \frac{1}{n}\right) \frac{n_c - 2 }{n_c - 3} \hat{\sigma}_{22\cdot 1}^2 - \frac{2}{n} \frac{\hat{\rho} \hat{\sigma}_1^{(c)} \hat{\sigma}_2 }{(\hat{\sigma}_1^{(c)})^2} \hat{\sigma}_1^2 + \frac{\hat{\sigma}_1^2}{n}
\end{align*}

The exact distribution of $T_{LT}$ is rather complicated and Little suggests to approximate it by a $t$-reference distribution with $n_c - 1$ degrees of freedom, i.e. the test is given by  $\varphi_{LT} := \mathds{1}\{  | T_{LT} | > t_{n_c - 1, 1-\alpha/2 } \}$ for some level $\alpha \in (0,1)$.\\

In addition, the non-parametric combination test\cite{qi2018testing} proposed by Qi et al., is based upon a linear combination of the sign and the Wilcoxon Mann-Whitney test statistics:
\begin{equation}
T_{NC}=T_s+T_{m},
\end{equation}
where $T_s=\frac{1}{n_c}\sum_{i=1}^{n_c}\phi(X_{1i}^{(c)},X_{2i}^{(c)})$ and $T_{m}=\frac{1}{n_cn_u}\sum_{j=1}^{n_u}\sum_{k=1}^{n_c}\phi(X_{1j}^{(i)},X_{2k}^{(c)})$ with\\
\begin{center}
$\phi(X_1,X_2)=\begin{cases}
1&\text{if} \;  X>Y\\
1/2&\text{if} \; X=Y\\
0&\text{otherwise.}
\end{cases}$
\end{center}

It is proposed \cite{qi2018testing} to approximate the null distribution of $T_{NC}$ by a normal distribution with mean $1$ and variance estimated by $\widehat{Var}(T_{NC})=\frac{1}{n_c}+\frac{n_c+n_u+1}{12n_cn_u}+\widehat{Cov}(T_s,T_{m})$,
where\\ $\widehat{Cov}(T_s,T_{m})=\frac{1}{n_c^2n_u}\sum_{i=1}^{n_c}\sum_{j=1}^{n_u}\mathds{1}\{X_{1i}^{(c)}>X_{2i}^{(c)},  X_{1j}^{(i)}>X_{2j}^{(c)}\} - \frac{1}{n_c}T_{s}T_{m}$.

\begin{table*}[ht]
	
	\caption{Type-I error simulation results ($\alpha=0.05$) of the tests for different distributions under varying correlation values ($\rho$) with sample sizes $(n_c,n_u)\in\{(10,10),(30,10),(10,30)\}$ and homoscedastic covariance matrix $\Sigma_1$ under the MCAR framework. For each setting, the values closest to the prescribed level are printed in {\bf bold} and values exceeding the upper limit ($6.8\%$) of the $99\%$ binomial interval are in {\color{red} \bf red} colour. 
		\label{Type1errorhomoMCAR}}
	{\tabcolsep=4pt
		\begin{tabular*}{1.09\linewidth}{lc ccccc c ccccc  c ccccc   cc}
			Dist & $\rho$ & \multicolumn{5}{c}{$(10,10)$}  && \multicolumn{5}{c}{$(30,10)$} && \multicolumn{5}{c}{$(10,30)$} \\ \cline{3-7} \cline{9-13} \cline{15-19}
			&& $T_{W}^*$ & $T_{A}^*$ & $T_{M}^*$ & $T_{NC}$ & $T_{LT}$
			&& $T_{W}^*$ & $T_{A}^*$ & $T_{M}^*$ & $T_{NC}$ & $T_{LT}$ 
			&& $T_{W}^*$ & $T_{A}^*$ & $T_{M}^*$ & $T_{NC}$ & $T_{LT}$\\
			Normal&-0.9& {\bf 5.4}&5.5&{\bf 5.4}&{\color{red} \bf 6.8}&4.4&&{\bf 5.2}&{\bf 5.2}&{\bf 5.2}&5.6&5.5&&5.5&{\bf 5.4}&6.7&{\color{red} \bf 6.8}&4.0\\
			&-0.5&{\bf 5.1}&5.4&5.7&6.5&4.9&&{\bf 5.0}&5.2&5.1&5.5&5.3&&5.5&5.6&6.4&6.7&{\bf 5.1}\\
			&-0.1&{\bf 5.4}&5.6&{\bf 5.4}&6.6&5.6&&5.6&{\bf 5.3}&5.4&5.4&5.6&&{\bf 5.3}&5.6&6.1&{\color{red} \bf 6.8}&6.0\\
			&0.1&{\bf 5.2}&5.5&5.5&6.4&6.1&&{\bf 5.0}&4.5&4.5&4.9&5.3&&{\bf 5.2}&5.5&5.8&6.5&6.1\\
			&0.5&5.1&{\bf 5.0}&4.8&6.0&{\color{red} \bf 6.8}&&{\bf 4.9}&5.2&{\bf 4.9}&5.6&5.6&&{\bf 5.0}&5.1&4.9&6.2&6.5\\
			&0.9&{\bf 5.3}&{\bf 5.3}&4.6&5.8&{\color{red} \bf 12.3}&&4.8&{\bf 5.0}&4.6&4.8&{\color{red} \bf 12.0}&&{\bf 5.4}&{\bf 5.4}&4.3&6.1&{\color{red} \bf 8.0}\\
			\\
			Laplace&-0.9&4.6&{\bf {\bf 4.9}}&5.4&6.6&4.6&&4.4&4.8&4.5&5.2&{\bf 4.9}&&4.6&4.7&6.0&{\color{red} \bf 6.8}&3.1\\
			&-0.5&4.3&4.8&{\bf 5.0}&6.5&4.6&&4.4&4.7&{\bf 5.0}&{ 5.1}&{ 5.1}&&4.3&{\bf 5.0}&5.8&6.7&4.3\\
			&-0.1&4.3&{\bf 5.0}&4.8&6.5&5.4&&4.0&4.7&4.7&{\bf 5.2}&5.4&&4.6&{\bf 4.9}&5.3&6.4&5.3\\
			&0.1&4.4&{4.7}&4.6&6.6&{\bf 5.2}&&4.5&{ 4.9}&{\bf 5.0}&5.4&5.4&&4.3&{\bf 4.8}&5.3&6.4&5.6\\
			&0.5&4.3&{\bf 4.5}&4.4&6.2&6.4&&4.3&4.3&4.5&{\bf 5.2}&{\bf 5.2}&&4.3&4.5&{\bf 4.9}&6.6&5.7\\
			&0.9&4.2&{\bf 5.2}&4.0&6.1&{\color{red} \bf 13.3}&&4.5&4.7&4.2&{\bf 5.0}&{\color{red} \bf14.3}&&4.3&{\bf 5.1}&4.0&6.1&{\color{red} \bf 9.4}\\
			\\
			Exp&-0.9&4.5&4.2&{\bf 5.2}&6.4&4.2&&5.3&{\bf 5.0}&6.0&5.7&5.4&&{\bf 4.6}&4.2&6.1&6.7&3.9\\
			&-0.5&5.2&{ 4.9}&{\bf 5.0}&6.6&{\bf 5.0}&&{\color{red} \bf 6.9}&4.7&6.4&5.2&{\bf 5.0}&&{\bf 5.0}&4.7&6.4&6.2&5.4\\
			&-0.1&{\bf 4.8}&4.4&4.2&6.1&5.8&&{\color{red} \bf7.0}&4.7&6.2&5.2&{\bf 5.1}&&{\color{red} \bf7.3}&{\bf 5.9}&{\color{red} \bf7.5}&{\color{red} \bf7.0}&{\color{red} \bf7.8}\\
			&0.1&5.4&{\bf 4.9}&4.5&6.4&{\color{red} \bf7.2}&&{\color{red} \bf 6.8}&{\bf 4.9}&6.4&5.4&5.7&&{\color{red} \bf7.5}&{\bf 5.7}&{\color{red} \bf7.1}&6.4&{\color{red} \bf8.1}\\
			&0.5&{\bf 5.7}&{\bf 4.3}&4.2&6.5&{\color{red} \bf 9.4}&&5.8&{\bf 5.0}&5.6&5.4&6.4&&{\color{red} \bf8.4}&{\bf 6.0}&{\color{red} \bf 6.8}&6.1&{\color{red} \bf10.8}\\
			&0.9&6.4&{\bf 4.8}&3.8&6.1&{\color{red} \bf15.3}&&{\bf 5.0}&{ 5.1}&4.6&{ 4.9}&{\color{red} \bf15.6}&&{\color{red} \bf10.0}&6.6&{\bf 4.9}&6.4&{\color{red} \bf16.2}\\
			\\
			Chisq&-0.9&{\bf 5.2}&5.4&5.5&6.4&4.3&&5.3&{\bf 5.0}&5.2&5.5&5.2&&{\bf 5.0}&5.4&6.6&6.6&3.7\\
			&-0.5&4.6&{\bf 5.1}&5.2&6.3&4.7&&{ 5.1}&{ 4.9}&{\bf 5.0}&5.2&4.7&&{\bf 4.9}&5.2&6.0&6.4&4.6\\
			&-0.1&{\bf 5.4}&5.6&5.5&6.7&5.9&&{ 5.1}&4.6&{\bf 5.0}&5.3&5.2&&{\bf 5.5}&5.7&6.2&6.6&5.6\\
			&0.1&5.2&5.2&{\bf 5.0}&6.3&5.5&&5.2&4.8&{\bf 5.0}&{\bf 5.0}&5.4&&{\bf 5.8}&{\bf 5.8}&6.0&6.4&6.5\\
			&0.5&{\bf 4.9}&4.8&4.7&6.2&6.3&&{\bf 5.0}&4.7&4.8&{ 5.1}&5.6&&5.7&{\bf 5.0}&5.3&6.0&{\color{red} \bf 6.8}\\
			&0.9&5.6&{\bf 5.3}&4.5&5.7&{\color{red} \bf12.8}&&5.4&5.6&{\bf 5.2}&5.4&{\color{red} \bf12.6}&&5.8&{\bf 5.3}&4.5&6.0&{\color{red} \bf8.9}\\\hline		
		\end{tabular*} }
	\end{table*}

	\begin{table*}[ht]
		
		\caption{Type-I error simulation results ($\alpha=0.05$) of the tests for different distributions under varying correlation values ($\rho$) with sample sizes $(n_c,n_u)\in\{(10,10),(30,10),(10,30)\}$ and heteroscedastic covariance matrix $\Sigma_2$ under the MCAR framework. For each setting, the values closest to the prescribed level are printed in {\bf bold} and values exceeding the upper limit ($6.8\%$) of the $99\%$ binomial interval are in {\color{red} \bf red} colour. 
			\label{Type1errorhetroMCAR}}
		{\tabcolsep=4pt
			\begin{tabular*}{1.09\linewidth}{lc ccccc c ccccc  c ccccc   cc}
				Dist & $\rho$ & \multicolumn{5}{c}{$(10,10)$}  && \multicolumn{5}{c}{$(30,10)$} && \multicolumn{5}{c}{$(10,30)$} \\ \cline{3-7} \cline{9-13} \cline{15-19}
				&& $T_{W}^*$ & $T_{A}^*$ & $T_{M}^*$ & $T_{NC}$ & $T_{LT}$
				&& $T_{W}^*$ & $T_{A}^*$ & $T_{M}^*$ & $T_{NC}$ & $T_{LT}$ 
				&& $T_{W}^*$ & $T_{A}^*$ & $T_{M}^*$ & $T_{NC}$ & $T_{LT}$\\
				Normal&-0.9&5.5&5.5&5.7&{\color{red} \bf 7.0}&{\bf 4.7}&&{ 4.8}&{ 5.1}&{\bf 5.0}&4.6&{ 5.2}&&{ 5.5}&{\bf 5.4}&{\color{red} \bf 7.0}&{\color{red} \bf 7.8}&4.1\\
				&-0.5&{\bf 5.1}&5.6&5.8&{\color{red} \bf7.2}&{\bf 5.1}&&{ 5.2}&{ 5.2}&{ 5.2}&{\bf 5.0}&{ 5.3}&&{\bf 5.5}&5.6&6.7&{\color{red} \bf 7.4}&5.6\\
				&-0.1&{\bf 5.5}&5.8&5.7&{\color{red} \bf 7.1}&6.0&&{ 4.8}&{\bf 4.9}&{ 4.8}&4.4&{\bf 5.1}&&{\bf 5.4}&5.7&6.4&{\color{red} \bf 7.8}&6.4\\
				&0.1&{\bf 5.2}&5.5&5.6&{\color{red} \bf 6.8}&6.2&&4.6&{ 4.8}&{\bf 5.0}&{ 4.7}&5.5&&{\bf 5.2}&5.6&6.0&{\color{red} \bf 7.2}&6.6\\
				&0.5&5.4&{\bf 5.2}&{\bf 5.2}&6.4&{\color{red} \bf 7.0}&&{ 5.2}&{\bf 5.0}&{\bf 5.0}&{ 4.9}&5.4&&{\bf 5.0}&5.6&5.8&{\color{red} \bf 7.4}&6.7\\
				&0.9&{\bf 5.4}&{\bf 5.4}&6.1&6.4&{\color{red} \bf12.4}&&{\bf 5.1}&{\bf 4.9}&{ 5.2}&{ 4.8}&{\color{red} \bf 11.0}&&{\bf 5.0}&6.0&{\color{red} \bf 7.2}&{\color{red} \bf 6.9}&{\color{red} \bf 7.7}\\
				\\
				Laplace&-0.9&4.6&{\bf 4.8}&5.6&{\color{red} \bf 7.2}&{\bf 4.8}&&4.5&{ 4.7}&{ 4.8}&{ 4.8}&{\bf 5.1}&&{\bf4.6}&{\bf4.6}&6.3&{\color{red} \bf 7.5}&3.3\\
				&-0.5&4.5&{\bf 4.8}&{\bf 5.2}&{\color{red} \bf 6.8}&{\bf 4.8}&&4.4&{ 4.8}&{ 5.2}&{ 4.7}&{\bf 5.0}&&4.2&{\bf 4.8}&6.1&{\color{red} \bf 7.4}&4.5\\
				&-0.1&4.5&{\bf 5.0}&{ 5.2}&6.7&5.7&&4.1&{\bf 5.0}&{\bf 5.0}&{ 4.9}&{ 5.3}&&4.5&\bf {4.8}&5.5&{\color{red} \bf 7.0}&5.6\\
				&0.1&4.6&{ 4.8}&{\bf 4.9}&{\color{red} \bf 7.0}&5.4&&4.4&{\bf 4.9}&{ 5.2}&{ 4.7}&5.4&&4.3&{\bf 5.1}&5.8&{\color{red} \bf 7.2}&5.8\\
				&0.5&4.2&{\bf4.6}&{\bf4.6}&6.7&6.4&&4.2&4.6&4.5&{ 4.7}&{\bf 5.0}&&4.0&{\bf 4.7}&5.8&{\color{red} \bf 7.2}&6.0\\
				&0.9&4.1&{\bf 5.3}&{\bf 5.3}&{\color{red} \bf 6.8}&{\color{red} \bf 13.9}&&4.4&{ 4.7}&{\bf 4.8}&4.5&{\color{red} \bf14.8}&&4.0&{\bf 5.2}&6.6&{\color{red} \bf 6.9}&{\color{red} \bf 8.7}\\
				\\
				Exp&-0.9&4.6&{\bf 5.2}&6.2&{\color{red} \bf 9.1}&{\bf 5.2}&&{\bf 5.3}&5.5&6.4&{\color{red} \bf 9.1}&5.7&&{5.1}&{ 5.1}&{\color{red} \bf 7.3}&{\color{red} \bf 9.5}&{\bf 5.0}\\
				&-0.5&{\bf 5.3}&6.0&6.2&{\color{red} \bf 9.4}&6.0&&6.6&{\bf 4.9}&6.3&{\color{red} \bf 9.2}&5.5&&{\bf6.3}&6.4&{\color{red} \bf 8.2}&{\color{red} \bf 9.7}&{\color{red} \bf 7.5}\\
				&-0.1&{\bf5.4}&6.2&5.6&{\color{red} \bf 9.9}&{\color{red} \bf 7.3}&&6.5&{\bf 4.6}&6.0&{\color{red} \bf 10.8}&5.8&&{\color{red} \bf 8.0}&{\color{red} \bf 7.9}&{\color{red} \bf 9.3}&{\color{red} \bf 11.1}&{\color{red} \bf 9.8}\\
				&0.1&6.4&{\color{red} \bf 7.0}&{\bf6.1}&{\color{red} \bf 10.0}&{\color{red} \bf 8.9}&&6.2&{\bf4.6}&6.0&{\color{red} \bf 11.4}&6.2&&{\color{red} \bf 8.3}&{\color{red} \bf 7.9}&{\color{red} \bf 8.7}&{\color{red} \bf 10.9}&{\color{red} \bf 9.8}\\
				&0.5&{\color{red} \bf 6.9}&{\bf 6.4}&6.5&{\color{red} \bf 10.5}&{\color{red} \bf 11.0}&&6.3&{\bf 4.7}&6.1&{\color{red} \bf 12.2}&{\color{red} \bf 7.2}&&{\color{red} \bf 9.3}&{\color{red} \bf 8.4}&{\color{red} \bf 9.0}&{\color{red} \bf 11.2}&{\color{red} \bf 12.1}\\
				&0.9&{\color{red} \bf 8.1}&{\bf 5.8}&{\color{red} \bf 8.8}&{\color{red} \bf 12.2}&{\color{red} \bf 18.2}&&{\color{red} \bf 6.9}&{\bf4.5}&{\color{red} \bf 7.0}&{\color{red} \bf 17.6}&{\color{red} \bf 17.6}&&{\color{red} \bf 10.2}&{\color{red} \bf 8.2}&{\color{red} \bf 11.5}&{\color{red} \bf 13.6}&{\color{red} \bf 15.8}\\
				\\
				Chisq&-0.9&{\bf 5.2}&5.4&5.8&{\color{red} \bf 7.1}&4.4&&5.4&{ 5.2}&{\bf 5.0}&{ 4.9}&{ 5.3}&&{\bf 5.3}&5.4&{\color{red} \bf 7.0}&{\color{red} \bf 7.6}&3.8\\
				&-0.5&4.6&{ 5.3}&{ 5.3}&{\color{red} \bf 6.9}&{\bf 5.0}&&{5.2}&{ 4.7}&{ 4.8}&{\bf 5.0}&{ 5.1}&&{\bf 4.9}&5.5&6.6&{\color{red} \bf 7.4}&{ 5.2}\\
				&-0.1&{\bf5.5}&6.0&5.9&{\color{red} \bf 7.2}&6.6&&{5.1}&{4.9}&{\bf 5.0}&{ 4.9}&5.4&&{\bf5.6}&5.8&6.5&{\color{red} \bf 7.7}&6.2\\
				&0.1&{\bf 5.0}&5.4&{ 5.1}&6.6&5.9&&{5.2}&{\bf 5.0}&{4.9}&{ 4.8}&5.6&&{\bf5.8}&5.9&6.5&{\color{red} \bf 7.3}&{\color{red} \bf 7.0}\\
				&0.5&{\bf 5.0}&{\bf5.5}&5.4&{\color{red} \bf 6.9}&{\color{red} \bf 6.8}&&{ 5.1}&{ 4.9}&{\bf 5.0}&{ 5.1}&5.6&&5.8&{\bf5.5}&6.0&{\color{red} \bf 7.2}&{\color{red} \bf 7.2}\\
				&0.9&5.8&{\bf 5.3}&6.3&{\color{red} \bf 7.1}&{\color{red} \bf 13.0}&&5.5&{ 5.2}&5.7&{\bf 5.0}&{\color{red} \bf 11.9}&&{\bf5.5}&{\bf5.5}&{\color{red} \bf 7.4}&{\color{red} \bf 7.4}&{\color{red} \bf 8.3}\\\hline
								\end{tabular*} }
	\end{table*}

\section{Simulation Study}
\label{SimStudy}
In this section, we investigate the finite sample behavior of the methods described in Sections \ref{SecPBtsrtp} and \ref{Secpooling} in extensive simulations. All procedures were studied with respect to their 
\begin{enumerate}[(i)]
	\item type-I-error rate control at level $\alpha = 5\%$ and their
	\item  power to detect deviations from the null hypothesis. 
\end{enumerate}

Small to moderate sized paired data samples were generated from the model
\begin{center} 
	$\mathbf{X}_j = \boldsymbol{\Sigma}^{\frac{1}{2}} \boldsymbol{\epsilon}_j + \boldsymbol{\mu},\quad  j=1,... ,n ,$
\end{center}
where $\boldsymbol{\epsilon}_j = [\epsilon_{1j}, \epsilon_{2j}]^\top$ is an i.i.d. bivariate random vector with mutually independent components and $E(\boldsymbol{\epsilon_1})=0$ and $Cov(\boldsymbol{\epsilon_1})=I_2$.

Different choices of symmetric as well as skewed residuals are considered such as standardized normal, exponential, Laplace and the $\chi^2$ distribution with $df = 30$ degrees of freedom. For the covariance matrix $\boldsymbol{\Sigma}$, we considered the choices  
\begin{center}
	$ \boldsymbol{\Sigma_1} =  \begin{bmatrix*}[c]
	1 & \rho \\
	\rho & 1
	\end{bmatrix*} \text{ and } \boldsymbol{\Sigma_2} =  \begin{bmatrix*}[c]
	1 & \sqrt{2}\rho \\
	\sqrt{2}\rho & 2
	\end{bmatrix*} \textbf{}.$
\end{center} 
with varying correlation factor $\rho \in (-1,1)$, representing a homoscedastic and a heteroscedastic covariance setting, respectively. 
The sample sizes were chosen as $(n_c,n_u)\in \{(10,10),(30,10),(10,30)\}$ under a MCAR mechanism and $n\in \{10,20,30\}$ under a MAR mechanism. 

For each scenario, we generated missings as described below:
For the {\it MCAR mechanism}, missing values are inserted randomly to  the second component of the bivariate vector $\mathbf{X}_{j}$ until a fixed amount of missing values of size $n_u$ for  the second component is achieved.\\

\begin{table*}[ht]
	
	\caption{Type-I error simulation results ($\alpha=0.05$) of the tests for different distributions under varying correlation values ($\rho$), different total sample sizes $n\in\{10,20,30\}$ (numbers of subjects) and homoscedastic covariance matrix $\Sigma_1$ under the MAR framework. For each setting, the values closest to the prescribed level are printed in {\bf bold} and values exceeding the upper limit ($6.8\%$) of the $99\%$ binomial interval are in {\color{red} \bf red} colour. 
		\label{Table:Type1MARhomo}}
	{\tabcolsep=4pt
		\begin{tabular*}{1.09\linewidth}{lc ccccc c ccccc  c ccccc   cc}
			Dist & $\rho$ & \multicolumn{5}{c}{$n=10$}  && \multicolumn{5}{c}{$n=20$} && \multicolumn{5}{c}{$n=30$} \\ \cline{3-7} \cline{9-13} \cline{15-19}
			&& $T_{W}^*$ & $T_{A}^*$ & $T_{M}^*$ & $T_{NC}$ & $T_{LT}$
			&& $T_{W}^*$ & $T_{A}^*$ & $T_{M}^*$ & $T_{NC}$ & $T_{LT}$ 
			&& $T_{W}^*$ & $T_{A}^*$ & $T_{M}^*$ & $T_{NC}$ & $T_{LT}$\\
			Normal&-0.9&{ 5.4}&3.7&{\bf 5.2}&6.4&3.6&&{ 5.4}&{\bf 5.1}&5.7&5.8&5.5&&{\bf 5.1}&{ 5.4}&6.0&5.7&5.5\\
			&-0.5&{\bf 4.8}&3.9&4.4&6.4&{ 4.6}&&{5.3}&{\bf 4.9}&5.5&5.6&{5.2}&&{ 5.2}&{\bf 5.1}&5.8&{ 5.3}&{\bf 5.1}\\
			&-0.1&{\bf 4.6}&4.1&4.4&6.2&5.7&&{ 5.1}&{ 4.7}&{\bf 5.0}&5.5&5.5&&{\bf 5.0}&{\bf 5.0}&{ 5.1}&{ 5.3}&{ 5.2}\\
			&0.1&{\bf 4.7}&4.3&4.3&6.7&6.2&&{4.9}&4.3&{\bf 5.0}&{5.1}&5.9&&{\bf 5.2}&4.4&{ 5.3}&{5.3}&5.5\\
			&0.5&{\bf 4.7}&4.2&4.1&6.3&{\color{red} \bf7.3}&&{ 5.2}&{\bf 5.0}&{ 5.1}&5.5&6.4&&{ 5.3}&{ 4.8}&{\bf 5.0}&{ 4.8}&5.9\\
			&0.9&{\bf 4.9}&{\bf 4.9}&4.3&6.5&{\color{red} \bf 12.5}&&{\bf 5.0}&{4.8}&{ 4.6}&{ 5.3}&{\color{red} \bf14.2}&&{ 5.1}&{ 5.1}&{ 4.8}&{\bf 5.0}&{\color{red} \bf 12.6}\\
			\\
			Laplace&-0.9&4.2&3.2&{\bf 5.1}&6.1&3.2&&4.3&{ 4.9}&6.0&{5.3}&{\bf 5.0}&&{ 4.9}&{ 5.1}&6.5&{5.3}&{\bf5.0}\\
			&-0.5&3.1&3.6&3.5&{\bf 6.2}&3.9&&4.2&{ 4.6}&5.5&5.5&{\bf 5.0}&&4.4&{\bf 5.0}&6.0&{ 5.3}&{\bf5.0}\\
			&-0.1&3.1&3.2&3.1&6.3&{\bf 4.8}&&3.9&3.9&{ 4.6}&{\bf 5.1}&{ 4.7}&&4.4&4.3&{ 4.9}&{\bf 5.0}&{\bf5.0}\\
			&0.1&3.2&2.7&3.0&6.3&{\bf 5.6}&&4.1&3.7&{ 4.6}&5.6&{\bf 5.3}&&{ 4.6}&4.4&{ 5.3}&{\bf 5.2}&5.6\\
			&0.5&3.5&3.1&3.0&{\bf6.3}&{\color{red} \bf7.8}&&4.5&4.0&4.2&{\bf 5.2}&5.6&&4.5&3.6&4.2&{\bf 4.8}&{ 5.3}\\
			&0.9&3.6&3.9&3.3&{\bf5.6}&{\color{red} \bf 11}&&4.0&4.1&3.8&{\bf 4.9}&{\color{red} \bf14.4}&&4.5&4.4&4.3&{\bf 4.6}&{\color{red} \bf14.4}\\
			\\
			Exp&-0.9&{\bf4.3}&2.8&5.6&6.1&3.2&&5.6&{\bf 5.0}&{\color{red} \bf7.8}&{\bf 5.0}&{ 5.3}&&6.0&{\bf 5.1}&{\color{red} \bf8.7}&5.5&{ 5.3}\\
			&-0.5&{\bf 4.8}&3.0&4.2&6.6&4.1&&{\color{red} \bf7.6}&{\bf 4.9}&{\color{red} \bf7.4}&5.7&{ 5.4}&&{\color{red} \bf8.9}&{ 5.3}&{\color{red} \bf8.7}&{\bf 5.2}&{5.3}\\
			&-0.1&4.3&2.7&3.4&6.5&{\bf 5.2}&&{\color{red} \bf7.4}&4.2&6.4&{ 5.4}&{\bf 5.2}&&{\color{red} \bf8.8}&{\bf 5.4}&{\color{red} \bf8.0}&5.6&{\bf 5.4}\\
			&0.1&{\bf4.0}&2.6&3.4&6.5&6.2&&{\color{red} \bf7.1}&4.3&6.5&{\bf5.6}&5.9&&{\color{red} \bf8.1}&{ 5.4}&{\color{red} \bf7.6}&5.5&{\bf 5.1}\\
			&0.5&{\bf3.5}&2.7&3.2&{\bf6.5}&{\color{red} \bf10.1}&&5.7&{\bf 4.8}&5.7&5.7&{\color{red} \bf 6.9 }&&6.5&6.1&6.5&{\bf5.6}&{\bf5.6}\\
			&0.9&4.3&{\bf 4.7}&4.2&5.8&{\color{red} \bf10.2}&&{\bf 5.0}&5.7&{ 5.2}&{5.2}&{\color{red} \bf13.4}&&5.6&{\color{red} \bf 6.9 }&5.7&{\bf 5.2}&{\color{red} \bf 15.3}\\
			\\
			Chisq&-0.9&{\bf 4.6}&3.7&4.5&5.6&3.7&&{\bf 5.1}&{ 5.3}&5.9&5.9&5.5&&{\bf 5.0}&{ 5.2}&6.0&{5.4}&{5.3}\\
			&-0.5&{\bf4.5}&3.7&4.4&6.3&4.3&&{\bf 5.1}&{ 4.8}&{5.4}&{ 5.4}&{ 5.3}&&{5.4}&{\bf 4.8}&5.5&{ 5.3}&{ 5.3}\\
			&-0.1&{\bf 4.7}&3.8&4.4&6.6&5.7&&{ 5.3}&{ 4.7}&{ 5.3}&{\bf 5.2}&5.8&&5.7&{\bf 5.1}&5.7&{5.4}&5.6\\
			&0.1&{\bf 4.8}&3.7&4.1&6.1&6.1&&{\bf 5.1}&4.1&{ 4.7}&{ 5.3}&5.5&&{ 5.3}&{ 4.8}&{ 5.3}&{\bf 5.0}&{ 5.2}\\
			&0.5&{\bf4.2}&3.8&3.6&6.0&{\color{red} \bf7.3}&&{\bf 4.9}&4.4&{ 4.8}&{ 5.3}&6.1&&{ 4.9}&{ 4.7}&{ 4.8}&{ 4.9}&{\bf5.0}\\
			&0.9&4.5&{\bf 4.6}&3.8&5.5&{\color{red} \bf12.8}&&{\bf 4.9}&{ 5.3}&4.5&{\bf 4.9}&{\color{red} \bf 14.7}&&{ 5.2}&{ 5.2}&{\bf 4.9}&{ 5.2}&{\color{red} \bf 13.0}\\\hline				\end{tabular*} }
\end{table*}

\begin{table*}[ht]
	
	\caption{Type-I error simulation results ($\alpha=0.05$) of the tests for different distributions under varying correlation values ($\rho$), different total sample sizes $n\in\{10,20,30\}$ (numbers of subjects)  and heteroscedastic covariance matrix $\Sigma_2$ under the MAR framework. For each setting, the values closest to the prescribed level are printed in {\bf bold} and values exceeding the upper limit ($6.8\%$) of the $99\%$ binomial interval are in {\color{red} \bf red} colour. 
		\label{Table:Type1MARhetro}}
	{\tabcolsep=4pt
		\begin{tabular*}{1.09\linewidth}{lc ccccc c ccccc  c ccccc   cc}
			Dist & $\rho$ & \multicolumn{5}{c}{$n=10$}  && \multicolumn{5}{c}{$n=20$} && \multicolumn{5}{c}{$n=30$} \\ \cline{3-7} \cline{9-13} \cline{15-19}
			&& $T_{W}^*$ & $T_{A}^*$ & $T_{M}^*$ & $T_{NC}$ & $T_{LT}$
			&& $T_{W}^*$ & $T_{A}^*$ & $T_{M}^*$ & $T_{NC}$ & $T_{LT}$ 
			&& $T_{W}^*$ & $T_{A}^*$ & $T_{M}^*$ & $T_{NC}$ & $T_{LT}$\\
		
		Normal&-0.9&{\bf 5.1}&4.1&{\bf 4.9}&6.3&4.0&&{ 5.2}&5.4&5.9&{\bf 5.1}&5.5&&5.4&5.6&6.5&{\bf 5.2}&5.6\\
		&-0.5&{\bf 5.1}&4.4&{ 5.2}&{\color{red} \bf6.8}&{ 5.2}&&{\bf 5.1}&{ 5.3}&5.9&{ 5.3}&5.4&&4.6&{\bf 5.2}&5.8&{\bf 4.8}&{\bf 5.2}\\
		&-0.1&{\bf 4.9}&4.4&4.5&6.3&6.0&&{\bf 5.1}&{ 5.2}&5.5&{ 5.2}&5.6&&{ 5.1}&{\bf 5.0}&5.4&4.5&{ 5.3}\\
		&0.1&{\bf 5.2}&4.5&4.5&6.6&6.5&&{5.3}&{\bf 5.1}&5.5&5.4&6.5&&{\bf 5.1}&{ 4.7}&{ 5.2}&4.5&{\bf 5.1}\\
		&0.5&{\bf4.4}&4.0&4.0&6.3&{\color{red} \bf 7.3}&&{\bf 5.1}&{\bf 4.9}&{ 5.2}&{ 5.2}&6.0&&{ 5.2}&{ 4.8}&{ 5.2}&{\bf 5.0}&5.9\\
		&0.9&4.6&{ 4.8}&{\bf 5.0}&{\color{red} \bf 7.0}&{\color{red} \bf 14.7}&&{\bf 4.9}&{ 4.7}&{ 5.3}&{ 5.2}&{\color{red} \bf 14.1}&&{\bf 5.0}&{\bf 5.0}&{ 5.3}&4.6&{\color{red} \bf11.8}\\
		\\
		Laplace&-0.9&4.0&3.4&{\bf 5.2}&6.3&3.6&&4.2&{4.9}&{\color{red} \bf6.8}&{4.8}&{\bf 5.0}&&4.4&{\bf 5.1}&{\color{red} \bf 7.4}&{ 4.8}&{ 5.3}\\
		&-0.5&3.2&3.6&4.1&6.0&{\bf4.2}&&4.5&{\bf 5.0}&6.4&5.4&{5.2}&&{ 4.7}&{\bf 4.9}&6.7&{\bf 4.9}&{ 4.8}\\
		&-0.1&3.1&3.7&4.0&6.5&{\bf5.4}&&3.9&4.5&{ 5.3}&{5.3}&{\bf 5.2}&&{ 4.7}&{\bf 5.0}&6.1&{ 4.8}&{ 5.1}\\
		&0.1&3.1&2.8&3.0&5.9&{\bf5.7}&&4.2&4.3&{ 5.1}&{\bf 5.0}&5.5&&4.3&{\bf 5.0}&5.9&{ 4.9}&5.8\\
		&0.5&3.4&3.4&3.4&{\bf6.5}&{\color{red} \bf 8.1}&&4.2&4.0&4.5&{\bf 5.1}&6.0&&4.4&4.6&{\bf 5.1}&4.6&5.6\\
		&0.9&3.3&{\bf4.1}&3.9&6.6&{\color{red} \bf 13.1}&&4.3&{\bf 4.9}&{\bf 5.1}&{5.2}&{\color{red} \bf 16.2}&&{ 4.7}&{\bf 4.9}&5.5&{\bf 4.9}&{\color{red} \bf 15.3}\\
		\\
		Exp&-0.9&3.3&4.0&6.0&{\color{red} \bf 7.4}&{\bf4.3}&&{\bf 4.7}&5.7&{\color{red} \bf 9.0}&{\color{red} \bf 7.4}&6.0&&{\bf 5.3}&5.5&{\color{red} \bf 8.6}&{\color{red} \bf 7.7}&5.6\\
		&-0.5&4.2&4.1&4.1&{\color{red} \bf 7.8}&{\bf5.4}&&6.5&{\bf5.5}&{\color{red} \bf 7.0}&{\color{red} \bf 7.2}&6.4&&{\color{red} \bf 8.4}&{\bf5.8}&{\color{red} \bf 8.5}&{\color{red} \bf 8.2}&6.2\\
		&-0.1&{\bf4.5}&3.5&3.8&{\color{red} \bf8.2}&{\color{red} \bf 7.0}&&{\color{red} \bf6.9}&{\bf 4.7}&6.2&{\color{red} \bf 7.4}&6.2&&{\color{red} \bf 8.6}&{\bf5.6}&{\color{red} \bf 7.9}&{\color{red} \bf 9.0}&6.3\\
		&0.1&{\bf3.9}&3.5&3.5&{\color{red} \bf 8.6}&{\color{red} \bf 8.5}&&6.3&{\bf4.4}&6.1&{\color{red} \bf 7.7}&6.7&&{\color{red} \bf 7.8}&{\bf 5.2}&{\color{red} \bf 7.2}&{\color{red} \bf 8.8}&5.9\\
		&0.5&{\bf4.6}&2.9&4.2&{\color{red} \bf 7.8}&{\color{red} \bf 11.0}&&6.3&{\bf4.5}&6.3&{\color{red} \bf 8.1}&{\color{red} \bf 8.4}&&{\color{red} \bf 6.9}&{\bf 5.2}&{\color{red} \bf 7.0}&{\color{red} \bf 9.3}&{\color{red} \bf 7.3}\\
		&0.9&{\color{red} \bf 7.8}&{\bf4.0}&{\color{red} \bf 7.8}&{\color{red} \bf 8.7}&{\color{red} \bf 15.9}&&{\color{red} \bf 7.7}&{\bf4.5}&{\color{red} \bf 8.1}&{\color{red} \bf 8.6}&{\color{red} \bf 17.5}&&{\color{red} \bf 8.1}&{\bf5.4}&{\color{red} \bf 7.7}&{\color{red} \bf 10.4}&{\color{red} \bf 17.2}\\
		\\
		Chisq&-0.9&{\bf 5.1}&4.2&{ 5.3}&6.1&4.3&&{\bf 5.8}&{\bf 5.8}&6.6&{\bf 5.8}&5.9&&{ 5.2}&5.5&6.6&{\bf 5.0}&5.6\\
		&-0.5&{ 4.8}&4.5&{\bf 5.1}&6.2&{\bf 5.1}&&{\bf 5.1}&{4.7}&5.5&{\bf 5.1}&{ 5.2}&&{ 5.3}&{ 5.1}&6.0&{\bf 5.0}&{\bf5.0}\\
		&-0.1&4.5&4.3&{\bf 4.7}&6.5&6.3&&{ 5.3}&{\bf 5.2}&5.7&{\bf 5.2}&5.6&&6.1&5.5&6.1&{\bf5.4}&5.8\\
		&0.1&{\bf4.3}&4.0&4.0&6.3&6.6&&{\bf 5.0}&{4.8}&{ 5.2}&{5.2}&5.8&&{\bf 5.1}&{\bf 5.1}&5.5&{\bf 4.9}&5.6\\
		&0.5&{\bf 4.7}&3.9&4.3&6.5&{\color{red} \bf 7.6}&&5.4&{\bf 4.8}&{ 5.3}&5.6&6.5&&5.5&{ 4.8}&5.5&{\bf 5.1}&5.8\\
		&0.9&{ 4.9}&4.5&{\bf 5.0}&{\color{red} \bf 7.0}&{\color{red} \bf 15}&&{ 5.3}&{\bf 4.8}&5.4&5.6&{\color{red} \bf 15.2}&&5.7&{\bf 5.1}&5.6&{\bf 5.1}&{\color{red} \bf 13.3}\\\hline					\end{tabular*} }
\end{table*}

For the {\it MAR mechanism}, the probability of being missing on the  the second component of  $\mathbf{X}_{j}$ is based on the corresponding value on the first component in the following way: first, we divide $\mathbf{X}$ into three groups based on their first component values corresponding to a $2\sigma-$rule: the first group is given by $\{\vX_j=(X_{1j},X_{2j}): X_{1j}\in (-\infty, -2\whsigma_1), j=1,.., n\}$, the second by $\{\vX_j: X_{1j}\in (-2\whsigma_1,2\whsigma_1), j=1,.., n\}$ and the last group by $\{\vX_j: X_{1j}\in  (2\whsigma_1,\infty), j=1,.., n\}$, where $\whsigma_1$ is the estimated sample variance from all first components. Then, we randomly insert missing values on the second component based on the following missing percentages: $15\%$ for group one and three and $30\%$ for the second group 
.\\

In order to assess the power of all methods, we set $\boldsymbol{\mu} = [\delta, 0]^\top$ with shift parameter $\delta \in \{ 0, 1/2, 1 \}$. All simulations were operated by means of the statistical computing environment $\textsf{R}$ based on $n_{sim} = 10,000$ Monte-Carlo runs and $B=1,000$ bootstrap runs (in case of the three bootstrapped methods based upon $T_W^*,T_A^*$, and $T_M^*$). The algorithm for the computation of the $p$-value of the parametric bootstrap tests is as follows:
\begin{enumerate}
	\item For the given incomplete paired data,  calculate the observed test statistic, say $T$. 
	\item Estimate the covariance matrix $\boldsymbol{\Gamma} $ by $\hat{\boldsymbol{\Gamma} }$.
	
	\item Generate a bootstrap sample $\mathbf{X}^*_{j}= (X_{1j}^{*}, X_{2j}^{*})$ from $N(\boldsymbol{0},\hat{\boldsymbol{\Gamma} })$, $j=1,..., n$.
	\item Insert missing values { in a MCAR or MAR manner} to  the second component of the vector $\mathbf{X}^*_{j}$  resulting in 
	$\mathbf{X}^{*(c)}_{j}$ and $\mathbf{X}^{*(i)}_k$ where $j=1,\dots,n_c,$ $k=1,\dots,n_u$.
		
	\item Calculate the value of the test statistic for the bootstrapped sample $T^*$. 
	\item Repeat the Steps $3$ and $4$ independently $B=1,000$ times and collect the observed test statistic values in ${T^*_{b}, b=1,.....,B}$.
	\item Finally, estimate the bootstrap $p$-value as   $p-value=\frac{\sum_{b=1}^{B}I(T^*_{b}>= T)}{B}$. 
\end{enumerate}

{\bf Type-I-Error Results.}
Simulation results of type-I error level of the studied procedures under the MCAR framework for different sample sizes and for homoscedastic as well as heteroscedastic settings are summarized in Table \ref{Type1errorhomoMCAR}  and Table \ref{Type1errorhetroMCAR}  respectively.\\ 
It can be readily seen 
that the suggested bootstrap approaches based upon $T_W^*, T_A^*$ and $T_M^*$ tend to result in quite accurate type-I error rate control under homoscedasticity as well as heteroscedasticity and over the whole range of correlation factors for most settings. Only in two cases; First, in case of the negative unbalanced sample size $(10,30)$, particularly under heteroscedasticity, the bootstrapped MATS ($T_M^*$) is not recommended due to it's liberal behavior. However, in such case, the other two suggested bootstrapped tests $T_W^*$, and $T_A^*$ are controlling type-I error rate accurately. Secondly, in case of the skewed exponential distribution, the control is not adequate and a liberal behavior is observed. However, in this case, all the other chosen procedures also failed to control type-I error rate for the underlying sample sizes, { which are indicated in bold red through all tables. Specifically}, in the case of homoscedasticity, and a balanced sample size  $(10,10)$, our three suggested tests still result in accurate test decisions. For a positive balanced sample size $(30,10)$, the bootstrapped ANOVA ($T_A^*$) still controls type-I error rate accurately under homoscedastic as well heteroscedastic settings.  It has even the best control of type-I error rate among all considered methods that are identified by bold entries in the table. \\

In contrast, the other tests ($T_{NC}$, $T_{LT}$) do not control type-I error level constantly
under heteroscedasticity or even under homoscedasticity in { all of the considered} sample sizes. It can also be seen from Table \ref{Type1errorhomoMCAR}  and Table \ref{Type1errorhetroMCAR} that the nonparametric combination test $T_{NC}$, controls type-I error quite accurately in the case of larger numbers of complete pairs  $(n_c=30)$, but it shows liberal behavior for smaller numbers of complete pairs $(n_c=10)$. This test { turns} very liberal in the case of heteroscedasticity. Moreover, the test that is based on the maximum likelihood estimator $T_{LT}$ tends to result in a very liberal decision in the case of { smaller numbers} of complete pairs together with positive correlation factors $\rho$. 
{For larger numbers of complete pairs, it leads to an accurate type-I error rate control for correlation factors being less than  $0.5$.} This behavior of the test does not depend on the homoscedasticity assumption. \\

 It was also interesting to discover the type-I error rate control of the tests under {\it similar attributes to the breast cancer gene study data} which reflects data sets with a few pairs and  large amount of unpaired portions. Simulation results for the type-I error rate of the studied procedures for $(n_c=16,n_u=74)$ sample sizes  are presented in Table S.3 and Table S.4 in the supplement. { The correlation $\rho$ in Table S.4  is estimated based on the data } It can be easily seen from Table S.3 and Table S.4 that the bootstrap tests are robust under large amounts of missing observations and control type-I error rate accurately, especially the bootstrapped tests $T_W^*$, and $T_A^*$. Except in the case of skewed distribution. The alternative approaches $T_{NC}$ and $T_{LT}$ have acceptable control under homoscedasticity, but, under the exponential distribution, { they turned} very liberal under heteroscedasticity.\\

Simulation results of the type-I error level of the studied procedures under the MAR framework for different sample sizes and {covariance structures} are summarized in Table \ref{Table:Type1MARhomo}  and \ref{Table:Type1MARhetro} respectively.
{ There, it can be seen} that for moderate to large sample sizes ($n\in\{20,30\}$), the bootstrapped ANOVA $T^*_{A}$, the bootstrapped Wald $T_W^*$ and the nonparametric combination test $T_{NC}$ exhibit a fairly good type-I-error rate control for almost all considered scenarios under homoscedasticity as well as heteroscedasticity. Only in the case of the skewed exponential distribution, the control of $T_{W}$ and $T_{NC}$ is not adequate and liberal behavior is observed { which is marked with bold red through all tables.} In contrast, the bootstrapped MATS $T_{M}$ and Little's $T_{LT}$ tend to be  sensitive to the dependency structure in the data. In particular, $T_{M}$ exhibits a liberal behavior for negative correlations, while $T_{LT}$ { does the same} for positive correlations. For small sample sizes $(n = 10)$, the $T_{NC}$  test tends to be liberal in all considered situations while $T_{LT}$  performs well and is only liberal for positive correlations. In contrast, the bootstrapped tests $T^*_W$ and $T_M^*$ exhibit good type-I-error rate control for most settings except for the Laplace distribution. 
The bootstrapped ANOVA $T_{A}^*$ tends to be very conservative especially under heteroscedasticity.\\

{ }

{\bf Further Investigations on Type-I-Error}. In addition to the small and moderate sample size settings, we were also interested in studying type-I error rate control when {\it sample sizes increase}, while missing rates remain nearly unchanged. For moderate to large sample sizes, we considered the choices $(n_c,n_u)= k\cdot(1,1)+(10,10)$ and $(n_c,n_u)=k\cdot(10, 30)$, where $k$ ranges from $0-500$ (unbalanced case) and $1-50$ (balanced case), respectively. Figure $1$ and Figure $S.1$ (in the supplement) summarize the  type-I error rate ($\alpha=0.05$) for these settings. The results indicate that the nonparametric combination test by Qi et. al. \cite{qi2018testing} $T_{NC}$  controls type-I error rate quite accurate under symmetric distributions, however, it fails to control type-I error rate under skewed distributions. In fact, it gets even more  liberal with increasing sample sizes. In contrast, Little's test \cite{little1976inference} $T_{LT}$ 
tends to be highly liberal when small sample sizes like $(n_c,n_c)=(10,10)$ are present and less liberal with accurate type-I error control for larger sample sizes. Only the suggested bootstrap approaches $T_A^*,T_W^*$, and $T_{M}^* $ control type-I error rate accurately among all considered settings.\\

In order to cover the {\it effect of increasing missing rates}, we studied type-I error control for sample sizes of the form $(n_c,n_u)=((1-r)\cdot30,r\cdot30)$ with $r\in\{0.1,0.2,0.3,0.4,0.5,0.6,0.7,0.8\}$ covering missing rates  (among subjects) from $10\%$ to $80\%$. Figure \ref{fig:MCARMissRate} and Figure S.2 in the supplement summarize type-I error rate control for these settings under a homoscedastic and a heteroscedastic covariance structure,  respectively.  The results indicate that under homoscedasticity, the alternative approaches $T_{NC}$ and $T_{LT}$ tend to be slightly liberal. They move closer to the $0.05$ threshold for missing rates below $60\%$.  In contrast, under heteroscedasticity, $T_{LT}$ keeps the same behavior while $T_{NC}$ tends to be more sensitive to the
missing rates. In particular, it exhibits a  conservative or liberal behavior for lower and larger missing rates, respectively.  In contrast, the suggested bootstrap approaches tend to control type-I error rate more accurate over the range of missing rates $r$ for most settings. Only in  case of the skewed exponential distribution and missing rates greater than $50\%$, the control is not adequate. However, in this case all the other chosen procedures also failed to control the type-I error rate. Especially, Little's test $T_{LT}$ tends to be very liberal under the whole range of missing rates.\\

{\bf Power.} In addition to the type-I error rate control, we studied the power of the five  tests  for all considered settings. Due to the rather liberal behavior of the nonparametric combination test $T_{NC}$ and the maximum likelihood test $T_{LT}$  in the case of small number of complete pairs, their power functions are not really comparable to the others. Therefore, we present here the power simulation results for the case of large numbers of complete pairs. Hence, we consider positively balanced sample sizes $(n_c,n_u)=(30,10)$ and $n=20$ for MCAR and MAR mechanisms respectively. The Power simulation results for the other scenarios are included in the supplement. The power analysis results of the considered methods under MCAR and MAR frameworks involving homoscedastic as well as heteroscedastic settings are summarized in Table \ref{Powern3010homo} and \ref{Powern3010hetro} { for the MCAR mechanism}   and Table S.8 and  S.9 in the supplement  { for the MAR mechanism}. The entries that belong to very liberal tests have been coloured in red in the power tables.
It can be easily seen that the five tests have almost similar large power behavior under homoscedastic as well as heteroscedastic settings. Only in the heteroscedastic cases with skewed exponential distribution, the nonparametric combination test $T_{NC}$ shows  larger power than the others, which is  due to it's rather liberal behavior. {One should also notice} that the power behavior of each test varies based on the dependency structure of the data except for the bootstrapped ANOVA test ($T_A^*$).

\begin{figure}[h]
	\begin{center}
		
		\includegraphics[scale=0.8]{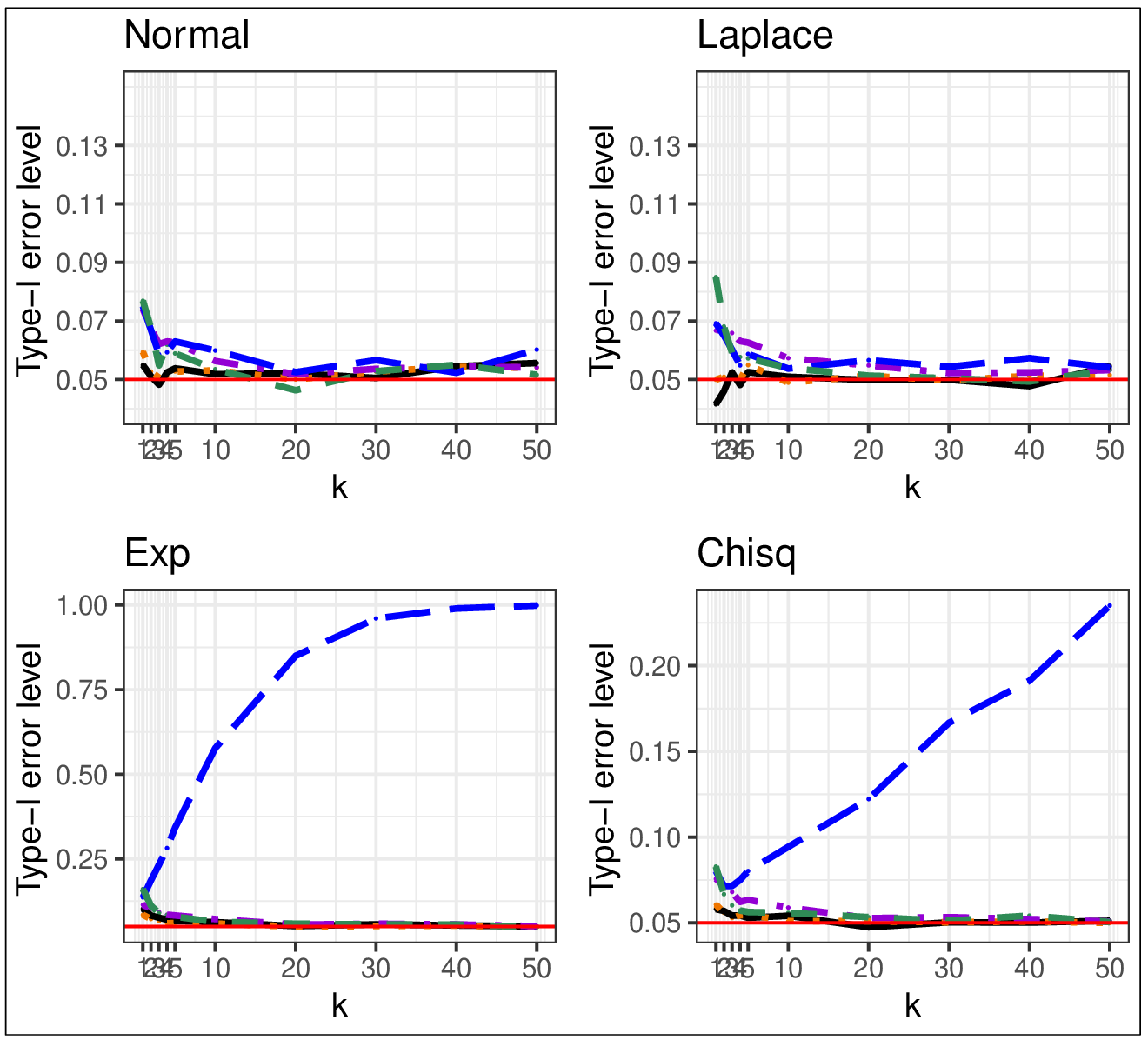}
		
	\end{center}
	\caption{Type-I error simulation results ($\alpha=0.05$) of the tests  $T_W^*$ $({\color{black}\textendash\textendash\textendash})$, $T_A^*$ $({\color{brown}\cdots})$, 	$T_M^*$ $({\color{violet}-\cdot-})$, $T_{NC}$ $({\color{blue} \textendash\textendash \quad \textendash\textendash})$,  and  $T_{LT}$ $({\color{ao(english)} -- })$, for different distributions under correlation factor ($\rho=0.9$)  and heteroscedastic covariance matrix $\Sigma_2$ for varying $k$ values multiplied by  $(n_c,n_u)=(10,30)$ under the MCAR framework.}
	\label{fig:kmultn}	
	
\end{figure}

	\begin{figure}[h]
		\begin{center}
			
			\includegraphics[scale=0.8]{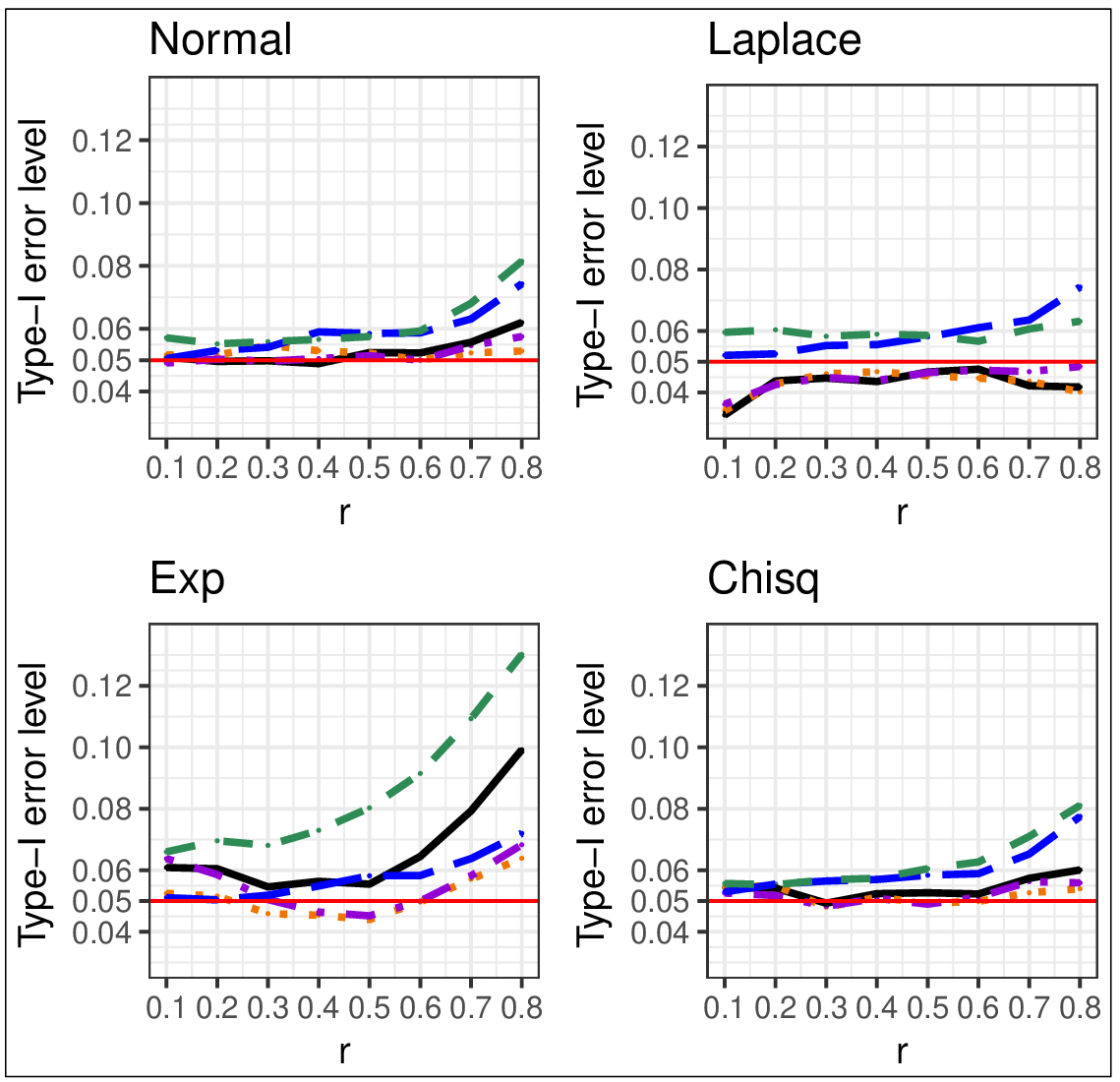}
			
		\end{center}
		\caption{Type-I error simulation results ($\alpha=0.05$) of the tests  $T_W^*$ $({\color{black}\textendash\textendash\textendash})$, $T_A^*$ $({\color{brown}\cdots})$, 	$T_M^*$ $({\color{violet}-\cdot-})$, $T_{NC}$ $({\color{blue} \textendash\textendash \quad \textendash\textendash})$,  and  $T_{LT}$ $({\color{ao(english)} -- })$, for different distributions under correlation factor ($\rho=0.5$) with sample size $(n=30)$  and homoscedastic covariance matrix $\Sigma_1$ for varying missing rates  $r\in\{0.1,0.2,0.3,0.4,0.5,0.6,0.7,0.8\}$ under the MCAR framework.}
		\label{fig:MCARMissRate}	

	\end{figure}

\section{Breast Cancer Study: Gene Expression Data}
	\label{Secdata}
	
The Cancer Genome Atlas (TCGA) project is a pilot project which was launched in 2005 with a financial support from the National Institutes of Health. It aims to 	understand the genetic basis of several types of human cancers through the application of  high-throughput genome analysis techniques. TCGA collects molecular information such as miRNA/mRNA expressions, protein expressions, weight of the sample as well as clinical data about the  patients. \\

A breast cancer study has been performed by TCGA to improve the ability of diagnosing, treating and preventing breast cancer through investigating the genetic basis of carcinoma. Their study consists of $1093$ breast cancer patients with Clinical and RNA sequencing records. Among them, there were 112 subjects that provided both, normal and tumor tissues. Here, we were interested in a subset of this data that contains patients with pathologic stage I. This subset contains a total of $n_c=16$ complete pairs and an unpaired sample for the patients who developed only tumor tissues of size
$n_u=74$. The data can be downloaded from Firehouse (www.gdac.broadinstitute.org).\\

\begin{figure}[h!]	
	\centering
	\fbox{\includegraphics[scale=0.2,width=0.5\linewidth]{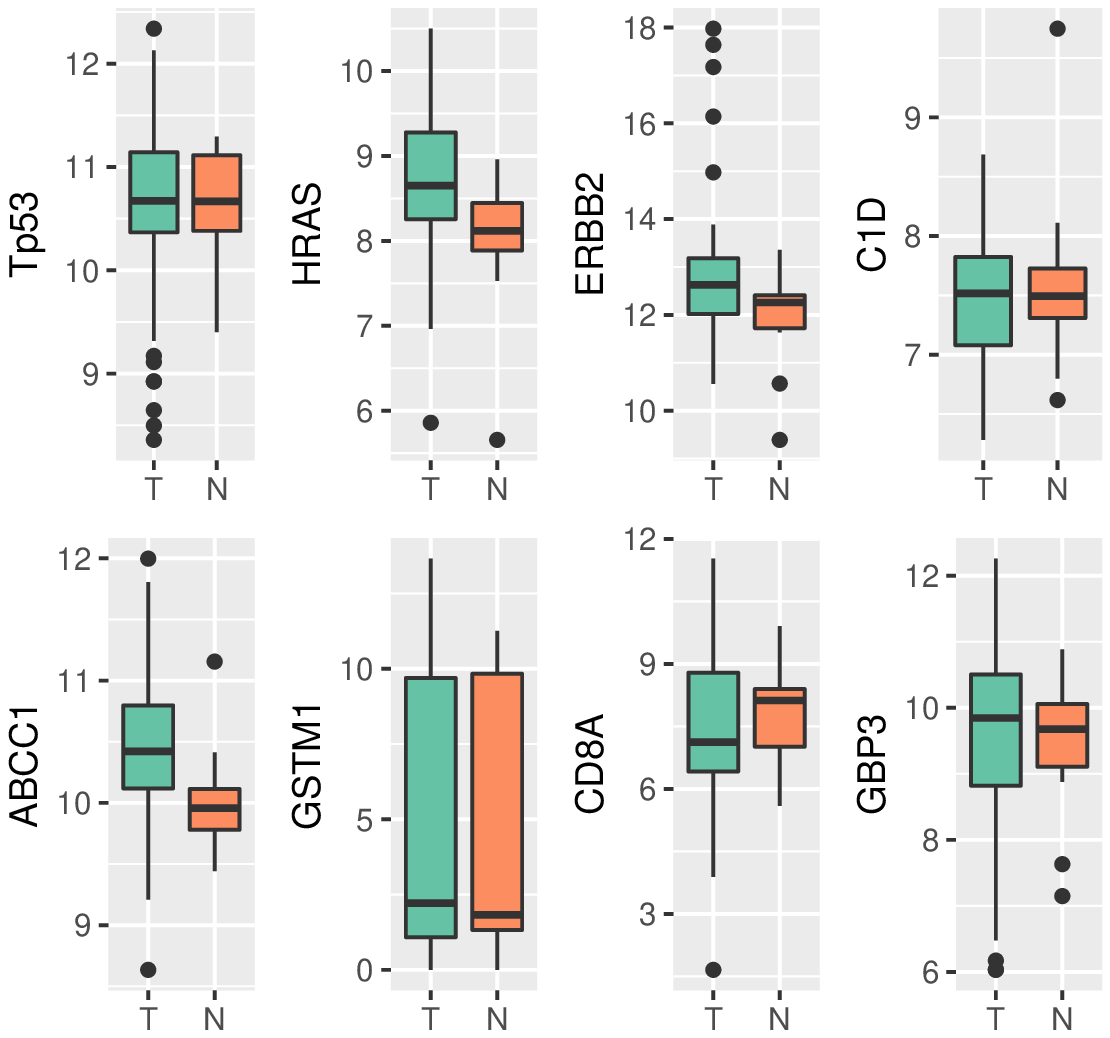}}
	\caption{ Profile of the Gene expression levels of the tumor and normal breast tissues \label{genes}  }
\end{figure}

Based on previous studies, six genes have been found to be significantly associated with breast cancer: \textbf{TP53, ABCC1, HRAS, GSTM1, ERBB2} and \textbf{CD8A}.\cite{finak2008stromal, harari2000molecular, munoz2007role, de2002genes} Another two genes; \textbf{C1D} and \textbf{GBP3} were under investigation although they did not show any significant relation towards breast cancer patients.\cite{qi2018testing} 
In this paper, we aim to test the hypothesis whether mean genetic expressions of the eight genes are significantly different between normal and tumor tissues for patients with early stage I breast cancer. Boxplots representing the characteristics of the eight genes are shown in Figure \ref{genes}.

We applied all  bootstrap testing methods $T_W^*, T_A^*$ and $T_M^*$ as well as the alternative approaches $T_{NC}, T_{LT}$ to detect the null hypothesis of equal
 means between normal and tumor tissues $(H_0 : \mu_1 = \mu_2)$ against the two-sided alternative $(H_1 : \mu_1 \neq \mu_2)$. The results are summarized in Table \ref{pvalue}.\\

It can bee seen from Table \ref{pvalue} that the  bootstrapped approaches $T_W^*, T_A^*$ and $T_M^*$ and Little's method ($T_{LT}$) identified three out of eight genes having significantly different genetic expressions in normal and tumor tissues; genes \textbf{ABCC1, HRAS}, and \textbf{ERBB2}. However, the nonparametric combination method $T_{NC}$ led to different results for the \textbf{ERBB2} gene.

			\begin{table*}[ht]%
				\centering
				   \captionsetup{justification=centering}
				\caption{Unadjusted two-sided P-values of  the breast cancer study.\label{pvalue}}%
				\begin{tabular*}{0.7\textwidth}{@{\extracolsep\fill}lccccc@{\extracolsep\fill}}
					\textbf{Gene}& $\boldsymbol{T}_{\mathbf{W}}^* $  & $\boldsymbol{T}_{\mathbf{A} }^*$ & $\boldsymbol{T}_{\mathbf{M} }^*$& $\boldsymbol{T}_\textbf{NC}$& $\boldsymbol{T}_\textbf{LT}$  \\
					\hline 
					\textbf{TP53}&0.9120&0.8560&0.8990&0.9539&0.8804\\
					\textbf{ABCC1}&0.0000&0.0010&0.0000&0.0030&0.0019\\
					\textbf{HRAS}&0.0050&0.0010&0.0030&0.0011&0.0034\\
					\textbf{GSTM1}&0.8080&0.8280&0.8300&0.6295&0.9202\\
					\textbf{ERBB2}&0.0390&0.0350&0.0140&0.0712&0.0180\\
					\textbf{CD8A}&0.4290&0.4800&0.4690&0.5545&0.4534\\
					\textbf{C1D}&0.7900&0.5550&0.6360&0.5869&0.5097\\
					\textbf{GBP3}&0.2030&0.2900&0.2270&0.1027&0.1294\\
					
					\bottomrule
				\end{tabular*}
				
			\end{table*}

\section{Discussion and outlook}
\label{Discussion}

	The problem of matched pairs with missing  values occurs frequently in practice. Most available procedures in the literature are not applicable when missing values occur in a single arm. One exception is the recent NCT approach by Qi et al. \cite{qi2018testing} who utilize a combination of the sign and Wilcoxon Mann-Whitney rank sum test. For homoscedastic settings with symmetric distributions, this approach can be recommended. If, however, the underlying assumptions are not true (e.g. in skewed heteroscedastic set-ups), the NCT may result in highly inflated type-I-errors or considerable power loss. To overcome these issues, we have provided resampling procedures that are not based on any parametric assumptions and use all observed information of the matched pairs design. They were shown to be asymptotically correct and robust under heteroscedasticity and skewed distributions. The tests were based on restructuring all observed information in a { test statistic of} quadratic form that can be either a Wald-type statistic (WTS), an ANOVA-type statistic (ATS), or a modified ANOVA-type statistic (MATS). Since WTS is well known {(from other situations \cite{vallejo2010analysis, konietschke2015parametric, smaga2017bootstrap}) }for being liberal, while ATS and MATS tend to be rather conservative or liberal for small to moderate sample sizes, we  improved their small sample behavior by an asymptotic model based bootstrap approach. The procedure's asymptotic validity { was also proven}.\\  
	
In an extensive simulation study, the type-I error rate control of the tests have been examined for symmetric and skewed distributions with homoscedastic and heteroscedastic covariance settings  under different missing mechanisms.  There, it was seen that the parametric bootstrap versions of WTS, ATS, and MATS  improve their small sample behaviour. In particular, our bootstrap tests have been shown to perform very well in { most of the} cases, even with larger amount of missingness, heteroscedastic covariance or skewed data. Only the type-I error control { for the exponential distribution}, particularly under heteroscedasticity, MCAR and small paired sample sizes with rather large unpaired portions $(n_c=10,n_u=30)$, is not maintained. In this setting, however, all other considered methods \cite{qi2018testing, little1976inference} also failed to control the type-I error rate. Regarding the individual performance of each bootstrapped test, the parametric bootstrap version of the ATS yielded the most robust results and is therefore recommended.\\ 

Furthermore, our simulation study exhibits that the bootstrap procedures' type-I-error control is not much affected by less stringent missing data mechanism { such as the} MAR. 	However, their power behaviors is quite affected.\\
	
	In order to simplify the application of our approaches, the three proposed bootstrap statistical methods have been implemented within the  {\fontfamily{qcr}\selectfont
			PBT}  function in the freely available R-package {\bf MissPair}. 
		It is available on GitHub (https://github.com/lubnaamro/MissPair) and will be available on the CRAN repository.\\
		
	Future research will be concerned with extending our procedures to multivariate settings (MANOVA).

\begin{acks}
	Burim Ramosaj and Markus Pauly acknowledge the support of the German Research Foundation (DFG). Lubna Amro's work  was also  supported  by  the  German  Academic  Exchange  Service  (DAAD)  under  the 	project: Research Grants - Doctoral Programmes in Germany, 2015/16 (No. 57129429). \\
	{\it Conflict of Interest}: None declared.
\end{acks}

\bibliographystyle{SageV}
\bibliography{reference}{}

\begin{table*}[ht]
\centering
	\caption{Power simulation results ($\alpha=0.05$) of the tests for different distributions under varying correlation values ($\rho$) with sample sizes $(n_c,n_u)=(30,10)$ and homoscedastic covariance matrix $\Sigma_1$ under the MCAR framework. Values of too liberal tests corresponding to {\color{red} \bf red} values in the $(30,10)-$column from Table \ref{Type1errorhomoMCAR} are printed in {\color{red} \bf red} colour. 
		\label{Powern3010homo}}
	{\tabcolsep=4pt
		\begin{tabular*}{0.8\textwidth}{lc ccccc  c ccccc}
			Dist & $\rho$ & \multicolumn{5}{c}{$\delta=0.5$} && \multicolumn{5}{c}{$\delta=1$} \\ \cline{3-7} \cline{9-13} 
			&& $T_{W}^*$ & $T_{A}^*$ & $T_{M}^*$ & $T_{NC}$ & $T_{LT}$ 
			&& $T_{W}^*$ & $T_{A}^*$ & $T_{M}^*$ & $T_{NC}$ & $T_{LT}$\\
			Normal&-0.9&24.7&34.1&32.1&32.1&35.7&&76&86.4&84.2&82.6&87.9\\
			&-0.5&28.7&36.9&36.2&34.9&40.1&&84.1&90.6&90&87.3&92.5\\
			&-0.1&35.6&40.2&42.5&40.5&48.7&&92.5&94.1&95.4&92.4&97.2\\
			&0.1&42.4&42.8&49.4&45.1&56.1&&96.5&95.9&97.9&95.3&98.9\\
			&0.5&64.7&44.6&68.9&58.1&78.1&&99.9&98.1&100&99.1&100\\
			&0.9&100&46.1&100&96.1&{\color{red} \bf100}&&100&99.6&100&100&{\color{red} \bf100}\\
			\\
			Laplace&-0.9&25.9&34.9&34.6&40.6&36.6&&78.4&86.9&86&90.8&87.9\\
			&-0.5&29.5&38.6&38.8&46.8&42&&85.3&90.1&90.6&94.5&92.1\\
			&-0.1&38.2&42.4&45.9&54.7&50.7&&92.8&93.4&95.4&97.2&96.6\\
			&0.1&44.2&44.2&51.6&59.1&57.5&&95.9&94.5&97.4&98.4&98.3\\
			&0.5&68.2&47.9&72&74.1&80&&99.7&96.9&99.8&99.7&99.9\\
			&0.9&99.9&49.4&99.9&97.9&{\color{red} \bf99.3}&&100&98.7&100&100&{\color{red} \bf100}\\
			\\
			Exp&-0.9&25.4&36&34.8&50.3&38.1&&77.4&86.8&85.9&94.3&87.5\\
			&-0.5&{\color{red} \bf27.9}&38.6&37&57.2&43.3&&{\color{red} \bf81.5}&90.2&89&96.6&91.5\\
			&-0.1&{\color{red} \bf37.4}&43.8&45&66.5&52.9&&{\color{red} \bf89.1}&93.8&93.6&98.1&95.3\\
			&0.1&{\color{red} \bf42.2}&44.5&48.9&70.4&58.6&&{\color{red} \bf93.3}&95.4&95.9&98.8&97.2\\
			&0.5&65.2&48.9&69&81.2&78.3&&99.2&98.3&99.5&99.9&99.8\\
			&0.9&99.8&51.1&99.7&98.1&{\color{red} \bf98.5}&&100&99.7&100&100&{\color{red} \bf100}\\
			\\
			Chisq&-0.9&23&33.1&30.5&31.8&34.9&&75.6&85.9&84.1&83.3&87.3\\
			&-0.5&28&37.1&35.6&36.6&41.2&&83.7&90.8&89.8&88.4&92.6\\
			&-0.1&34.3&39.9&41.8&41.1&49.4&&92.1&94.7&95.3&93&97.3\\
			&0.1&41.9&43.3&49&46.2&57&&95.6&96.1&97.5&95.4&98.8\\
			&0.5&65.7&45.8&69.9&60.7&79&&99.8&98.6&99.9&99.1&100\\
			&0.9&100&46&100&95.6&{\color{red} \bf99.9}&&100&99.8&100&100&{\color{red} \bf100}\\
			\hline
		\end{tabular*} }
	\end{table*}

\begin{table*}[ht]
	\centering
	\caption{Power simulation results ($\alpha=0.05$) of the tests for different distributions under varying correlation values ($\rho$) with sample sizes $(n_c,n_u)=(30,10)$ and heteroscedastic covariance matrix $\Sigma_2$ under the MCAR framework. Values of too liberal tests corresponding to {\color{red} \bf red} values in the $(30,10)-$column from Table \ref{Type1errorhetroMCAR} are printed in {\color{red} \bf red} colour.
		\label{Powern3010hetro}}
	{\tabcolsep=4pt
		\begin{tabular*}{0.8\textwidth}{lc ccccc  c ccccc}
			Dist & $\rho$ & \multicolumn{5}{c}{$\delta=0.5$} && \multicolumn{5}{c}{$\delta=1$} \\ \cline{3-7} \cline{9-13} 
			&& $T_{W}^*$ & $T_{A}^*$ & $T_{M}^*$ & $T_{NC}$ & $T_{LT}$ 
			&& $T_{W}^*$ & $T_{A}^*$ & $T_{M}^*$ & $T_{NC}$ & $T_{LT}$\\
Normal&-0.9&17.8&25.6&23.6&21.7&26.3&&58.8&72.3&69.1&65.0&73.1\\
&-0.5&19.9&27.3&26.1&23.4&28.7&&66.3&78.2&76.0&70.0&79.4\\
&-0.1&24.4&30.7&30.8&26.8&34.1&&76.4&83.8&83.8&77.5&87.4\\
&0.1&28.2&32.8&34.9&30.0&39.4&&83.6&87.0&88.7&82.1&92.3\\
&0.5&42.9&35.0&48.1&38.3&56.7&&96.6&91.4&97.7&91.6&98.9\\
&0.9&95.0&36.8&93.1&69.9&{\color{red} \bf96.3}&&100.0&96.9&100.0&99.8&{\color{red} \bf100.0}\\
\\
Laplace&-0.9&18.4&26.3&25.9&28.4&27.3&&61.4&73.8&72.0&77.0&74.2\\
&-0.5&20.6&28.3&28.2&32.9&29.5&&68.0&78.2&77.5&84.0&79.3\\
&-0.1&26.1&32.5&33.5&39.0&36.1&&77.3&83.4&84.5&89.4&87.2\\
&0.1&29.1&33.8&36.6&42.5&41.2&&83.7&86.1&88.6&92.4&91.1\\
&0.5&46.4&37.6&51.7&54.6&59.5&&96.0&91.6&97.2&97.4&98.3\\
&0.9&93.7&38.5&91.6&84.5&{\color{red} \bf92.9}&&100.0&95.7&100.0&99.9&{\color{red} \bf100.0}\\
\\
Exp&-0.9&21.2&29.5&29.0&{\color{red} \bf50.3}&31.0&&61.0&72.2&71.6&{\color{red} \bf87.9}&73.4\\
&-0.5&22.0&31.3&29.9&{\color{red} \bf55.5}&33.4&&64.1&77.1&75.6&{\color{red} \bf91.2}&78.2\\
&-0.1&28.6&36.0&35.8&{\color{red} \bf63.2}&40.7&&73.0&82.2&81.5&{\color{red} \bf93.8}&84.2\\
&0.1&31.2&36.2&37.5&{\color{red} \bf66.0}&43.4&&79.0&84.8&85.2&{\color{red} \bf95.4}&88.1\\
&0.5&47.7&39.7&51.4&{\color{red} \bf74.3}&{\color{red} \bf59.0}&&92.8&90.4&93.8&{\color{red} \bf98.2}&{\color{red} \bf96.4}\\
&0.9&{\color{red} \bf90.4}&41.3&{\color{red} \bf86.7}&{\color{red} \bf89.5}&{\color{red} \bf88.6}&&{\color{red} \bf100.0}&95.5&{\color{red} \bf99.9}&{\color{red} \bf99.9}&{\color{red} \bf99.5}\\
\\
Chisq&-0.9&17.8&25.4&23.2&24.9&26.1&&57.9&72.3&69.1&68.1&73.2\\
&-0.5&20.3&28.7&26.9&28.3&30.3&&65.4&78.6&75.9&74.4&79.6\\
&-0.1&23.7&30.6&30.2&31.2&35.0&&75.3&84.0&83.4&80.9&87.0\\
&0.1&29.0&33.9&35.6&35.5&41.2&&81.5&86.5&87.6&84.8&90.7\\
&0.5&44.6&36.1&49.6&45.4&58.0&&95.8&91.9&96.9&93.2&98.4\\
&0.9&93.5&36.8&90.6&73.6&{\color{red} \bf94.4}&&100.0&96.8&100.0&99.8&{\color{red} \bf100.0}\\
			\cline{1-13} 
		\end{tabular*} }
	\end{table*}

\begin{table*}[ht]
	\centering
	\caption{Power simulation results ($\alpha=0.05$) of the tests for different distributions under varying correlation values ($\rho$) with sample sizes $n=20$ and homoscedastic covariance matrix $\Sigma_1$ under the MAR framework. Values of too liberal tests corresponding to {\color{red} \bf red} values in the $(n=20)-$column from Table \ref{Table:Type1MARhomo} are printed in {\color{red} \bf red} colour.
		\label{Table1}}
	{\tabcolsep=4pt
			\begin{tabular*}{0.8\textwidth}{lc ccccc  c ccccc}
				Dist & $\rho$ & \multicolumn{5}{c}{$\delta=0.5$} && \multicolumn{5}{c}{$\delta=1$} \\ \cline{3-7} \cline{9-13} 
				&& $T_{W}^*$ & $T_{A}^*$ & $T_{M}^*$ & $T_{NC}$ & $T_{LT}$ 
				&& $T_{W}^*$ & $T_{A}^*$ & $T_{M}^*$ & $T_{NC}$ & $T_{LT}$\\
		Normal&-0.9&11.6&18.8&17.4&17.2&19.4&&35.7&55.8&49.7&52.6&57.8\\
		&-0.5&14.3&20.5&20.1&20&22.9&&43.6&60.2&56.4&56.6&64.8\\
		&-0.1&17.2&20.7&22&21.4&26.6&&55.6&64.7&65.5&63.7&75.4\\
		&0.1&20.6&21.9&25&24.4&32.2&&63.6&67.1&71.9&68.8&82\\
		&0.5&31.9&20.8&35&30.4&47.2&&88&72&90.5&82.1&95.8\\
		&0.9&93.1&19.9&92.8&68.8&{\color{red} \bf93.3}&&100&75&100&98.9&{\color{red} \bf100}\\
		\\
		Laplace&-0.9&12.7&20.9&21.8&22.6&22&&40.9&58.6&56.1&61.7&59.4\\
		&-0.5&14.3&21.6&22.9&25.5&23.8&&48.8&63.1&62.8&69&66.3\\
		&-0.1&18.1&23.6&25.9&30.5&29.4&&60.4&68.9&71.1&76.6&76.6\\
		&0.1&21.5&24.8&28.3&33.5&33.9&&67.8&71.3&76&80.2&81.8\\
		&0.5&35.4&25.6&38.8&42.5&50.5&&87.9&76.7&90.3&89.7&95\\
		&0.9&91.6&22.3&90.7&74.4&{\color{red} \bf89.6}&&100&79.1&100&99.1&{\color{red} \bf99.6}\\
		\\
		Exp&-0.9&11.1&21.8&{\color{red} \bf22.6}&26.8&22.4&&41.9&61&{\color{red} \bf60.7}&69.2&62.5\\
		&-0.5&{\color{red} \bf14.2}&24.5&{\color{red} \bf22.6}&31.4&26.8&&{\color{red} \bf45.6}&66.9&{\color{red} \bf65.4}&74&68.9\\
		&-0.1&{\color{red} \bf18.6}&25.2&23.6&36.7&33&&{\color{red} \bf57.1}&72&71&79.9&76.5\\
		&0.1&{\color{red} \bf22.4}&25.6&25.8&38.4&37.4&&{\color{red} \bf64.5}&75&75.4&83.4&82.1\\
		&0.5&35.1&23.4&35.1&44&{\color{red} \bf51.8}&&85.2&80.4&87.6&89.9&{\color{red} \bf93.8}\\
		&0.9&91.5&17.5&87.6&71.2&{\color{red} \bf88.8}&&100&85.2&100&99.4&{\color{red} \bf99.2}\\
		\\
		Chisq&-0.9&11.7&19.1&18.1&17.8&19.6&&34.8&56.7&50.7&53.2&58\\
		&-0.5&13.2&19.9&18.7&19.1&22.3&&42.4&60.2&56.3&57&64.8\\
		&-0.1&16.7&20.6&21.2&21.3&27.8&&54.9&66.6&66&64.7&76.1\\
		&0.1&19.6&20.8&23.3&23.3&31.5&&63.5&69.4&72.4&68.8&82.1\\
		&0.5&31.8&20.4&33.8&31.2&47.8&&87.1&73.8&89.6&81.5&95.4\\
		&0.9&92.2&17.2&91.9&65.4&{\color{red} \bf93.1}&&100&78.1&100&98.9&{\color{red} \bf100}\\
					\cline{1-13} 	
		\end{tabular*} }
	\end{table*}

	\begin{table*}[ht]
		\centering
		\caption{Power simulation results ($\alpha=0.05$) of the tests for different distributions under varying correlation values ($\rho$) with sample sizes $n=20$ and heteroscedastic covariance matrix $\Sigma_2$ under the MAR framework. Values of too liberal tests corresponding to {\color{red} \bf red} values in the $(n=20)-$column from Table \ref{Table:Type1MARhetro} are printed in {\color{red} \bf red} colour.
			\label{Powern1010hetro}}
		{\tabcolsep=4pt
				\begin{tabular*}{0.8\textwidth}{lc ccccc  c ccccc}
					Dist & $\rho$ & \multicolumn{5}{c}{$\delta=0.5$} && \multicolumn{5}{c}{$\delta=1$} \\ \cline{3-7} \cline{9-13} 
					&& $T_{W}^*$ & $T_{A}^*$ & $T_{M}^*$ & $T_{NC}$ & $T_{LT}$ 
					&& $T_{W}^*$ & $T_{A}^*$ & $T_{M}^*$ & $T_{NC}$ & $T_{LT}$\\
				Normal&-0.9&9.9&16&15.8&13.2&15.8&&24.7&42.5&37.7&37&43.4\\
				&-0.5&10.8&15.9&16.4&13.7&16.6&&29.7&46.4&43.5&40.8&48.7\\
				&-0.1&12.6&16.1&16.8&14.8&19.5&&36.8&49.3&48.7&44.6&56.2\\
				&0.1&14.1&16.6&18.2&16.3&22.1&&44.5&52.2&54.5&48.7&63.5\\
				&0.5&20.9&16.6&23.9&19.4&31.9&&66.4&56.7&71.2&60.1&82.9\\
				&0.9&61.5&14.5&53.1&33.1&{\color{red} \bf74.5}&&99.7&58.1&99.2&88.6&{\color{red} \bf99.2}\\
				\\
				Laplace&-0.9&10.2&16&{\color{red} \bf17.9}&16.2&16.4&&27.6&45&{\color{red} \bf42.9}&45.9&45.4\\
				&-0.5&11.3&17.3&18.9&19.1&18.3&&34.3&48.9&48.9&53.4&50\\
				&-0.1&12.6&18.4&20.2&21.8&21.4&&43.2&55&56.2&60.5&60.2\\
				&0.1&14.7&18.6&21&23&24.3&&47.9&56.4&59.5&63.1&65.4\\
				&0.5&23.3&19.3&26.7&28.3&36.3&&68.1&60.1&73&72.4&82.5\\
				&0.9&64.2&16.5&57.6&46.8&{\color{red} \bf74.3}&&98.9&62.2&97.8&91.3&{\color{red} \bf97}\\
				\\
				Exp&-0.9&11.5&19.9&{\color{red} \bf22.6}&{\color{red} \bf27.6}&20.2&&33.2&48.7&{\color{red} \bf49.8}&{\color{red} \bf57.7}&49.4\\
				&-0.5&13.7&22.1&{\color{red} \bf22.6}&{\color{red} \bf30.7}&23.2&&34.7&53.4&{\color{red} \bf53.9}&{\color{red} \bf62.2}&54.4\\
				&-0.1&{\color{red} \bf18}&23.5&23.2&{\color{red} \bf34.5}&27.6&&{\color{red} \bf43}&56.8&57&{\color{red} \bf67.3}&60.3\\
				&0.1&19.9&22.6&23.4&{\color{red} \bf34.9}&29.5&&48.9&59&60.3&{\color{red} \bf70.1}&65.2\\
				&0.5&29.8&21.3&29.7&{\color{red} \bf38.6}&{\color{red} \bf39.7}&&66.5&61.7&69.4&{\color{red} \bf75.7}&{\color{red} \bf79.1}\\
				&0.9&{\color{red} \bf66.5}&15.2&{\color{red} \bf54.2}&{\color{red} \bf49.9}&{\color{red} \bf72.9}&&{\color{red} \bf97.2}&63.5&{\color{red} \bf93.9}&{\color{red} \bf90.4}&{\color{red} \bf93.8}\\
				\\
				Chisq&-0.9&9.5&15.5&15.1&14.4&15.8&&25.6&43.2&39.4&39.7&44.1\\
				&-0.5&11&16.5&16.3&15.6&17.3&&30.5&47.1&44.2&43.3&49.1\\
				&-0.1&13&16.6&17.2&17&21.1&&38.4&51.5&50.9&49.1&57.9\\
				&0.1&15.1&17.3&18.6&17.9&23.6&&44.5&52.9&55&51.6&63.8\\
				&0.5&22.6&16.5&24.3&22.3&33.7&&65.6&56.5&69.4&61.1&81.4\\
				&0.9&63.3&13.2&53.4&35.7&{\color{red} \bf73.9}&&99.3&59&98&85.7&{\color{red} \bf98.5}\\
							\cline{1-13} 
			\end{tabular*} }
		\end{table*}

\end{document}


\definecolor{ao(english)}{rgb}{0.0, 0.5, 0.0}
\definecolor{bronze}{rgb}{0.8, 0.5, 0.2}
\definecolor{byzantine}{rgb}{0.74, 0.2, 0.64}

\runninghead{Amro, Pauly, and Ramosaj}

\title{Supplementary Materials for: Asymptotic based bootstrap approach for matched pairs with missingness in a single-arm.}

\author{Lubna Amro\affilnum{1}, Markus Pauly\affilnum{1} and Burim Ramosai\affilnum{1} }

\affiliation{\affilnum{1}Mathematical Statistics and Applications in Industry, Faculty of Statistics, Technical University of Dortmund, Germany}

\corrauth{Lubna Amro, Mathematical Statistics and Applications in Industry, Faculty of Statistics, Technical University of Dortmund, Germany.}

\email{lubna.amro@tu-dortmund.de}

\maketitle
  In this supplementary material, we present the proofs of the propositions and theorems in the paper. Further, we recall the definition of the different missing mechanisms and present additional type-I error and power simulation results of our suggested methods and the alternative approaches from Section 5 of the paper.

\section{Proofs}\label{Proofs}

\textit{Proof of Proposition 2.1:} \\
The results follow from the multivariate central limit theorem (CLT) and the low of large numbers, respectively. \\

\textit{Proof of Proposition 2.2:}\\
The stated convergence follows from Proposition 2.1 
, and an application of the continuous mapping theorem (CMT). \\

\textit{Proof of Theorem 3.1:}\\ 
It follows from Proposition 2.2 
that we have convergence in distribution $\vA\vZ_n \xrightarrow{\text{d}}N_2(\boldsymbol{0},\vA\vSigma \vA^T)$ as $n\rightarrow\infty$ under $H_0$. Hence, using the CMT, the quadratic form $\tilde{T}_W=(\vA\vZ_n )^\top(\vA\vSigma \vA^\top)^+(\vA\vZ_n)$ has asymptotically a central $\chi^2_f$ distribution with $f=rank(\vA)$ degrees of freedom. Moreover, as $\hat{\vSigma}_n$ is a consistent estimator for $\vSigma>0$, the result follows from Slutzky theorem, see, e.g., Konietschke et.al. \cite{konietschke2015parametric} for similar arguments .\\

\textit{Proof of Theorem 3.2:}\\ 
Applying again the CMT, it follows that $tr(\mathbf{A} \boldsymbol{\Sigma} \mathbf{A}^\top ) \cdot T_A=(A\vZ_n )^T(A\vZ_n)$ has asymptotically the same distribution as  $\sum_{i=1}^{2}\lambda_iY_i$ \cite{graybill1976theory, brunner2001nonparametric}. Then, the result follows from the invariance of the multivariate standard normal distribution under orthogonal transformations and the consistency of $\hat{\vSigma}_n$ by using Slutzky theorem.\\

\textit{Proof of Theorem 3.3:}\\
Following similar arguments as prescribed in Friedrich and Pauly\cite{friedrich2018mats}, we can obtain that $\hat{\mathbf{D}}_n=diag(( \mathbf{A} \hat{\boldsymbol{\Sigma}}_n \mathbf{A}^\top )^+_{ii})\xrightarrow{\text{p}} diag(( \mathbf{A} \vSigma \mathbf{A}^\top )^+_{ii})=\mathbf{D}$ as $\hat{\boldsymbol{\Sigma}}_n \overset{p}{\to} \vSigma >0 $. Thus, the result follows  from the representation theorem of quadratic forms \cite{rao1972generalized}. \\

\textit{Proof of Theorem 4.1:}\\ 
First, we apply the Multivariate Lindeberg-Feller Theorem (MLFT) to show that (given the data) $\sqrt{n}\bar{\vX}_.^{^*(c)}=\sqrt{n}[\bar{\vX}_{1.}^{^*(c)},\bar{\vX}_{2.}^{^*(c)}]$ converges in distribution to a normal distributed random variable.  
We start by checking the MLFT conditions:\\
\begin{align*}
 &A) \sum_{k=1}^{n_c}\mathbb{E}\Big( \frac{\sqrt{n}}{n_c}\vX_{k}^{*(c)}|\vX\Big)=\sum_{k=1}^{n_c}\frac{\sqrt{n}}{n_c}\mathbb{E}\Big( \vX_{k}^{*(c)}|\vX\Big)=0\\ 
&B) \sum_{k=1}^{n_c}Cov\Big(\frac{\sqrt{n}}{n_c}\vX_{k}^{*(c)}|\vX\Big)=\sum_{k=1}^{n_c}\frac{n}{n_c^2}Cov\Big(\vX_{k}^{*(c)}|\vX\Big)=\sum_{k=1}^{n_c}\frac{n}{n_c^2} \hat{\boldsymbol{\Gamma}}  \xrightarrow{\text{p}} \frac{1}{\kappa_1}\boldsymbol{\Gamma}\\ 
&C) \lim\limits_{n \to \infty} \sum_{k=1}^{n_c} \mathbb{E}\Big( \Big\|\frac{\sqrt{n}}{n_c}\vX_{k}^{*(c)}\Big\|^2 \cdot \mathds{1}\Big\{\Big \|\frac{\sqrt{n}}{n_c}\vX_{k}^{*(c)}\Big\|>\epsilon\Big\} |\vX \Big)\\
   &\quad=\lim\limits_{n \to \infty}\frac{n}{n_c^2}\sum_{k=1}^{n_c} \mathbb{E}\Big( \Big\|\vX_{k}^{*(c)}\Big\|^2 \cdot \mathds{1}\Big\{\Big \|\frac{\sqrt{n}}{n_c}\vX_{k}^{*(c)}\Big\|>\epsilon\Big\} |\vX  \Big)\\
 &\quad=\frac{1}{\kappa_1} \cdot  \lim\limits_{n \to \infty} \mathbb{E}\Big( \Big\|\vX_{k}^{*(c)}\Big\|^2 \cdot \mathds{1}\Big\{\Big \|\frac{\sqrt{n}}{n_c}\vX_{k}^{*(c)}\Big\|>\epsilon\Big\} |\vX   \Big)\\
 &\quad\leq \frac{1}{\kappa_1} \cdot  \lim\limits_{n \to \infty} \sqrt{\mathbb{E}\Big( \Big\|\vX_{k}^{*(c)}\Big\|^2 |\vX\Big)^2} \cdot \sqrt{\mathbb{E}\Big(\mathds{1}\Big\{\Big \|\frac{\sqrt{n}}{n_c}\vX_{k}^{*(c)}\Big\|>\epsilon\Big\} |\vX \Big)^2}\\ 
  \end{align*}
 
The last step follows from the Cauchy–Schwarz inequality. Now, the first term $\mathbb{E}\Big( \Big\|\vX_{k}^{*(c)}\Big\|^2 |\vX\Big)^2$ is asymptotically bounded, while the second term converges to zero in probability since \\  $\mathds{1}\Big\{\Big \|\frac{\sqrt{n}}{n_c}\vX_{k}^{*(c)}\Big\|>\epsilon\Big\}=1$ holds iff $\|\vX_{k}^{*(c)}\big\|>\frac{n_c}{\sqrt{n}}\epsilon=\frac{n_c}{n}\sqrt{n}\epsilon $. As $\frac{n_c}{n}\sqrt{n}\epsilon  \rightarrow \infty $ while  $\vX_{k}^{*(c)}\xrightarrow{d} N(0,\boldsymbol{\Gamma})$, it follows that the Lindeberg condition is satisfied (in probability). Thus, proves that the conditional distribution of  $\sqrt{n}\bar{\vX}_.^{^*(c)}$ given the data weakly converges to $\frac{1}{\kappa_1} N(0,\boldsymbol{\Gamma})$ in probability.\\

In a similar way , we proof that $\sqrt{n}\bar{\vX}_1^{^*(i)}$ given the data weakly converges to  $\frac{1}{\kappa_2}N(0,\sigma_1^2)$ in probability. \\

Now, due to the MCAR setting, $\vX^{c}$ is independent of $\vX^{(i)}$ and by using Slutzky, $\vZ_n^*=\sqrt{n}[\bar{X}_{1.}^{*(c)}, \bar{X}_{2.}^{*(c)}, \bar{X}_{1.}^{*(i)}]^\top$ converges in distribution to $N_3(\mathbf{0}, \boldsymbol{\Sigma})$ where, $\vSigma$ is defined in 
Section 6 in the paper. Following the same steps as in the proof of Theorem 3.1-3.3 
, this concludes the proof.

	\section{MCAR, MAR and MNAR}
	
	To explain the different missing schemes we define a missing indicator variable $\mathbf{R}_j = [R_{1j}, R_{2j}]^\top  \in \R ^2$, that identify what is known and what is missing, i.e.  $R_{ij} = 0$ if   $X_{ij}$ is missing and $R_{ij} = 1$ otherwise, $j = 1, \dots , n$. Then, Rubin defines the missing mechanism through a parametric distributional model on $\mathbf{R} = \{ \mathbf{R}_j \}_{j = 1}^n$ and classifies their presence through Missing Completely at Random (MCAR), Missing at Random (MAR) and Missing not at Random (MNAR) schemes\cite{rubin2004multiple}. 
	To describe this in our model let us denote the observed data as $\vX^{obs}=(\vX^{(c)},\vX^{(i)})$  and denote with $\vX^{mis}$ the missing observations.\\ 
	
	The data are said to be {\bf MCAR} if the probability of an observation being missing does not depend on observed or unobserved data, i.e. if $P(\mathbf{R}\mid \vX^{obs},\vX^{mis})=P(\mathbf{R})$.\\
	
	Data are said to be {\bf MAR} if the probability of missingness may depend on observed data but does not depend on unobserved data, $P(\mathbf{R}\mid \vX^{obs},\vX^{mis})=P(\mathbf{R}\mid \vX^{obs})$.\\ Finally, data are said to be {\bf MNAR}, if missingness does depend on the unobserved data. It can easily be seen that MAR includes MCAR as a special case. For more details about the different missing mechanisms, we refer to the monograph of Little and rubin (2014) \cite{a2014incomplete}. 

	\section{Type-I Error and Power Results }
In the sequel, we present some additional type-I error and power results of the Monte Carlo simulation study, that is described in detail in Section 6 of the paper, for testing $H_0^\mu$ for matched pairs with missingness in one arm under the MCAR, and MAR schemes.

\begin{table*}[ht]
	
	\caption{Type-I error simulation results ($\alpha=0.05$) of the tests for different distributions under varying correlation values ($\rho$), different sample sizes $n\in\{10,20,30\}$  and homoscedastic covariance matrix $\Sigma_1$ under the MAR framework.
		\label{Type1errorhetro}}
	{\tabcolsep=4pt
		\begin{tabular*}{1.09\linewidth}{lc ccccc c ccccc  c ccccc   cc}
			Dist & $\rho$ & \multicolumn{5}{c}{$n=10$}  && \multicolumn{5}{c}{$n=20$} && \multicolumn{5}{c}{$n=30$} \\ \cline{3-7} \cline{9-13} \cline{15-19}
			&& $T_{W}^*$ & $T_{A}^*$ & $T_{M}^*$ & $T_{NC}$ & $T_{LT}$
			&& $T_{W}^*$ & $T_{A}^*$ & $T_{M}^*$ & $T_{NC}$ & $T_{LT}$ 
			&& $T_{W}^*$ & $T_{A}^*$ & $T_{M}^*$ & $T_{NC}$ & $T_{LT}$\\
	Normal&-0.9&5.4&3.7&5.2&6.4&3.6&&5.4&5.1&5.7&5.8&5.5&&5.1&5.4&6&5.7&5.5\\
	&-0.5&4.8&3.9&4.4&6.4&4.6&&5.3&4.9&5.5&5.6&5.2&&5.2&5.1&5.8&5.3&5.1\\
	&-0.1&4.6&4.1&4.4&6.2&5.7&&5.1&4.7&5&5.5&5.5&&5&5&5.1&5.3&5.2\\
	&0.1&4.7&4.3&4.3&6.7&6.2&&4.9&4.3&5&5.1&5.9&&5.2&4.4&5.3&5.3&5.5\\
	&0.5&4.7&4.2&4.1&6.3&7.3&&5.2&5&5.1&5.5&6.4&&5.3&4.8&5&4.8&5.9\\
	&0.9&4.9&4.9&4.3&6.5&12.5&&5&4.8&4.6&5.3&14.2&&5.1&5.1&4.8&5&12.6\\
\\
	Laplace&-0.9&4.2&3.2&5.1&6.1&3.2&&4.3&4.9&6&5.3&5&&4.9&5.1&6.5&5.3&5\\
	&-0.5&3.1&3.6&3.5&6.2&3.9&&4.2&4.6&5.5&5.5&5&&4.4&5&6&5.3&5\\
	&-0.1&3.1&3.2&3.1&6.3&4.8&&3.9&3.9&4.6&5.1&4.7&&4.4&4.3&4.9&5&5\\
	&0.1&3.2&2.7&3&6.3&5.6&&4.1&3.7&4.6&5.6&5.3&&4.6&4.4&5.3&5.2&5.6\\
	&0.5&3.5&3.1&3&6.3&7.8&&4.5&4&4.2&5.2&5.6&&4.5&3.6&4.2&4.8&5.3\\
	&0.9&3.6&3.9&3.3&5.6&11&&4&4.1&3.8&4.9&14.4&&4.5&4.4&4.3&4.6&14.4\\
\\
	Exp&-0.9&4.3&2.8&5.6&6.1&3.2&&5.6&5&7.8&5&5.3&&6&5.1&8.7&5.5&5.3\\
	&-0.5&4.8&3&4.2&6.6&4.1&&7.6&4.9&7.4&5.7&5.4&&8.9&5.3&8.7&5.2&5.3\\
	&-0.1&4.3&2.7&3.4&6.5&5.2&&7.4&4.2&6.4&5.4&5.2&&8.8&5.4&8&5.6&5.4\\
	&0.1&4&2.6&3.4&6.5&6.2&&7.1&4.3&6.5&5.6&5.9&&8.1&5.4&7.6&5.5&5.1\\
	&0.5&3.5&2.7&3.2&6.5&10.1&&5.7&4.8&5.7&5.7&6.9&&6.5&6.1&6.5&5.6&5.6\\
	&0.9&4.3&4.7&4.2&5.8&10.2&&5&5.7&5.2&5.2&13.4&&5.6&6.9&5.7&5.2&15.3\\
\\
	Chisq&-0.9&4.6&3.7&4.5&5.6&3.7&&5.1&5.3&5.9&5.9&5.5&&5&5.2&6&5.4&5.3\\
	&-0.5&4.5&3.7&4.4&6.3&4.3&&5.1&4.8&5.4&5.4&5.3&&5.4&4.8&5.5&5.3&5.3\\
	&-0.1&4.7&3.8&4.4&6.6&5.7&&5.3&4.7&5.3&5.2&5.8&&5.7&5.1&5.7&5.4&5.6\\
	&0.1&4.8&3.7&4.1&6.1&6.1&&5.1&4.1&4.7&5.3&5.5&&5.3&4.8&5.3&5&5.2\\
	&0.5&4.2&3.8&3.6&6&7.3&&4.9&4.4&4.8&5.3&6.1&&4.9&4.7&4.8&4.9&5\\
	&0.9&4.5&4.6&3.8&5.5&12.8&&4.9&5.3&4.5&4.9&14.7&&5.2&5.2&4.9&5.2&13\\\hline				\end{tabular*} }
\end{table*}

\clearpage

\begin{table*}[ht]
	
	\caption{Type-I error simulation results ($\alpha=0.05$) of the tests for different distributions under varying correlation values ($\rho$), different sample sizes $n\in\{10,20,30\}$  and heteroscedastic covariance matrix $\Sigma_2$ under the MAR framework.
		\label{Type1errorhetro}}
	{\tabcolsep=4pt
		\begin{tabular*}{1.09\linewidth}{lc ccccc c ccccc  c ccccc   cc}
			Dist & $\rho$ & \multicolumn{5}{c}{$n=10$}  && \multicolumn{5}{c}{$n=20$} && \multicolumn{5}{c}{$n=30$} \\ \cline{3-7} \cline{9-13} \cline{15-19}
			&& $T_{W}^*$ & $T_{A}^*$ & $T_{M}^*$ & $T_{NC}$ & $T_{LT}$
			&& $T_{W}^*$ & $T_{A}^*$ & $T_{M}^*$ & $T_{NC}$ & $T_{LT}$ 
			&& $T_{W}^*$ & $T_{A}^*$ & $T_{M}^*$ & $T_{NC}$ & $T_{LT}$\\
	Normal&-0.9&5.1&4.1&4.9&6.3&4&&5.2&5.4&5.9&5.1&5.5&&5.4&5.6&6.5&5.2&5.6\\
	&-0.5&5.1&4.4&5.2&6.8&5.2&&5.1&5.3&5.9&5.3&5.4&&4.6&5.2&5.8&4.8&5.2\\
	&-0.1&4.9&4.4&4.5&6.3&6&&5.1&5.2&5.5&5.2&5.6&&5.1&5&5.4&4.5&5.3\\
	&0.1&5.2&4.5&4.5&6.6&6.5&&5.3&5.1&5.5&5.4&6.5&&5.1&4.7&5.2&4.5&5.1\\
	&0.5&4.4&4&4&6.3&7.3&&5.1&4.9&5.2&5.2&6&&5.2&4.8&5.2&5&5.9\\
	&0.9&4.6&4.8&5&7&14.7&&4.9&4.7&5.3&5.2&14.1&&5&5&5.3&4.6&11.8\\
	\\
	Laplace&-0.9&4&3.4&5.2&6.3&3.6&&4.2&4.9&6.8&4.8&5&&4.4&5.1&7.4&4.8&5.3\\
	&-0.5&3.2&3.6&4.1&6&4.2&&4.5&5&6.4&5.4&5.2&&4.7&4.9&6.7&4.9&4.8\\
	&-0.1&3.1&3.7&4&6.5&5.4&&3.9&4.5&5.3&5.3&5.2&&4.7&5&6.1&4.8&5.1\\
	&0.1&3.1&2.8&3&5.9&5.7&&4.2&4.3&5.1&5&5.5&&4.3&5&5.9&4.9&5.8\\
	&0.5&3.4&3.4&3.4&6.5&8.1&&4.2&4&4.5&5.1&6&&4.4&4.6&5.1&4.6&5.6\\
	&0.9&3.3&4.1&3.9&6.6&13.1&&4.3&4.9&5.1&5.2&16.2&&4.7&4.9&5.5&4.9&15.3\\
	\\
	Exp&-0.9&3.3&4&6&7.4&4.3&&4.7&5.7&9&7.4&6&&5.3&5.5&8.6&7.7&5.6\\
	&-0.5&4.2&4.1&4.1&7.8&5.4&&6.5&5.5&7&7.2&6.4&&8.4&5.8&8.5&8.2&6.2\\
	&-0.1&4.5&3.5&3.8&8.2&7&&6.9&4.7&6.2&7.4&6.2&&8.6&5.6&7.9&9&6.3\\
	&0.1&3.9&3.5&3.5&8.6&8.5&&6.3&4.4&6.1&7.7&6.7&&7.8&5.2&7.2&8.8&5.9\\
	&0.5&4.6&2.9&4.2&7.8&11&&6.3&4.5&6.3&8.1&8.4&&6.9&5.2&7&9.3&7.3\\
	&0.9&7.8&4&7.8&8.7&15.9&&7.7&4.5&8.1&8.6&17.5&&8.1&5.4&7.7&10.4&17.2\\
	\\
	Chisq&-0.9&5.1&4.2&5.3&6.1&4.3&&5.8&5.8&6.6&5.8&5.9&&5.2&5.5&6.6&5&5.6\\
	&-0.5&4.8&4.5&5.1&6.2&5.1&&5.1&4.7&5.5&5.1&5.2&&5.3&5.1&6&5&5\\
	&-0.1&4.5&4.3&4.7&6.5&6.3&&5.3&5.2&5.7&5.2&5.6&&6.1&5.5&6.1&5.4&5.8\\
	&0.1&4.3&4&4&6.3&6.6&&5&4.8&5.2&5.2&5.8&&5.1&5.1&5.5&4.9&5.6\\
	&0.5&4.7&3.9&4.3&6.5&7.6&&5.4&4.8&5.3&5.6&6.5&&5.5&4.8&5.5&5.1&5.8\\
	&0.9&4.9&4.5&5&7&15&&5.3&4.8&5.4&5.6&15.2&&5.7&5.1&5.6&5.1&13.3\\\hline				\end{tabular*} }
\end{table*}

	\begin{figure}[H]
		\begin{center}
			
			\includegraphics[scale=1]{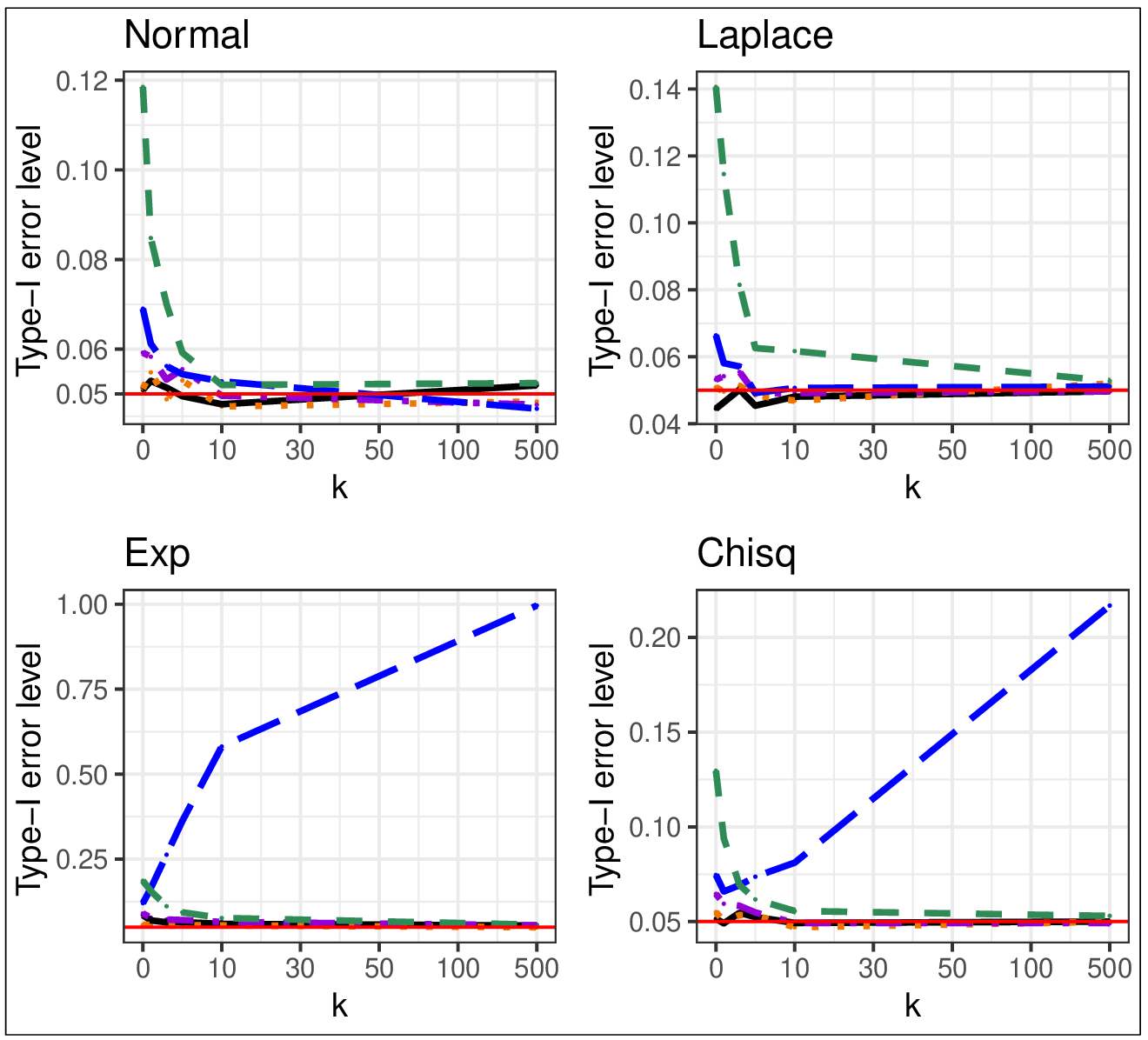}
			
		\end{center}
		\caption{Type-I error simulation results ($\alpha=0.05$) of the tests  $T_W^*$ $({\color{black}\textendash\textendash\textendash})$, $T_A^*$ $({\color{brown}\cdots})$, 	$T_M^*$ $({\color{violet}-\cdot-})$, $T_{NC}$ $({\color{blue} \textendash\textendash \quad \textendash\textendash})$,  and  $T_{LT}$ $({\color{ao(english)} -- })$, for different distributions under correlation value ($\rho=0.9$)  and heteroscedastic covariance matrix $\Sigma_2$ for varying $k$ values added to  $(n_c,n_u)=(10,10)$ under the MCAR framework.}
		\label{fig:multi}	

	\end{figure}

		\begin{figure}[H]
		\begin{center}
			
			\includegraphics[scale=1]{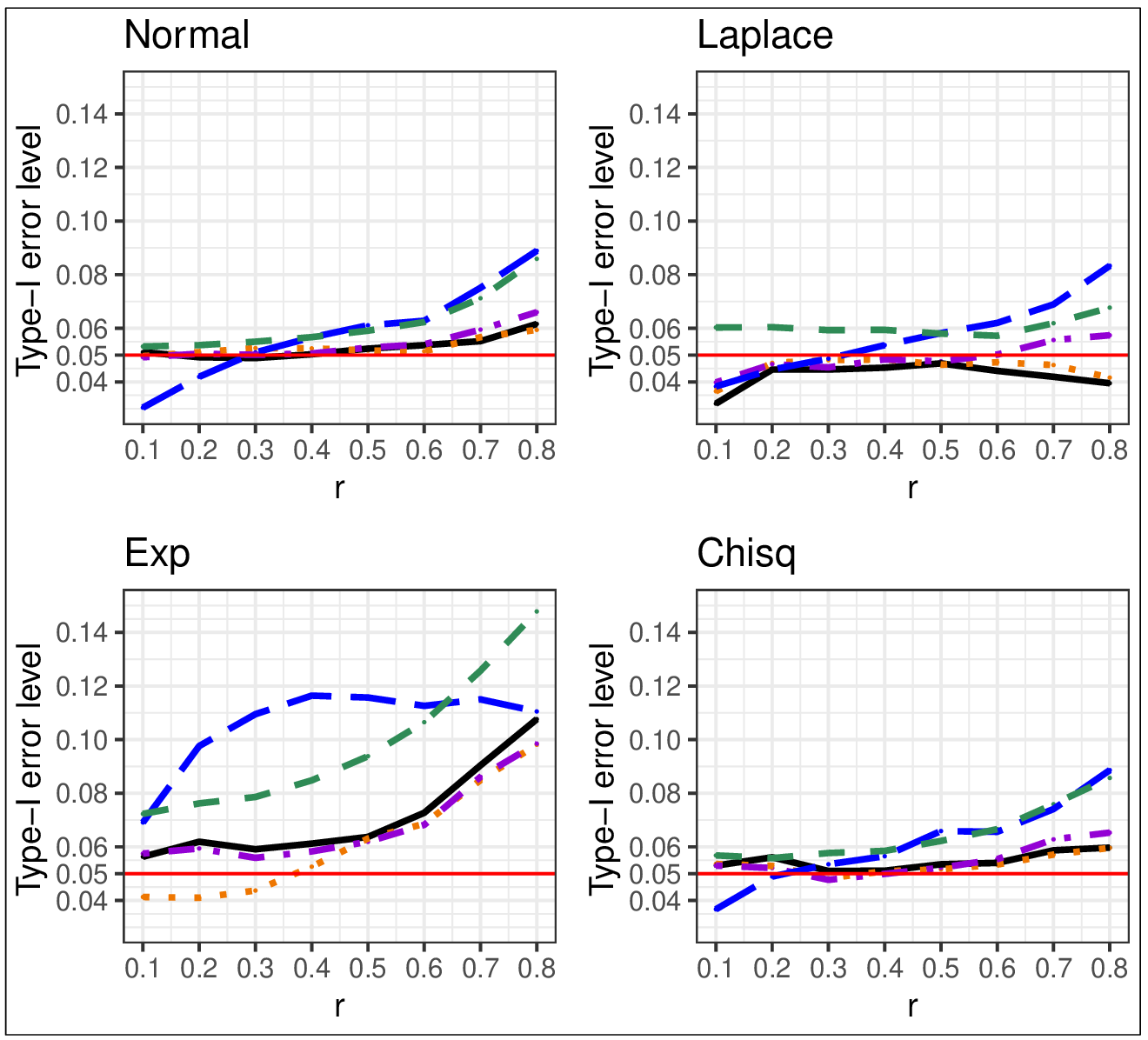}
			
		\end{center}
		\caption{Type-I error simulation results ($\alpha=0.05$) of the tests  $T_W^*$ $({\color{black}\textendash\textendash\textendash})$, $T_A^*$ $({\color{brown}\cdots})$, 	$T_M^*$ $({\color{violet}-\cdot-})$, $T_{NC}$ $({\color{blue} \textendash\textendash \quad \textendash\textendash})$,  and  $T_{LT}$ $({\color{ao(english)} -- })$, for different distributions under correlation value ($\rho=0.5$) with sample size $(n=30)$  and heteroscedastic covariance matrix $\Sigma_2$ for varying missing rates  $r\in\{0.1,0.2,0.3,0.4,0.5,0.6,0.7,0.8\}$ under the MCAR framework.}
		\label{fig:MCARMissRate}	

	\end{figure}

\begin{table*}[ht]
	\centering
	\caption{Type-I error simulation results ($\alpha=0.05$) of the tests for different distributions under varying correlation values ($\rho$) with sample sizes $(n_c,n_u)=(16,74)$ and homoscedastic and heteroscedastic covariance matrices $\Sigma_1$ and $\Sigma_2$ respectively under the MCAR framework.
		\label{Type1errorhetro}}
	{\tabcolsep=4pt
		\begin{tabular*}{0.8\textwidth}{lc ccccc c ccccc}
		Dist & $\rho$ & \multicolumn{5}{c}{$\Sigma_1$}  && \multicolumn{5}{c}{$\Sigma_2$}  \\ \cline{3-7} \cline{9-13} 
			&& $T_{W}^*$ & $T_{A}^*$ & $T_{M}^*$ & $T_{NC}$ & $T_{LT}$
			&& $T_{W}^*$ & $T_{A}^*$ & $T_{M}^*$ & $T_{NC}$ & $T_{LT}$\\
			Normal&-0.9&5.4&5&5.9&6&4.1&&5.3&5&6.1&6.8&4.3\\
			&-0.5&5.2&5.4&6&6.1&5.3&&5.2&5.2&6.1&6.7&5.6\\
			&-0.1&5.3&5.4&5.8&5.8&5.9&&5.2&5.5&5.9&6.6&6.3\\
			&0.1&4.8&5.2&5.6&5.6&5.8&&4.7&5.3&5.8&6.4&6.1\\
			&0.5&5.1&4.9&5&6&5.8&&5&5&5.6&6.7&6.1\\
			&0.9&5.1&5.7&4.6&5.7&6.5&&5.1&5.7&7.1&6.4&6.4\\
			\\
			Laplace&-0.9&4.8&5&6.3&6.5&3.7&&4.7&5.1&6.4&7.1&3.9\\
			&-0.5&4.7&4.9&5.9&6&4.7&&4.7&4.9&6&6.3&5.1\\
			&-0.1&4.4&5&5.9&5.9&5.2&&4.4&5.1&6.3&6.8&5.6\\
			&0.1&4.5&5&5.4&6.2&5.3&&4.5&4.9&5.8&6.7&5.4\\
			&0.5&4.1&4.5&5.2&5.1&5.2&&4&4.8&6&5.8&5.5\\
			&0.9&4.7&5.4&4.5&5.8&6.7&&4.6&5.5&7.6&6.9&6.4\\
			\\
			Exp&-0.9&4.8&4.7&6.1&6.2&4.4&&5.6&5.5&7.2&9.9&5.4\\
			&-0.5&6.4&5.2&7.1&6.4&6.1&&7&6.3&8.2&10.7&7.3\\
			&-0.1&7.7&5.7&7.8&6.2&7.5&&8.4&7&8.7&11.7&8.5\\
			&0.1&7.8&6&7.9&5.9&8&&8.3&7.3&9.2&12.1&9.1\\
			&0.5&8.8&6.5&7.9&5.5&10.4&&9.3&8.5&10&13.6&11\\
			&0.9&9.7&6.4&5.5&5.9&12.6&&9.4&7.7&11.6&16.5&12.2\\
			\\
			Chisq&-0.9&5.1&5.1&6.3&6.2&3.8&&5.1&5.3&6.8&6.9&4.1\\
			&-0.5&5.3&5.5&6.2&6.1&5.3&&5.3&5.4&6.2&6.8&5.6\\
			&-0.1&5.6&5.3&5.6&5.8&5.5&&5.4&5.2&5.8&7&5.8\\
			&0.1&5.2&5.6&6&5.9&6.1&&5.2&5.7&6.2&7.1&6.3\\
			&0.5&5.2&5.2&5.5&5.6&6.1&&5.3&5.3&6&6.9&6.5\\
			&0.9&5.4&5.6&4.5&5.5&6.7&&5.5&5.5&7.1&7.3&6.5\\
			 \hline
		\end{tabular*} }
\end{table*}

\begin{table*}[ht]
		\centering
	\caption{Type-I error simulation results ($\alpha=0.05$) of the tests for different distributions under varying correlation values ($\rho$) inspired from the dependency structure of the genes data example with sample sizes $(n_c,n_u)=(16,74)$ and homoscedastic and heteroscedastic covariance matrices $\Sigma_1$ and $\Sigma_2$ respectively under the MCAR framework.
		\label{Table2:Genes}}
	{\tabcolsep=4pt
		\begin{tabular*}{0.8\textwidth}{lc ccccc c ccccc}
			Dist & $\rho$ & \multicolumn{5}{c}{$\Sigma_1$}  && \multicolumn{5}{c}{$\Sigma_2$}  \\ \cline{3-7} \cline{9-13} 
			&& $T_{W}^*$ & $T_{A}^*$ & $T_{M}^*$ & $T_{NC}$ & $T_{LT}$
			&& $T_{W}^*$ & $T_{A}^*$ & $T_{M}^*$ & $T_{NC}$ & $T_{LT}$\\
Normal&-0.3&5.3&5.1&5.7&5.8&5.0&&5.3&5.1&5.9&6.7&5.4\\
&-0.1&5.2&5.4&5.7&6.0&5.4&&5.2&5.3&5.8&6.8&5.8\\
&0.3&5.4&5.4&5.5&5.7&6.2&&5.2&5.3&5.5&6.6&6.4\\
&0.4&4.8&5.0&5.4&5.7&5.9&&4.8&5.1&5.9&6.6&6.2\\
&0.7&5.1&4.8&4.9&6.0&6.0&&5.0&4.9&5.8&6.6&6.3\\
&0.8&5.1&5.2&5.1&5.7&6.2&&5.0&5.4&6.1&6.7&6.4\\
\\
Laplace&-0.3&4.7&5.1&6.1&6.3&5.2&&4.6&5.2&6.4&7.0&5.5\\
&-0.1&4.7&4.8&5.8&5.9&5.2&&4.7&4.9&6.0&6.3&5.3\\
&0.3&4.4&4.8&5.7&6.0&5.6&&4.4&5.1&6.3&7.0&5.8\\
&0.4&4.5&4.7&5.2&6.1&5.4&&4.5&4.8&5.8&6.7&5.6\\
&0.7&4.2&4.6&5.0&5.2&5.3&&4.2&4.7&6.3&5.9&5.6\\
&0.8&4.5&5.2&5.2&6.0&6.3&&4.6&5.5&6.7&6.9&6.1\\
\\
Exp&-0.3&7.1&5.3&7.8&6.2&7.3&&8.1&6.6&8.8&11.3&8.6\\
&-0.1&7.5&5.7&7.6&6.3&7.1&&7.9&7.0&8.6&11.7&8.0\\
&0.3&8.7&6.5&8.0&6.2&8.7&&9.1&7.8&9.2&12.5&9.6\\
&0.4&8.3&6.2&7.8&6.1&9.2&&8.8&7.9&9.3&12.5&10.2\\
&0.7&9.2&6.6&7.2&5.5&11.0&&9.4&8.6&10.2&14.5&11.4\\
&0.8&9.5&6.3&6.6&6.0&11.5&&9.7&8.0&10.3&15.1&11.8\\
\\
Chisq&-0.3&5.3&5.6&6.3&6.0&5.4&&5.3&5.6&6.4&7.1&5.5\\
&-0.1&5.2&5.4&5.9&6.0&5.8&&5.3&5.3&6.0&6.8&6.1\\
&0.3&5.6&5.1&5.5&5.7&5.9&&5.6&5.4&5.8&7.0&6.2\\
&0.4&5.4&5.5&5.7&5.7&6.2&&5.4&5.6&6.1&7.1&6.6\\
&0.7&5.2&5.2&5.2&5.6&6.2&&5.4&5.6&6.3&7.2&6.4\\
&0.8&5.3&5.2&5.0&5.6&6.2&&5.5&5.4&6.0&7.0&6.4\\
	\hline
		\end{tabular*} }
	\end{table*}

\begin{table*}[ht]
	
	\caption{Power simulation results ($\alpha=0.05$) of the tests for different distributions under varying correlation values ($\rho$) with sample sizes $(n_c,n_u)=(10,10)$ and homoscedastic covariance matrix $\Sigma_1$ under the MCAR framework.
		\label{Table1}}
	{\tabcolsep=4pt
		\begin{tabular*}{1.09\linewidth}{lc ccccc c ccccc  c ccccc   cc}
			Dist & $\rho$ & \multicolumn{5}{c}{$\delta=0$}  && \multicolumn{5}{c}{$\delta=0.5$} && \multicolumn{5}{c}{$\delta=1$} \\ \cline{3-7} \cline{9-13} \cline{15-19}
			&& $T_{W}^*$ & $T_{A}^*$ & $T_{M}^*$ & $T_{NC}$ & $T_{LT}$
			&& $T_{W}^*$ & $T_{A}^*$ & $T_{M}^*$ & $T_{NC}$ & $T_{LT}$ 
			&& $T_{W}^*$ & $T_{A}^*$ & $T_{M}^*$ & $T_{NC}$ & $T_{LT}$\\
		Normal&-0.9&5.4&5.5&5.4&6.8&4.4&&13.6&15.5&16.9&18.2&17&&42.2&46.1&50.5&49.3&53.5\\
		&-0.5&5.1&5.4&5.7&6.5&4.9&&13.5&17.6&18&19.4&18.1&&43.3&52.6&53.8&53.1&54.9\\
		&-0.1&5.4&5.6&5.4&6.6&5.6&&15.8&21.1&20.7&22.7&21.6&&51.4&63.7&62.6&62.2&63.8\\
		&0.1&5.2&5.5&5.5&6.4&6.1&&17.6&22.6&21.8&23.6&23.4&&55&67.7&65.5&65.9&68.2\\
		&0.5&5.1&5&4.8&6&6.8&&25.7&27.5&29.8&31.8&35.2&&75.9&80.5&82.8&81.7&86.4\\
		&0.9&5.3&5.3&4.6&5.8&12.3&&81.1&30.2&80.5&70.5&85.9&&100&93.7&100&99.3&99.9\\
		\\
		Laplace&-0.9&4.6&4.9&5.4&6.6&4.6&&14.4&15.8&18.2&21.8&17.7&&45.2&47.8&53.6&57&54.3\\
		&-0.5&4.3&4.8&5&6.5&4.6&&15.9&19.9&21.1&25.4&20.5&&49.2&56.8&59.3&64.8&58.1\\
		&-0.1&4.3&5&4.8&6.5&5.4&&17&22&22.2&28.3&22.3&&55.4&65.4&65.6&71.9&65.4\\
		&0.1&4.4&4.7&4.6&6.6&5.2&&20.3&25.6&25.5&33&27.1&&61&70.7&70.6&77.2&71.3\\
		&0.5&4.3&4.5&4.4&6.2&6.4&&29.8&30.8&34.2&41.6&39.8&&78.5&81.1&83.6&87.1&86.3\\
		&0.9&4.2&5.2&4&6.1&13.3&&83.3&33.6&81.1&76.1&83.7&&99.9&91.6&99.7&99.1&99.3\\
		\\
		Exp&-0.9&4.5&4.2&5.2&6.4&4.2&&13.9&19.3&21.2&27.1&20.9&&47&51.6&56&63.3&58.2\\
		&-0.5&5.2&4.9&5&6.6&5&&16.5&24.4&23.6&31.8&25&&48.2&58.8&59.4&68&62\\
		&-0.1&4.8&4.4&4.2&6.1&5.8&&19.6&28.7&25.8&36.5&29.5&&54.3&66.2&64.6&73.7&66.5\\
		&0.1&5.4&4.9&4.5&6.4&7.2&&22.6&30.8&27.5&37.6&33&&58&69.8&67.9&76.2&69.3\\
		&0.5&5.7&4.3&4.2&6.5&9.4&&32.4&35.8&36&45.9&43&&73.9&78.4&79.4&84.6&82.9\\
		&0.9&6.4&4.8&3.8&6.1&15.3&&80.7&37.4&78.2&73.8&81.9&&99.9&89&99.2&98.1&98.1\\
		\\
		Chisq&-0.9&5.2&5.4&5.5&6.4&4.3&&12.6&15.6&17.1&17.7&17.1&&40.7&46&50.2&49&53.6\\
		&-0.5&4.6&5.1&5.2&6.3&4.7&&13.8&19.4&19.1&20&20.1&&43.6&54.5&55.1&54&56.8\\
		&-0.1&5.4&5.6&5.5&6.7&5.9&&16.5&22.5&20.8&22.5&22.7&&49.7&63&61&61.2&63.3\\
		&0.1&5.2&5.2&5&6.3&5.5&&18&24.9&22.8&25.2&25.8&&55.8&68.7&66.2&66.6&68.9\\
		&0.5&4.9&4.8&4.7&6.2&6.3&&26.2&29.7&30.8&32.2&37&&74.8&80.2&81.5&80.2&85.2\\
		&0.9&5.6&5.3&4.5&5.7&12.8&&80.1&32.2&80.3&69.2&85.1&&100&92.6&100&98.8&99.9\\\hline	
		\end{tabular*} }
	\end{table*}

\begin{table*}[ht]
	
	\caption{Power simulation results ($\alpha=0.05$) of the tests for different distributions under varying correlation values ($\rho$) with sample sizes $(n_c,n_u)=(10,10)$ and heteroscedastic covariance matrix $\Sigma_2$ under the MCAR framework.
		\label{Powern1010hetro}}
	{\tabcolsep=4pt
		\begin{tabular*}{1.09\linewidth}{lc ccccc c ccccc  c ccccc   cc}
			Dist & $\rho$ & \multicolumn{5}{c}{$\delta=0$}  && \multicolumn{5}{c}{$\delta=0.5$} && \multicolumn{5}{c}{$\delta=1$} \\ \cline{3-7} \cline{9-13} \cline{15-19}
			&& $T_{W}^*$ & $T_{A}^*$ & $T_{M}^*$ & $T_{NC}$ & $T_{LT}$
			&& $T_{W}^*$ & $T_{A}^*$ & $T_{M}^*$ & $T_{NC}$ & $T_{LT}$ 
			&& $T_{W}^*$ & $T_{A}^*$ & $T_{M}^*$ & $T_{NC}$ & $T_{LT}$\\
			
	Normal&-0.9&5.5&5.5&5.7&7&4.7&&10.8&11.9&13.0&14.4&13.0&&29.9&32.0&35.3&35.0&38.6\\
	&-0.5&5.1&5.6&5.8&7.2&5.1&&10.3&13.2&13.7&15.2&13.4&&29.1&36.6&38.5&38.3&39.2\\
	&-0.1&5.5&5.8&5.7&7.1&6&&11.5&15.5&15.6&17.0&16.0&&33.1&44.4&43.9&44.4&44.4\\
	&0.1&5.2&5.5&5.6&6.8&6.2&&12.4&16.0&15.9&17.5&16.9&&35.6&47.3&46.5&46.5&47.5\\
	&0.5&5.4&5.2&5.2&6.4&7&&16.9&18.8&20.1&21.9&23.6&&51.7&57.1&60.1&59.7&64.3\\
	&0.9&5.4&5.4&6.1&6.4&12.4&&48.4&20.3&43.3&38.4&61.4&&97.1&68.3&95.1&86.8&97.8\\
	\\
	Laplace&-0.9&4.6&4.8&5.6&7.2&4.8&&11.2&12.0&14.1&17.1&14.1&&32.4&33.8&38.0&43.1&40.3\\
	&-0.5&4.5&4.8&5.2&6.8&4.8&&11.6&14.9&15.8&19.4&15.1&&34.0&40.7&43.5&50.1&43.2\\
	&-0.1&4.5&5&5.2&6.7&5.7&&12.2&15.7&16.1&20.9&16.0&&36.8&47.4&47.8&55.3&47.5\\
	&0.1&4.6&4.8&4.9&7&5.4&&14.3&18.5&18.4&24.1&19.5&&41.8&52.5&52.4&60.3&52.8\\
	&0.5&4.2&4.6&4.6&6.7&6.4&&19.2&21.4&23.4&30.6&27.2&&57.2&61.0&64.6&70.7&68.2\\
	&0.9&4.1&5.3&5.3&6.8&13.9&&54.4&22.6&50.0&50.1&65.1&&96.6&70.6&92.6&89.0&94.8\\
	\\
	Exp&-0.9&4.6&5.2&6.2&9.1&5.2&&14.5&18.1&19.8&27.0&19.5&&38.2&40.6&43.8&53.4&46.8\\
	&-0.5&5.3&6&6.2&9.4&6&&15.3&21.9&22.0&31.4&22.2&&37.1&46.0&47.4&57.8&49.3\\
	&-0.1&5.4&6.2&5.6&9.9&7.3&&17.4&24.7&23.2&34.3&25.3&&40.8&52.3&51.2&62.6&52.9\\
	&0.1&6.4&7&6.1&10&8.9&&19.3&26.7&24.9&35.2&27.5&&44.0&54.4&53.2&64.2&54.2\\
	&0.5&6.9&6.4&6.5&10.5&11&&25.9&29.7&30.3&40.4&33.8&&56.5&61.3&62.7&71.8&65.5\\
	&0.9&8.1&5.8&8.8&12.2&18.2&&56.4&30.5&52.4&55.2&64.4&&94.0&69.0&86.6&85.6&90.3\\
	\\
	Chisq&-0.9&5.2&5.4&5.8&7.1&4.4&&10.8&12.6&13.8&15.7&14.1&&30.0&33.1&35.9&36.6&39.7\\
	&-0.5&4.6&5.3&5.3&6.9&5&&10.8&15.2&15.6&17.1&15.8&&30.8&39.6&40.7&41.2&41.5\\
	&-0.1&5.5&6&5.9&7.2&6.6&&13.0&17.5&16.8&19.0&17.8&&33.6&45.7&44.5&46.6&45.5\\
	&0.1&5&5.4&5.1&6.6&5.9&&13.7&18.7&17.9&20.6&19.5&&38.0&49.9&48.6&50.2&50.2\\
	&0.5&5&5.5&5.4&6.9&6.8&&18.2&21.6&22.1&24.9&25.8&&52.1&58.3&60.8&60.9&64.7\\
	&0.9&5.8&5.3&6.3&7.1&13&&50.7&23.5&46.0&41.8&61.5&&96.2&68.3&92.3&84.9&95.7\\\hline
			
		\end{tabular*} }
	\end{table*}

\begin{table*}[ht]
	
	\caption{Power simulation results ($\alpha=0.05$) of the tests for different distributions under varying correlation values ($\rho$) with sample sizes $(n_c,n_u)=(10,30)$ and homoscedastic covariance matrix $\Sigma_1$ under the MCAR framework.
		\label{Table1}}
	{\tabcolsep=4pt
		\begin{tabular*}{1.09\linewidth}{lc ccccc c ccccc  c ccccc   cc}
			Dist & $\rho$ & \multicolumn{5}{c}{$\delta=0$}  && \multicolumn{5}{c}{$\delta=0.5$} && \multicolumn{5}{c}{$\delta=1$} \\ \cline{3-7} \cline{9-13} \cline{15-19}
			&& $T_{W}^*$ & $T_{A}^*$ & $T_{M}^*$ & $T_{NC}$ & $T_{LT}$
			&& $T_{W}^*$ & $T_{A}^*$ & $T_{M}^*$ & $T_{NC}$ & $T_{LT}$ 
			&& $T_{W}^*$ & $T_{A}^*$ & $T_{M}^*$ & $T_{NC}$ & $T_{LT}$\\
		Normal&-0.9&5.5&5.4&6.7&6.8&4&&22.2&15.9&21.9&19.3&25.9&&70.1&46.4&62.4&52.2&77.8\\
		&-0.5&5.5&5.6&6.4&6.7&5.1&&18.4&18.4&23.3&21&23.7&&60.4&55.2&65.4&58.4&70.5\\
		&-0.1&5.3&5.6&6.1&6.8&6&&19.1&23&25&23.8&25.6&&59.6&66.2&70.1&66&71\\
		&0.1&5.2&5.5&5.8&6.5&6.1&&20.3&26.1&26.3&26.8&27.8&&62.8&73.5&73.9&71.5&74.8\\
		&0.5&5&5.1&4.9&6.2&6.5&&27.2&34.5&34.1&35.4&37.5&&78.9&88&86.8&86&88.7\\
		&0.9&5.4&5.4&4.3&6.1&8&&80.6&46.6&78.9&77.7&89.1&&100&98.6&99.9&99.8&100\\
		\\
		Laplace&-0.9&4.6&4.7&6&6.8&3.1&&23&16.3&23.4&23.8&26.9&&72.8&49.5&65.7&61.1&78.8\\
		&-0.5&4.3&5&5.8&6.7&4.3&&20.3&19.8&25.1&26.9&25.8&&64.2&57.8&68.4&67.6&72\\
		&-0.1&4.6&4.9&5.3&6.4&5.3&&21&25.2&28.3&32.2&28.3&&63.9&69&73.5&76.6&73.6\\
		&0.1&4.3&4.8&5.3&6.4&5.6&&22.8&28.5&30.8&35.2&30.3&&66.6&74.9&76.9&80.6&76.4\\
		&0.5&4.3&4.5&4.9&6.6&5.7&&32.2&39.5&40.3&47.1&43.3&&81.7&87.3&87&91&88.7\\
		&0.9&4.3&5.1&4&6.1&9.4&&84.2&49.7&79.5&82.6&88.6&&99.9&96.6&99.3&99.7&100\\
		\\
		Exp&-0.9&4.6&4.2&6.1&6.7&3.9&&25.8&20.4&28.4&29.7&33&&72.9&53.2&66.1&66.5&80.4\\
		&-0.5&5&4.7&6.4&6.2&5.4&&26.1&26&31.8&32.6&34.2&&63.1&60.3&67.2&70.6&73.8\\
		&-0.1&7.3&5.9&7.5&7&7.8&&29&32.6&35.2&39&37.4&&63.1&69.4&71.8&76.4&72.8\\
		&0.1&7.5&5.7&7.1&6.4&8.1&&30.3&36.3&36.8&42.5&38.4&&64.2&72.9&73.6&79.4&72.8\\
		&0.5&8.4&6&6.8&6.1&10.8&&37.1&44.6&43.9&50.5&45.8&&75.7&83&83.3&87.8&83\\
		&0.9&10&6.6&4.9&6.4&16.2&&80.8&52.7&77.6&77.9&86.5&&99.9&92.7&98.2&98.9&99.6\\
		\\
		Chisq&-0.9&5&5.4&6.6&6.6&3.7&&21.2&15.2&21.5&18.7&26.3&&70.2&45.4&62.5&51.9&78.5\\
		&-0.5&4.9&5.2&6&6.4&4.6&&19.6&19.1&24.2&21.5&26.1&&60.7&55.9&65.6&58.4&71.6\\
		&-0.1&5.5&5.7&6.2&6.6&5.6&&20.4&24.7&26.5&24.7&27.6&&59.3&66.4&69.6&65.3&70.5\\
		&0.1&5.8&5.8&6&6.4&6.5&&22.4&28.7&28.8&27.8&29.9&&62.9&72.9&73.5&71.1&74.4\\
		&0.5&5.7&5&5.3&6&6.8&&29.6&36.6&36&36.6&39.7&&77.8&86.5&85.7&84.2&86.9\\
		&0.9&5.8&5.3&4.5&6&8.9&&80.2&47.4&78&75.4&88.2&&100&97&99.8&99.6&100\\\hline
		\end{tabular*} }
	\end{table*}

\begin{table*}[ht]
	
	\caption{Power simulation results ($\alpha=0.05$) of the tests for different distributions under varying correlation values ($\rho$) with sample sizes $(n_c,n_u)=(10,30)$ and heteroscedastic covariance matrix $\Sigma_2$ under the MCAR framework.
		\label{Powern3010hetro}}
	{\tabcolsep=4pt
		\begin{tabular*}{1.09\linewidth}{lc ccccc c ccccc  c ccccc   cc}
			Dist & $\rho$ & \multicolumn{5}{c}{$\delta=0$}  && \multicolumn{5}{c}{$\delta=0.5$} && \multicolumn{5}{c}{$\delta=1$} \\ \cline{3-7} \cline{9-13} \cline{15-19}
			&& $T_{W}^*$ & $T_{A}^*$ & $T_{M}^*$ & $T_{NC}$ & $T_{LT}$
			&& $T_{W}^*$ & $T_{A}^*$ & $T_{M}^*$ & $T_{NC}$ & $T_{LT}$ 
			&& $T_{W}^*$ & $T_{A}^*$ & $T_{M}^*$ & $T_{NC}$ & $T_{LT}$\\

		Normal&-0.9&5.5&5.4&7&7.8&4.1&&15.6&12.0&16.2&15.5&18.3&&50.9&31.3&42.7&37.2&59.5\\
		&-0.5&5.5&5.6&6.7&7.4&5.6&&12.5&13.5&16.5&16.2&16.3&&38.1&37.7&44.3&41.6&48.4\\
		&-0.1&5.4&5.7&6.4&7.8&6.4&&12.7&15.8&17.1&18.1&17.5&&37.1&44.8&47.7&46.8&48.5\\
		&0.1&5.2&5.6&6&7.2&6.6&&13.5&17.2&18.3&19.7&18.7&&38.8&50.3&51.4&51.5&51.6\\
		&0.5&5&5.6&5.8&7.4&6.7&&16.8&21.1&21.6&23.6&23.9&&51.5&62.0&62.3&63.8&65.8\\
		&0.9&5&6&7.2&6.9&7.7&&49.8&25.9&43.4&42.6&62.6&&97.7&75.7&92.0&89.8&99.1\\
		\\
		Laplace&-0.9&4.6&4.6&6.3&7.5&3.3&&16.4&12.4&17.1&18.4&18.7&&54.3&35.0&46.2&46.5&61.7\\
		&-0.5&4.2&4.8&6.1&7.4&4.5&&13.4&14.3&17.8&20.1&18.0&&43.0&41.0&48.5&52.0&52.8\\
		&-0.1&4.5&4.8&5.5&7&5.6&&13.9&17.1&19.4&23.5&19.2&&42.0&49.8&53.5&59.8&52.8\\
		&0.1&4.3&5.1&5.8&7.2&5.8&&14.5&19.4&21.0&25.6&20.7&&44.2&54.4&56.8&63.7&55.3\\
		&0.5&4&4.7&5.8&7.2&6&&20.1&24.9&26.5&32.8&28.5&&58.3&66.7&67.4&75.4&70.4\\
		&0.9&4&5.2&6.6&6.9&8.7&&56.8&28.4&49.9&53.9&66.8&&97.1&77.7&89.9&91.7&97.9\\
		\\
		Exp&-0.9&5.1&5.1&7.3&9.5&5&&22.3&18.6&24.1&29.6&27.2&&58.4&41.5&50.6&56.7&65.1\\
		&-0.5&6.3&6.4&8.2&9.7&7.5&&21.4&22.4&26.5&31.9&27.6&&48.5&46.8&52.3&59.3&58.2\\
		&-0.1&8&7.9&9.3&11.1&9.8&&23.2&27.2&29.1&37.1&30.0&&48.0&54.1&56.3&65.2&57.3\\
		&0.1&8.3&7.9&8.7&10.9&9.8&&23.9&29.0&30.1&38.7&30.6&&48.6&56.8&58.2&68.0&57.1\\
		&0.5&9.3&8.4&9&11.2&12.1&&27.9&34.3&34.9&44.7&34.8&&56.8&64.9&66.2&76.2&65.6\\
		&0.9&10.2&8.2&11.5&13.6&15.8&&57.7&37.5&53.9&59.0&66.1&&96.4&73.4&84.2&88.7&96.2\\
		\\
		Chisq&-0.9&5.3&5.4&7&7.6&3.8&&16.0&12.2&16.6&16.3&19.6&&50.6&31.8&42.3&38.6&60.5\\
		&-0.5&4.9&5.5&6.6&7.4&5.2&&14.5&14.9&18.0&18.7&19.2&&40.5&39.6&46.6&43.9&51.8\\
		&-0.1&5.6&5.8&6.5&7.7&6.2&&14.7&17.8&19.2&20.6&19.7&&38.7&46.3&48.9&49.1&49.7\\
		&0.1&5.8&5.9&6.5&7.3&7&&15.6&20.3&20.9&22.5&21.4&&41.3&52.2&53.1&53.8&52.9\\
		&0.5&5.8&5.5&6&7.2&7.2&&18.9&24.5&25.1&27.5&26.5&&52.2&62.8&63.0&65.6&65.5\\
		&0.9&5.5&5.5&7.4&7.4&8.3&&52.0&28.2&45.8&44.8&63.1&&97.1&73.3&89.0&87.5&98.5\\\hline

		\end{tabular*} }
	\end{table*}


\begin{table*}[ht]
	
	\caption{Power simulation results ($\alpha=0.05$) of the tests for different distributions under varying correlation values ($\rho$) with sample sizes $n=10$ and homoscedastic covariance matrix $\Sigma_1$ under the MAR framework.
		\label{Table1}}
	{\tabcolsep=4pt
		\begin{tabular*}{1.09\linewidth}{lc ccccc c ccccc  c ccccc   cc}
			Dist & $\rho$ & \multicolumn{5}{c}{$\delta=0$}  && \multicolumn{5}{c}{$\delta=0.5$} && \multicolumn{5}{c}{$\delta=1$} \\ \cline{3-7} \cline{9-13} \cline{15-19}
			&& $T_{W}^*$ & $T_{A}^*$ & $T_{M}^*$ & $T_{NC}$ & $T_{LT}$
			&& $T_{W}^*$ & $T_{A}^*$ & $T_{M}^*$ & $T_{NC}$ & $T_{LT}$ 
			&& $T_{W}^*$ & $T_{A}^*$ & $T_{M}^*$ & $T_{NC}$ & $T_{LT}$\\
		Normal&-0.9&5.4&3.7&5.2&6.4&3.6&&6.2&5.6&6.2&8.6&5.5&&10.2&19.2&13.8&24.1&20.3\\
		&-0.5&4.8&3.9&4.4&6.4&4.6&&6.6&6.6&7.3&11.3&8.2&&13.5&20.9&17.7&28&24.3\\
		&-0.1&4.6&4.1&4.4&6.2&5.7&&6.8&7.6&7.9&12.1&11.1&&17.1&24.5&21.7&33.6&32.1\\
		&0.1&4.7&4.3&4.3&6.7&6.2&&7.8&8.8&8.9&14&13.6&&20.8&26.7&25.2&36.9&39.1\\
		&0.5&4.7&4.2&4.1&6.3&7.3&&12.5&10.4&12.9&18.7&23.3&&37.6&31.3&40.8&50.2&60.7\\
		&0.9&4.9&4.9&4.3&6.5&12.5&&48.8&10.6&46.9&39.9&65.6&&97.1&35.9&97.2&80.7&94.9\\
		\\
		Laplace&-0.9&4.2&3.2&5.1&6.1&3.2&&5.4&5.5&7.1&10.1&5.3&&10.5&20.1&16.3&29.9&21.9\\
		&-0.5&3.1&3.6&3.5&6.2&3.9&&5.4&7.1&8&13.4&8.4&&14.1&23.2&21&35.3&26.1\\
		&-0.1&3.1&3.2&3.1&6.3&4.8&&6.5&8.5&8.6&16.2&11.8&&20.4&28.2&27.2&42.3&35.5\\
		&0.1&3.2&2.7&3&6.3&5.6&&7.4&9.2&9.3&18.1&14.9&&24.9&30.8&30.6&45.9&42\\
		&0.5&3.5&3.1&3&6.3&7.8&&13&10.2&13.7&23.1&26&&42.3&35.7&44.5&57.4&62.1\\
		&0.9&3.6&3.9&3.3&5.6&11&&51.3&11.2&47.6&44&66.4&&95.4&38.7&94.5&81.6&89.2\\
		\\
		Exp&-0.9&4.3&2.8&5.6&6.1&3.2&&3.5&5.7&7.2&13.2&6.2&&10.3&23.3&19.5&35.1&24.6\\
		&-0.5&4.8&3&4.2&6.6&4.1&&4.1&8.6&6.2&17.1&11.7&&14.5&31.8&24.9&45.1&34.9\\
		&-0.1&4.3&2.7&3.4&6.5&5.2&&6.9&10.5&7.3&21.4&18.3&&23&40&31.2&54&47.9\\
		&0.1&4&2.6&3.4&6.5&6.2&&8.3&10.9&8.4&23.2&21.8&&27.7&40.6&33.7&55.2&52.7\\
		&0.5&3.5&2.7&3.2&6.5&10.1&&15.6&11&14.1&26.9&32.8&&45.8&44.3&46.5&62&68.8\\
		&0.9&4.3&4.7&4.2&5.8&10.2&&56.3&8.7&51&43.2&69.1&&95.4&42.9&92.8&84.3&89.9\\
		\\
		Chisq&-0.9&4.6&3.7&4.5&5.6&3.7&&5.3&5.3&6.4&8.8&5.3&&9.7&19.1&14.3&25&20.7\\
		&-0.5&4.5&3.7&4.4&6.3&4.3&&6.3&7.8&7.2&11.3&9.4&&12&21.4&16.6&27.9&25.3\\
		&-0.1&4.7&3.8&4.4&6.6&5.7&&6.6&7.9&7.5&13&12.9&&16.7&26.4&21.9&34.1&35.2\\
		&0.1&4.8&3.7&4.1&6.1&6.1&&7.8&8.5&8.5&13.7&15&&21.1&28.5&25.7&39&42.3\\
		&0.5&4.2&3.8&3.6&6&7.3&&12.1&9.3&12.2&18.4&24.2&&39&33.4&42.3&50.4&64\\
		&0.9&4.5&4.6&3.8&5.5&12.8&&49.9&10.2&46.8&38.7&67.1&&97.2&35&96.8&80.6&94.6\\\hline	
		\end{tabular*} }
	\end{table*}

	\begin{table*}[ht]
		
		\caption{Power simulation results ($\alpha=0.05$) of the tests for different distributions under varying correlation values ($\rho$) with sample sizes $n=10$ and heteroscedastic covariance matrix $\Sigma_2$ under the MAR framework.
			\label{Powern1010hetro}}
		{\tabcolsep=4pt
			\begin{tabular*}{1.09\linewidth}{lc ccccc c ccccc  c ccccc   cc}
				Dist & $\rho$ & \multicolumn{5}{c}{$\delta=0$}  && \multicolumn{5}{c}{$\delta=0.5$} && \multicolumn{5}{c}{$\delta=1$} \\ \cline{3-7} \cline{9-13} \cline{15-19}
				&& $T_{W}^*$ & $T_{A}^*$ & $T_{M}^*$ & $T_{NC}$ & $T_{LT}$
				&& $T_{W}^*$ & $T_{A}^*$ & $T_{M}^*$ & $T_{NC}$ & $T_{LT}$ 
				&& $T_{W}^*$ & $T_{A}^*$ & $T_{M}^*$ & $T_{NC}$ & $T_{LT}$\\
				
			Normal&-0.9&5.1&4.1&4.9&6.3&4&&5.2&4.9&5.5&7&4.7&&8.5&15.5&10.5&18.9&16.4\\
			&-0.5&5.1&4.4&5.2&6.8&5.2&&5.8&6.4&6.5&8.9&7.2&&9.6&16.5&13.5&20.6&18.2\\
			&-0.1&4.9&4.4&4.5&6.3&6&&6.3&7.6&7.4&10.6&10.2&&12.3&19.1&16.9&23.9&24.1\\
			&0.1&5.2&4.5&4.5&6.6&6.5&&6.8&8.2&8.1&11.1&11.7&&14.9&20.9&19.3&26.4&28.8\\
			&0.5&4.4&4&4&6.3&7.3&&9.4&8.3&10&13.3&17.1&&24.6&23.8&28.7&34.4&44.3\\
			&0.9&4.6&4.8&5&7&14.7&&25.4&8.8&22.6&22.6&49.2&&75.8&26.2&70&58.8&83.1\\
			\\
			Laplace&-0.9&4&3.4&5.2&6.3&3.6&&4.8&5&6.4&8.3&4.9&&8.1&16&12.5&22.8&17.4\\
			&-0.5&3.2&3.6&4.1&6&4.2&&4.4&6&6.4&10.6&7.2&&10.4&18.8&17.3&27.3&20.9\\
			&-0.1&3.1&3.7&4&6.5&5.4&&5&7.2&7&13&10&&13.4&20.8&20&31.8&25.9\\
			&0.1&3.1&2.8&3&5.9&5.7&&6.3&8.3&8&14.4&12.4&&17&23.9&23&35&31.8\\
			&0.5&3.4&3.4&3.4&6.5&8.1&&9.4&9.5&10.9&17.5&20.4&&29.2&27.9&32.9&42.8&49\\
			&0.9&3.3&4.1&3.9&6.6&13.1&&29.7&10.1&27.4&28.4&51.9&&75.8&28.8&71&62.1&78.8\\
			\\
			Exp&-0.9&3.3&4&6&7.4&4.3&&4.3&7.5&8.9&14.4&7.5&&10.3&20.6&17.3&30.4&21.9\\
			&-0.5&4.2&4.1&4.1&7.8&5.4&&5.5&10.8&8.7&18.1&12.9&&13.1&28.1&23.5&37.8&30\\
			&-0.1&4.5&3.5&3.8&8.2&7&&8.5&13.5&10.7&21.6&18.6&&19.5&34.4&28.8&43.4&39.6\\
			&0.1&3.9&3.5&3.5&8.6&8.5&&10.9&14.6&12.5&23.5&22.3&&23.3&35.5&30.5&44.7&43\\
			&0.5&4.6&2.9&4.2&7.8&11&&16.3&13.3&16&24.9&30.1&&37.7&38&40&49.5&56.4\\
			&0.9&7.8&4&7.8&8.7&15.9&&43&11.4&35.7&30.7&59.6&&77&33.9&67.8&60.9&78.4\\
			\\
			Chisq&-0.9&5.1&4.2&5.3&6.1&4.3&&5.2&5.4&5.4&8.2&5&&7.9&15.8&10.5&19&16.7\\
			&-0.5&4.8&4.5&5.1&6.2&5.1&&5.5&7.2&6.8&10.4&8.3&&9.9&18.6&14.8&22.1&20.5\\
			&-0.1&4.5&4.3&4.7&6.5&6.3&&7.2&8.8&7.9&12.1&12.1&&12.8&21.2&17.9&25.7&26.2\\
			&0.1&4.3&4&4&6.3&6.6&&7.4&8.5&8.1&12.6&13.3&&15.9&23.3&21.1&28.6&32\\
			&0.5&4.7&3.9&4.3&6.5&7.6&&10.8&10&11.3&15.3&20.2&&27.3&26.5&31.4&36.5&47.7\\
			&0.9&4.9&4.5&5&7&15&&29&9.3&25.2&23.3&52.2&&75.7&27.1&68.6&56.5&83.1\\\hline
				
			\end{tabular*} }
		\end{table*}


\begin{table*}[ht]
	
	\caption{Power simulation results ($\alpha=0.05$) of the tests for different distributions under varying correlation values ($\rho$) with sample sizes $n=20$ and homoscedastic covariance matrix $\Sigma_1$ under the MAR framework.
		\label{Table1}}
	{\tabcolsep=4pt
		\begin{tabular*}{1.09\linewidth}{lc ccccc c ccccc  c ccccc   cc}
			Dist & $\rho$ & \multicolumn{5}{c}{$\delta=0$}  && \multicolumn{5}{c}{$\delta=0.5$} && \multicolumn{5}{c}{$\delta=1$} \\ \cline{3-7} \cline{9-13} \cline{15-19}
			&& $T_{W}^*$ & $T_{A}^*$ & $T_{M}^*$ & $T_{NC}$ & $T_{LT}$
			&& $T_{W}^*$ & $T_{A}^*$ & $T_{M}^*$ & $T_{NC}$ & $T_{LT}$ 
			&& $T_{W}^*$ & $T_{A}^*$ & $T_{M}^*$ & $T_{NC}$ & $T_{LT}$\\
		Normal&-0.9&5.4&5.1&5.7&5.8&5.5&&11.6&18.8&17.4&17.2&19.4&&35.7&55.8&49.7&52.6&57.8\\
		&-0.5&5.3&4.9&5.5&5.6&5.2&&14.3&20.5&20.1&20&22.9&&43.6&60.2&56.4&56.6&64.8\\
		&-0.1&5.1&4.7&5&5.5&5.5&&17.2&20.7&22&21.4&26.6&&55.6&64.7&65.5&63.7&75.4\\
		&0.1&4.9&4.3&5&5.1&5.9&&20.6&21.9&25&24.4&32.2&&63.6&67.1&71.9&68.8&82\\
		&0.5&5.2&5&5.1&5.5&6.4&&31.9&20.8&35&30.4&47.2&&88&72&90.5&82.1&95.8\\
		&0.9&5&4.8&4.6&5.3&14.2&&93.1&19.9&92.8&68.8&93.3&&100&75&100&98.9&100\\
		\\
		Laplace&-0.9&4.3&4.9&6&5.3&5&&12.7&20.9&21.8&22.6&22&&40.9&58.6&56.1&61.7&59.4\\
		&-0.5&4.2&4.6&5.5&5.5&5&&14.3&21.6&22.9&25.5&23.8&&48.8&63.1&62.8&69&66.3\\
		&-0.1&3.9&3.9&4.6&5.1&4.7&&18.1&23.6&25.9&30.5&29.4&&60.4&68.9&71.1&76.6&76.6\\
		&0.1&4.1&3.7&4.6&5.6&5.3&&21.5&24.8&28.3&33.5&33.9&&67.8&71.3&76&80.2&81.8\\
		&0.5&4.5&4&4.2&5.2&5.6&&35.4&25.6&38.8&42.5&50.5&&87.9&76.7&90.3&89.7&95\\
		&0.9&4&4.1&3.8&4.9&14.4&&91.6&22.3&90.7&74.4&89.6&&100&79.1&100&99.1&99.6\\
		\\
		Exp&-0.9&5.6&5&7.8&5&5.3&&11.1&21.8&22.6&26.8&22.4&&41.9&61&60.7&69.2&62.5\\
		&-0.5&7.6&4.9&7.4&5.7&5.4&&14.2&24.5&22.6&31.4&26.8&&45.6&66.9&65.4&74&68.9\\
		&-0.1&7.4&4.2&6.4&5.4&5.2&&18.6&25.2&23.6&36.7&33&&57.1&72&71&79.9&76.5\\
		&0.1&7.1&4.3&6.5&5.6&5.9&&22.4&25.6&25.8&38.4&37.4&&64.5&75&75.4&83.4&82.1\\
		&0.5&5.7&4.8&5.7&5.7&6.9&&35.1&23.4&35.1&44&51.8&&85.2&80.4&87.6&89.9&93.8\\
		&0.9&5&5.7&5.2&5.2&13.4&&91.5&17.5&87.6&71.2&88.8&&100&85.2&100&99.4&99.2\\
		\\
		Chisq&-0.9&5.1&5.3&5.9&5.9&5.5&&11.7&19.1&18.1&17.8&19.6&&34.8&56.7&50.7&53.2&58\\
		&-0.5&5.1&4.8&5.4&5.4&5.3&&13.2&19.9&18.7&19.1&22.3&&42.4&60.2&56.3&57&64.8\\
		&-0.1&5.3&4.7&5.3&5.2&5.8&&16.7&20.6&21.2&21.3&27.8&&54.9&66.6&66&64.7&76.1\\
		&0.1&5.1&4.1&4.7&5.3&5.5&&19.6&20.8&23.3&23.3&31.5&&63.5&69.4&72.4&68.8&82.1\\
		&0.5&4.9&4.4&4.8&5.3&6.1&&31.8&20.4&33.8&31.2&47.8&&87.1&73.8&89.6&81.5&95.4\\
		&0.9&4.9&5.3&4.5&4.9&14.7&&92.2&17.2&91.9&65.4&93.1&&100&78.1&100&98.9&100\\\hline	
		\end{tabular*} }
	\end{table*}

	\begin{table*}[ht]
		
		\caption{Power simulation results ($\alpha=0.05$) of the tests for different distributions under varying correlation values ($\rho$) with sample sizes $n=20$ and heteroscedastic covariance matrix $\Sigma_2$ under the MAR framework.
			\label{Powern1010hetro}}
		{\tabcolsep=4pt
			\begin{tabular*}{1.09\linewidth}{lc ccccc c ccccc  c ccccc   cc}
				Dist & $\rho$ & \multicolumn{5}{c}{$\delta=0$}  && \multicolumn{5}{c}{$\delta=0.5$} && \multicolumn{5}{c}{$\delta=1$} \\ \cline{3-7} \cline{9-13} \cline{15-19}
				&& $T_{W}^*$ & $T_{A}^*$ & $T_{M}^*$ & $T_{NC}$ & $T_{LT}$
				&& $T_{W}^*$ & $T_{A}^*$ & $T_{M}^*$ & $T_{NC}$ & $T_{LT}$ 
				&& $T_{W}^*$ & $T_{A}^*$ & $T_{M}^*$ & $T_{NC}$ & $T_{LT}$\\
				
				Normal&-0.9&5.2&5.4&5.9&5.1&5.5&&9.9&16&15.8&13.2&15.8&&24.7&42.5&37.7&37&43.4\\
				&-0.5&5.1&5.3&5.9&5.3&5.4&&10.8&15.9&16.4&13.7&16.6&&29.7&46.4&43.5&40.8&48.7\\
				&-0.1&5.1&5.2&5.5&5.2&5.6&&12.6&16.1&16.8&14.8&19.5&&36.8&49.3&48.7&44.6&56.2\\
				&0.1&5.3&5.1&5.5&5.4&6.5&&14.1&16.6&18.2&16.3&22.1&&44.5&52.2&54.5&48.7&63.5\\
				&0.5&5.1&4.9&5.2&5.2&6&&20.9&16.6&23.9&19.4&31.9&&66.4&56.7&71.2&60.1&82.9\\
				&0.9&4.9&4.7&5.3&5.2&14.1&&61.5&14.5&53.1&33.1&74.5&&99.7&58.1&99.2&88.6&99.2\\
				\\
				Laplace&-0.9&4.2&4.9&6.8&4.8&5&&10.2&16&17.9&16.2&16.4&&27.6&45&42.9&45.9&45.4\\
				&-0.5&4.5&5&6.4&5.4&5.2&&11.3&17.3&18.9&19.1&18.3&&34.3&48.9&48.9&53.4&50\\
				&-0.1&3.9&4.5&5.3&5.3&5.2&&12.6&18.4&20.2&21.8&21.4&&43.2&55&56.2&60.5&60.2\\
				&0.1&4.2&4.3&5.1&5&5.5&&14.7&18.6&21&23&24.3&&47.9&56.4&59.5&63.1&65.4\\
				&0.5&4.2&4&4.5&5.1&6&&23.3&19.3&26.7&28.3&36.3&&68.1&60.1&73&72.4&82.5\\
				&0.9&4.3&4.9&5.1&5.2&16.2&&64.2&16.5&57.6&46.8&74.3&&98.9&62.2&97.8&91.3&97\\
				\\
				Exp&-0.9&4.7&5.7&9&7.4&6&&11.5&19.9&22.6&27.6&20.2&&33.2&48.7&49.8&57.7&49.4\\
				&-0.5&6.5&5.5&7&7.2&6.4&&13.7&22.1&22.6&30.7&23.2&&34.7&53.4&53.9&62.2&54.4\\
				&-0.1&6.9&4.7&6.2&7.4&6.2&&18&23.5&23.2&34.5&27.6&&43&56.8&57&67.3&60.3\\
				&0.1&6.3&4.4&6.1&7.7&6.7&&19.9&22.6&23.4&34.9&29.5&&48.9&59&60.3&70.1&65.2\\
				&0.5&6.3&4.5&6.3&8.1&8.4&&29.8&21.3&29.7&38.6&39.7&&66.5&61.7&69.4&75.7&79.1\\
				&0.9&7.7&4.5&8.1&8.6&17.5&&66.5&15.2&54.2&49.9&72.9&&97.2&63.5&93.9&90.4&93.8\\
				\\
				Chisq&-0.9&5.8&5.8&6.6&5.8&5.9&&9.5&15.5&15.1&14.4&15.8&&25.6&43.2&39.4&39.7&44.1\\
				&-0.5&5.1&4.7&5.5&5.1&5.2&&11&16.5&16.3&15.6&17.3&&30.5&47.1&44.2&43.3&49.1\\
				&-0.1&5.3&5.2&5.7&5.2&5.6&&13&16.6&17.2&17&21.1&&38.4&51.5&50.9&49.1&57.9\\
				&0.1&5&4.8&5.2&5.2&5.8&&15.1&17.3&18.6&17.9&23.6&&44.5&52.9&55&51.6&63.8\\
				&0.5&5.4&4.8&5.3&5.6&6.5&&22.6&16.5&24.3&22.3&33.7&&65.6&56.5&69.4&61.1&81.4\\
				&0.9&5.3&4.8&5.4&5.6&15.2&&63.3&13.2&53.4&35.7&73.9&&99.3&59&98&85.7&98.5\\\hline
				
			\end{tabular*} }
		\end{table*}
		

\begin{table*}[ht]
	
	\caption{Power simulation results ($\alpha=0.05$) of the tests for different distributions under varying correlation values ($\rho$) with sample sizes $n=30$ and homoscedastic covariance matrix $\Sigma_1$ under the MAR framework.
		\label{Table1}}
	{\tabcolsep=4pt
		\begin{tabular*}{1.09\linewidth}{lc ccccc c ccccc  c ccccc   cc}
			Dist & $\rho$ & \multicolumn{5}{c}{$\delta=0$}  && \multicolumn{5}{c}{$\delta=0.5$} && \multicolumn{5}{c}{$\delta=1$} \\ \cline{3-7} \cline{9-13} \cline{15-19}
			&& $T_{W}^*$ & $T_{A}^*$ & $T_{M}^*$ & $T_{NC}$ & $T_{LT}$
			&& $T_{W}^*$ & $T_{A}^*$ & $T_{M}^*$ & $T_{NC}$ & $T_{LT}$ 
			&& $T_{W}^*$ & $T_{A}^*$ & $T_{M}^*$ & $T_{NC}$ & $T_{LT}$\\
			Normal&-0.9&5.1&5.4&6&5.7&5.5&&18.2&27.3&27.1&24.6&27.7&&59.9&75.3&72.9&69.8&76\\
			&-0.5&5.2&5.1&5.8&5.3&5.1&&21.6&29.5&29&26.6&31.2&&68.1&80.5&79.1&75.4&83.1\\
			&-0.1&5&5&5.1&5.3&5.2&&26.2&31.4&33.4&30.3&38.9&&79.3&85.2&86.4&82&90.3\\
			&0.1&5.2&4.4&5.3&5.3&5.5&&31.5&32.6&37.6&33.5&44.4&&86.8&88.1&91&86.6&94.6\\
			&0.5&5.3&4.8&5&4.8&5.9&&49.9&33.9&54.1&43.7&64.5&&98.3&93.1&98.8&95.2&99.6\\
			&0.9&5.1&5.1&4.8&5&12.6&&99.5&32&99.5&87&99&&100&97.1&100&100&100\\
			\\
			Laplace&-0.9&4.9&5.1&6.5&5.3&5&&19.1&27.3&29.6&30.4&28&&64.7&77.4&77.7&81.1&77.7\\
			&-0.5&4.4&5&6&5.3&5&&23.6&31.6&33.9&35.9&33.1&&71.7&81.4&81.9&86.6&82.8\\
			&-0.1&4.4&4.3&4.9&5&5&&29.1&35.1&37.9&42.6&40.1&&82.2&87.2&88.5&92&90.4\\
			&0.1&4.6&4.4&5.3&5.2&5.6&&33.5&36.9&41.6&46.2&46.3&&87.3&88.9&91.5&93.3&93.6\\
			&0.5&4.5&3.6&4.2&4.8&5.3&&53&40.3&57.4&58.9&66.3&&97.8&93.1&98.4&97.8&99.2\\
			&0.9&4.5&4.4&4.3&4.6&14.4&&98.8&38.1&98.7&91.5&96.9&&100&96.4&100&100&100\\
			\\
			Exp&-0.9&6&5.1&8.7&5.5&5.3&&18.6&30.2&32.6&39.2&30.2&&61.6&75.8&77.4&85.1&76.4\\
			&-0.5&8.9&5.3&8.7&5.2&5.3&&22.4&33.3&33.2&44.4&35.3&&68.1&81.9&82.2&89.7&83\\
			&-0.1&8.8&5.4&8&5.6&5.4&&28.5&35.8&35.6&51&42.3&&76.5&85.8&86.3&93.1&88.4\\
			&0.1&8.1&5.4&7.6&5.5&5.1&&33&36.7&38.7&53.4&47.5&&82.8&88.3&89.4&94.9&92\\
			&0.5&6.5&6.1&6.5&5.6&5.6&&52.9&38.6&54.1&63&66.5&&96&93&96.8&98&98.7\\
			&0.9&5.6&6.9&5.7&5.2&15.3&&98.4&32.2&97.4&90&96&&100&97.6&100&100&99.9\\
			\\
			Chisq&-0.9&5&5.2&6&5.4&5.3&&17.9&27.2&26.4&24.3&27.6&&58.8&75&73&70.9&75.9\\
			&-0.5&5.4&4.8&5.5&5.3&5.3&&20.7&29.5&29.1&26.9&31.9&&67&80.4&79&75.6&82.6\\
			&-0.1&5.7&5.1&5.7&5.4&5.6&&26.1&31.2&33&31.2&39.2&&79.1&86&86.6&83.2&90.1\\
			&0.1&5.3&4.8&5.3&5&5.2&&31.3&32.2&36.9&32.9&45&&85.5&88.4&90.5&86.7&94\\
			&0.5&4.9&4.7&4.8&4.9&5&&49.4&32.5&52.2&43.2&64.6&&98.1&93.8&98.6&95.2&99.6\\
			&0.9&5.2&5.2&4.9&5.2&13&&99.3&29.1&99.2&85&98.9&&100&97.6&100&100&100\\\hline	
		\end{tabular*} }
	\end{table*}

	\begin{table*}[ht]
		
		\caption{Power simulation results ($\alpha=0.05$) of the tests for different distributions under varying correlation values ($\rho$) with sample sizes $n=30$ and heteroscedastic covariance matrix $\Sigma_2$ under the MAR framework.
			\label{Powern1010hetro}}
		{\tabcolsep=4pt
			\begin{tabular*}{1.09\linewidth}{lc ccccc c ccccc  c ccccc   cc}
				Dist & $\rho$ & \multicolumn{5}{c}{$\delta=0$}  && \multicolumn{5}{c}{$\delta=0.5$} && \multicolumn{5}{c}{$\delta=1$} \\ \cline{3-7} \cline{9-13} \cline{15-19}
				&& $T_{W}^*$ & $T_{A}^*$ & $T_{M}^*$ & $T_{NC}$ & $T_{LT}$
				&& $T_{W}^*$ & $T_{A}^*$ & $T_{M}^*$ & $T_{NC}$ & $T_{LT}$ 
				&& $T_{W}^*$ & $T_{A}^*$ & $T_{M}^*$ & $T_{NC}$ & $T_{LT}$\\
				
			Normal&-0.9&5.4&5.6&6.5&5.2&5.6&&14&21.1&21.6&17.8&21.1&&43.8&60.3&58.5&52.5&60.5\\
			&-0.5&4.6&5.2&5.8&4.8&5.2&&14.9&21.5&21.8&18.1&21.9&&49.3&64.9&63.7&55.8&66\\
			&-0.1&5.1&5&5.4&4.5&5.3&&17.7&23&24.4&20&26.6&&59.6&69.8&70.5&62.2&74.3\\
			&0.1&5.1&4.7&5.2&4.5&5.1&&20.8&23.6&26.2&21.7&29.9&&66.7&72.8&75.6&66.3&80.5\\
			&0.5&5.2&4.8&5.2&5&5.9&&31.5&24.6&35&26.7&44.1&&87.2&78.3&89.7&79.1&94.4\\
			&0.9&5&5&5.3&4.6&11.8&&83.4&21.7&76.4&48.5&88.3&&100&85.1&100&98.1&100\\
			\\
			Laplace&-0.9&4.4&5.1&7.4&4.8&5.3&&14.6&21.7&24.8&22.3&21.7&&47.1&61.5&62.2&63.7&61.7\\
			&-0.5&4.7&4.9&6.7&4.9&4.8&&16.8&24.3&27&26&24.6&&53.8&67.4&68.2&71.9&67.7\\
			&-0.1&4.7&5&6.1&4.8&5.1&&20.2&26.5&29.2&31&28.7&&61.9&72&73.5&77.3&74.8\\
			&0.1&4.3&5&5.9&4.9&5.8&&22&26.6&29.8&32.1&31.9&&69.4&75&77.9&81.6&81.2\\
			&0.5&4.4&4.6&5.1&4.6&5.6&&33.8&28.1&38.4&39.9&46.8&&87.2&81.1&89.7&89.6&93.7\\
			&0.9&4.7&4.9&5.5&4.9&15.3&&82.9&25.2&77.1&66.7&85.4&&100&85.7&99.9&98.9&99.7\\
			\\
			Exp&-0.9&5.3&5.5&8.6&7.7&5.6&&16.6&24.5&28.3&38.2&24.6&&48.1&60.5&63.2&75.8&61.4\\
			&-0.5&8.4&5.8&8.5&8.2&6.2&&19.1&27.9&29.1&42.9&28.4&&50.3&66.7&67.1&80.8&67.5\\
			&-0.1&8.6&5.6&7.9&9&6.3&&24&30.4&31.2&48.8&33.8&&59.3&71.4&72.1&84.8&74.1\\
			&0.1&7.8&5.2&7.2&8.8&5.9&&26.9&30.5&32.7&50.3&37&&65.1&73.6&75.6&86.9&78.2\\
			&0.5&6.9&5.2&7&9.3&7.3&&39.3&29.5&40.2&55.8&49.1&&83&77.9&84.5&91.1&90.7\\
			&0.9&8.1&5.4&7.7&10.4&17.2&&81.4&22.5&70.3&70.9&81.7&&99.8&84.3&99.1&98.4&98.2\\
			\\
			Chisq&-0.9&5.2&5.5&6.6&5&5.6&&13.9&20.9&21.3&19.4&20.6&&42.4&58.6&57.1&55.2&59.2\\
			&-0.5&5.3&5.1&6&5&5&&16.4&23.2&23.5&21.7&23.9&&48.8&65.1&63.7&60.5&66.2\\
			&-0.1&6.1&5.5&6.1&5.4&5.8&&19.3&24.6&25.6&23.8&28.4&&58.1&69.4&69.3&65.5&74.1\\
			&0.1&5.1&5.1&5.5&4.9&5.6&&22.5&24.7&27.6&25.4&32.5&&66.2&72.9&74.8&70.5&79.9\\
			&0.5&5.5&4.8&5.5&5.1&5.8&&33.1&24.4&35.9&30.9&45.4&&85.3&78.9&88&80.8&93.2\\
			&0.9&5.7&5.1&5.6&5.1&13.3&&82.8&20.6&74.5&51.9&86.5&&100&85.2&99.9&97.2&99.8\\\hline
				
			\end{tabular*} }
		\end{table*}


\begin{figure}[p] 
	\begin{center}
		
		\includegraphics[scale=0.6]{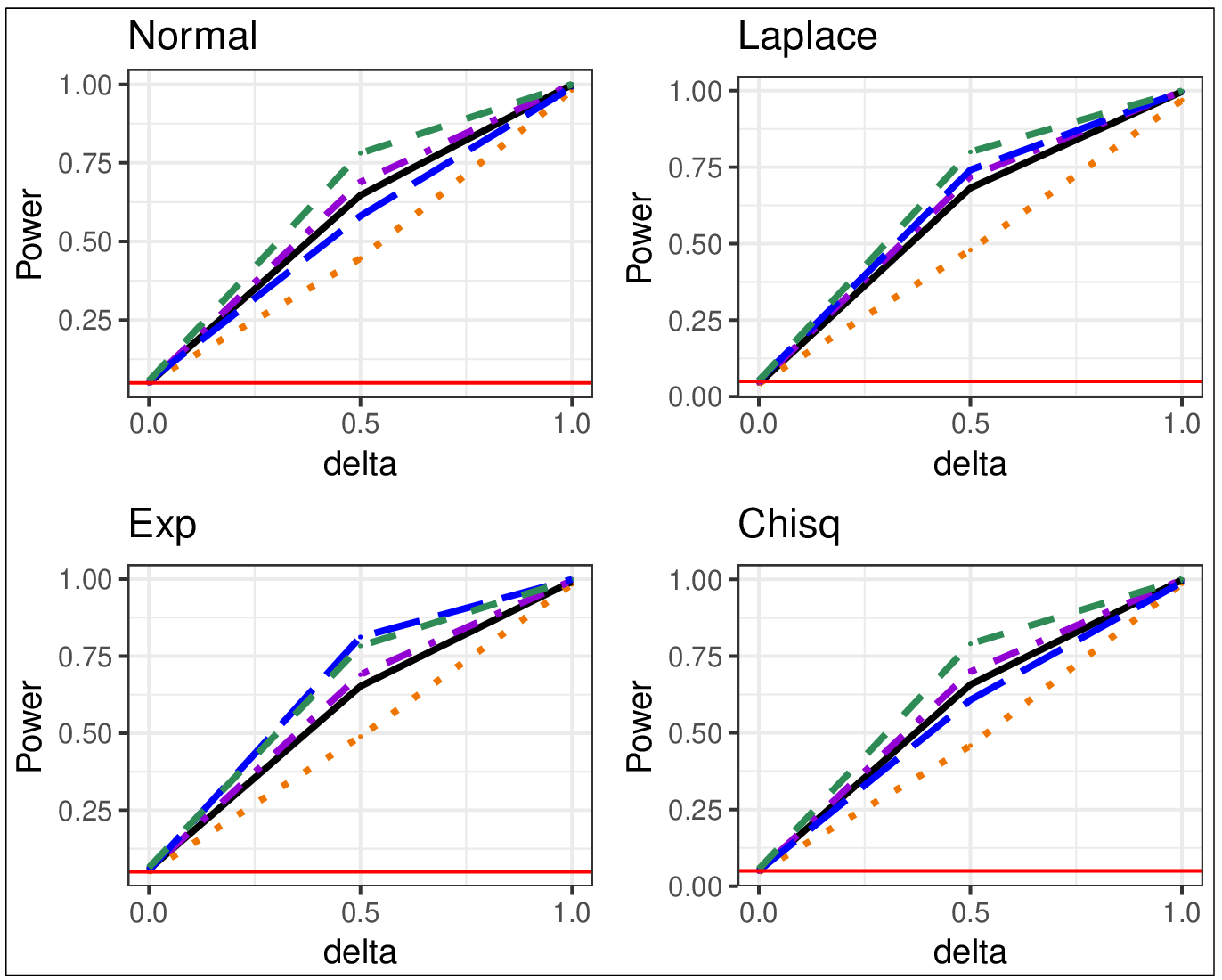}
		
	\end{center}
	\caption{Power simulation results ($\alpha=0.05$) of the tests  $T_W^*$ $({\color{black}\textendash\textendash\textendash})$, $T_A^*$ $({\color{brown}\cdots})$, 	$T_M^*$ $({\color{violet}-\cdot-})$, $T_{NC}$ $({\color{blue} \textendash\textendash \quad \textendash\textendash})$,  and  $T_{LT}$ $({\color{ao(english)} -- })$ for different distributions under  correlation value ($\rho=0.5$) with sample sizes $(nc,nu)=(30,10)$ and homoscedastic covariance matrix $\Sigma_1$ under the MCAR framework.}
	\label{fig:multi}	
	
\end{figure}

\begin{figure}[h]
	\begin{center}
		
		\includegraphics[scale=0.6]{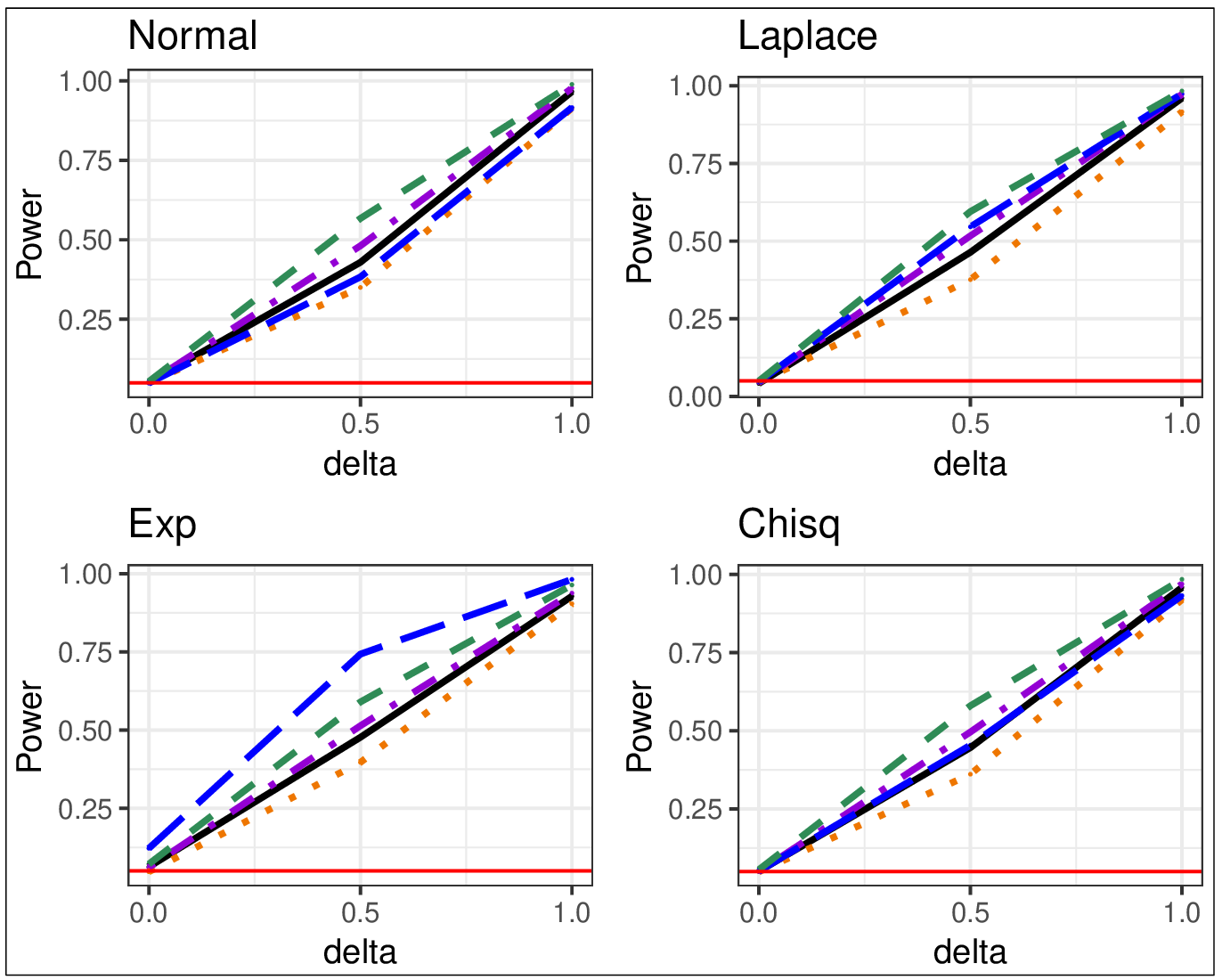}
		
	\end{center}
	\caption{Power simulation results ($\alpha=0.05$) of the tests  $T_W^*$ $({\color{black}\textendash\textendash\textendash})$, $T_A^*$ $({\color{brown}\cdots})$, 	$T_M^*$ $({\color{violet}-\cdot-})$, $T_{NC}$ $({\color{blue} \textendash\textendash \quad \textendash\textendash})$,  and  $T_{LT}$ $({\color{ao(english)} -- })$ for different distributions under  correlation value ($\rho=0.5$) with sample sizes $(nc,nu)=(30,10)$ and heteroscedastic covariance matrix $\Sigma_2$ under the MCAR framework.}
	\label{fig:multi}	
	
\end{figure}

				\begin{figure}[h]
					\begin{center}
						
						\includegraphics[scale=0.6]{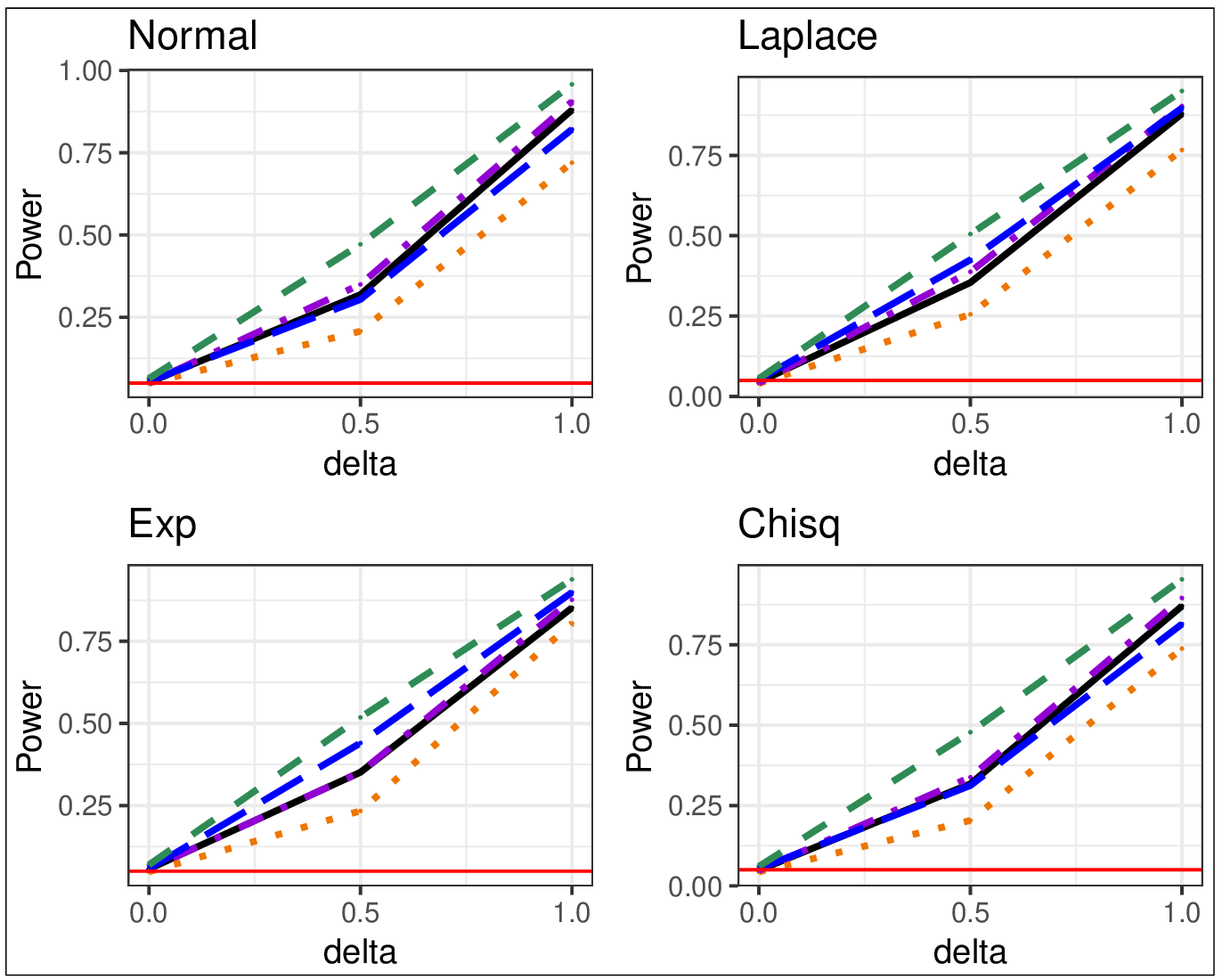}
						
					\end{center}
					\caption{
						Power simulation results ($\alpha=0.05$) of the tests  $T_W^*$ $({\color{black}\textendash\textendash\textendash})$, $T_A^*$ $({\color{brown}\cdots})$, 	$T_M^*$ $({\color{violet}-\cdot-})$, $T_{NC}$ $({\color{blue} \textendash\textendash \quad \textendash\textendash})$,  and  $T_{LT}$ $({\color{ao(english)} -- })$ for different distributions under  correlation value ($\rho=0.5$) with sample sizes $n=20$ and homoscedastic covariance matrix $\Sigma_1$ under the MAR framework.}
					\label{fig:multi}	
					
				\end{figure}

				\begin{figure}[h]
					\begin{center}
						
						\includegraphics[scale=0.6]{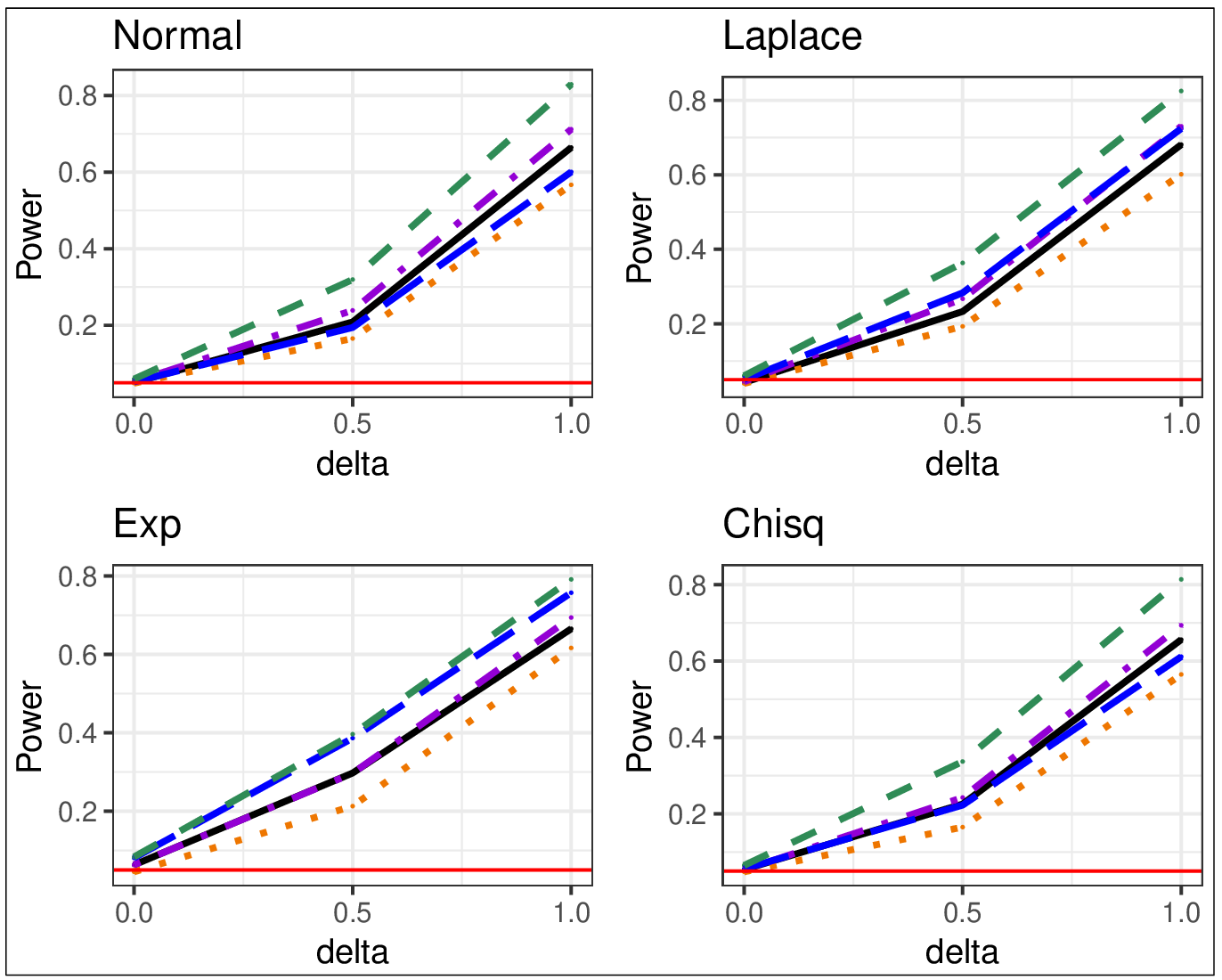}
						
					\end{center}
					\caption{
						Power simulation results ($\alpha=0.05$) of the tests  $T_W^*$ $({\color{black}\textendash\textendash\textendash})$, $T_A^*$ $({\color{brown}\cdots})$, 	$T_M^*$ $({\color{violet}-\cdot-})$, $T_{NC}$ $({\color{blue} \textendash\textendash \quad \textendash\textendash})$,  and  $T_{LT}$ $({\color{ao(english)} -- })$ for different distributions under  correlation value ($\rho=0.5$) with sample sizes $n=20$ and heteroscedastic covariance matrix $\Sigma_2$ under the MAR framework.}
					\label{fig:multi}	
					
				\end{figure}

\clearpage

	\begin{table*}[ht]%
		\centering
						   \captionsetup{justification=centering}

		\caption{Adjusted two-sided P-values of  the breast cancer study based on Holm's method.\label{Adjpvalue}}%
		\begin{tabular*}{0.7\textwidth}{@{\extracolsep\fill}lccccc@{\extracolsep\fill}}
			\textbf{Gene}& $\boldsymbol{T}_{\mathbf{W}}^* $  & $\boldsymbol{T}_{\mathbf{A}}^*$ & $\boldsymbol{T}_{\mathbf{M} }^*$& $\boldsymbol{T}_\textbf{NC}$& $\boldsymbol{T}_\textbf{LT}$ \\
			\hline 
			\textbf{TP53}&1&1&1&1&1\\
			\textbf{ABCC1}&0&0.008&0&0.0152&0.021\\
			\textbf{HRAS}&0.035&0.008&0.021&0.0238&0.0088\\
			\textbf{GSTM1}&1&1&1&1&1\\
			\textbf{ERBB2}&0.234&0.21&0.084&0.108&0.4272\\
			\textbf{CD8A}&1&1&1&1&1\\
			\textbf{C1D}&1&1&1&1&1\\
			\textbf{GBP3}&1&1&1&0.647&0.5135\\
			
			\bottomrule
		\end{tabular*}
		
	\end{table*}

\bibliographystyle{SageV}
\bibliography{reference}{}